\documentclass[11pt]{article}%
\usepackage{amsfonts}
\usepackage{xcolor}
\colorlet{linkequation}{green}
\usepackage{amsmath}%
\usepackage{amssymb}%
\usepackage{graphicx}
\usepackage{float}
\usepackage{longtable}
\usepackage[utf8]{inputenc}
\usepackage{hyperref}
\usepackage{natbib} 
\usepackage{mathdots}
\usepackage{upgreek}
\usepackage{amsmath}
\usepackage{algorithmicx}
\usepackage[ruled,vlined]{algorithm2e}
\SetAlgoCaptionSeparator{ }

\usepackage{enumerate}
\usepackage{fullpage}
\usepackage{authblk}

\newcommand{\fref}[1]{Fig.~\ref{#1}}

\newcommand{\tref}[1]{Table~\ref{#1}}
\newcommand{\sref}[1]{Section~\ref{#1}}

\newcommand*{\cref}[1]{%
  \begingroup
    \hypersetup{
      linkcolor=linkequation,
      linkbordercolor=linkequation,
    }%
    \ref{#1}%
  \endgroup
}

\providecommand{\U}[1]{\protect\rule{.1in}{.1in}}
\newtheorem{theorem}{Theorem}

\newtheorem{lemma}{Lemma}

\newtheorem{proposition}{Proposition}

\newenvironment{proof}[1][Proof]{\textbf{#1.} }{\ \rule{0.5em}{0.5em}}
\begin{document}

\title{Low-Dimensional Reconciliation for Continuous-Variable Quantum Key Distribution}
\author[1,2,3]{Laszlo Gyongyosi\footnote{Email: \href{mailto:l.gyongyosi@soton.ac.uk}{l.gyongyosi@soton.ac.uk}}}
\author[2]{Sandor Imre}
\affil[1]{School of Electronics and Computer Science, University of Southampton, Southampton, SO17 1BJ, UK}
\affil[2]{Department of Networked Systems and Services, Budapest University of Technology and Economics, Budapest, H-1117 Hungary}
\affil[3]{MTA-BME Information Systems Research Group, Hungarian Academy of Sciences, Budapest, H-1051 Hungary}
\date{}

\maketitle

\vspace{-0.5cm}
\begin{abstract}
We propose an efficient logical layer-based reconciliation method for continuous-variable quantum key distribution (CVQKD) to extract binary information from correlated Gaussian variables. We demonstrate that by operating on the raw-data level, the noise of the quantum channel can be corrected in the low-dimensional (scalar) space and the reconciliation can be extended to arbitrary dimensions. The CVQKD systems allow an unconditionally secret communication over standard telecommunication networks. To exploit the real potential of CVQKD a robust reconciliation technique is needed. It is currently unavailable, which makes it impossible to reach the real performance of the CVQKD protocols. The reconciliation is a post-processing step separated from the transmission of quantum states, which is aimed to derive the secret key from the raw data. The reconciliation process of correlated Gaussian variables is a complex problem that requires either tomography in the physical layer that is intractable in a practical scenario, or high-cost calculations in the multidimensional spherical space with strict dimensional limitations. To avoid these issues we define the low-dimensional reconciliation. We prove that the error probability of one-dimensional reconciliation is zero in any practical CVQKD scenario, and provides unconditional security. The results allow to significantly improve the currently available key rates and transmission distances of CVQKD.
\end{abstract}

\section{Introduction}
\label{sec1}
 The QKD (Quantum Key Distribution) systems represent one of the most important practical applications of quantum information theory [\cref{r1}-\cref{r11}], [\cref{r49}-\cref{r53}]. The QKD schemes allow to establish an unconditionally secret communication between distant parties by exploiting the fundamental attributes of quantum mechanics [\cref{r10}-\cref{r14}], [\cref{r34}-\cref{r42}], [\cref{r46}-\cref{r53}]. The QKD protocols can be classified into three main classes [\cref{r1}-\cref{r11}], [\cref{r49}-\cref{r53}]: DVQKD (Discrete-Variable), CVQKD (Continuous-Variable) and DPR-QKD (Differential Phase Reference) systems. The firstly introduced QKD protocols were based on discrete variables, such as photon polarization. Since the polarization of single photons cannot be encoded and decoded efficiently because of the technological limitations of current physical devices, the CVQKD systems were proposed. In a CVQKD system, the information is encoded on continuous variables by a Gaussian modulation, such as in the position or momentum quadratures of coherent states. In comparison to DVQKD, the modulation and decoding of continuous variables does not require specialized devices and can be implemented efficiently by standard technologies that are available and in widespread use. The CVQKD systems also provide higher secret key rates and higher communication distances. The CVQKD protocols can be further classified into \textit{one-way} and \textit{two-way} systems. In a one-way CVQKD system, Alice, the sender transmits her continuous variables to the receiver, Bob, over a quantum channel [\cref{r9}-\cref{r11}]. In a two-way system, Bob starts the communication, Alice adds her internal secret to the received message, and this is then sent back to Bob (e.g., one mode of the coupled beam that is outputted from a beamsplitter is transmitted back to Bob). The two-way CVQKD systems were introduced for practical reasons to exceed the limitations of one-way CVQKD, such as low key rates and short communication distances [\cref{r1}-\cref{r13}]. The two-way CVQKD protocols exploit the benefits of multiple channel uses and allow the leak of only lower valuable information to the eavesdropper. On the other hand, the achievable distances of one-way CVQKD can be extended by efficient channel-estimation methods [\cref{r36}], which is important since the one-way protocol currently is still the focus of the research owing to the easy experimental implementation.

The CVQKD schemes use continuous-variable Gaussian modulation which provably provides optimal key rates against collective attacks at finite-size block lengths [\cref{r1}-\cref{r11}] and also maximizes the mutual information between Alice and Bob. The security of CVQKD has also been proven against collective attacks in the asymptotic regime with infinite block sizes, and against arbitrary attacks in the finite-size regime [\cref{r9}, \cref{r13}], [\cref{r39}-\cref{r40}]. One of the most critical points in regard to CVQKD is the post-processing [\cref{r1}-\cref{r11}], [\cref{r47}]. The post-processing is aimed to correct the errors of the quantum channel that are cumulated in the raw data. The raw data is a correlated binary bitstring at Alice's and Bob's side, generated by the random quadrature measurements at the parties. Each quadrature measurement results in a unit in the raw data. The raw data itself is not a secret key; it consists only of the results of the random quadrature measurements. The secret key is a uniformly distributed long binary string that will be combined with the raw data elements, and will be added to the picture only in the stage of logical layer manipulations. The logical layer-based post-processing phase uses purely classical tools: precisely a classical-authenticated communication channel and classical error-correction algorithms. This method basically does the same in the logical layer as the tomography does in the physical layer, and it consists of two main phases: the reconciliation procedure with several error-correction steps, and privacy amplification. Without loss of generality, the aim of reconciliation is to extract as much valuable information from the correlated raw data as possible and to generate an error-free key between Alice and Bob. The privacy amplification operates on the shared, error-corrected common secret to extract the final key between the parties, and the aim of this phase is to reduce to zero the possible knowledge of an eavesdropper from the elements of the key. The implementation of tomography in the physical layer is a complex problem, and it is intractable in a practical scenario. But, well-characterized solutions can be proposed in the logical layer for the same purpose of giving an analogous, and also more valuable answer to the reconciliation of correlated Gaussian variables than the physical-layer tomography ever could. The theoretical background that makes the logical layer-based reconciliation possible also allow us to view the noisy physical quantum channel as a binary Gaussian channel in the logical layer [\cref{r1}-\cref{r13}]. This has the immediate consequence that very efficient binary error-correction tools can be integrated from the world of traditional communication theory into CVQKD---which would not be available for the physical-layer tomography to extract binary information from the correlated Gaussian variables.

The raw data shared over the quantum channel is noisy, and this must be corrected to distill the final secret key. Since a large amount of raw data bits have to be shared between the parties, the complexity of the post-processing phase is a critical point in CVQKD protocols, and it has to be in order to be as low as possible. The existing logical layer-based solutions require high-complexity calculations in the high-dimensional spherical space for the reconciliation of Gaussian variables [\cref{r9}-\cref{r11}]. Since a complex reconciliation is so undesirable, the aim is to find a more efficient solution in the logical layer. A slice method is a different reconciliation approach, which is also used in the current reconciliation steps of CVQKD for short distances, and can be implemented without spherical operations [\cref{r41}]. Basically, the error correction in the reconciliation phase consists of two phases: First, the binary-channel codes (such as LDPC -- Low Density Parity Check, turbo codes, polar codes, etc. [\cref{r22}-\cref{r35}]) that are used for the transmission of the classical bits in the reconciliation phase are corrected. Second, the real Gaussian noise on the received raw-data vector must be corrected, which noise arises from the effect of the quantum channel (i.e., from Eve's optimal Gaussian attack, which is considered in CVQKD protocols [\cref{r1}-\cref{r11}]). In this work we focus on the second phase of reconciliation, which has crucial role in CVQKD, since this phase makes it possible to correct the errors incurred on the quantum channel and to share an error-free key between Alice and Bob. Since the raw data is formulated by continuous real numbers resulted from quadrature measurements at the parties, the reconciliation problem is analogous to the well-known subject of binary-channel coding that operates on binary-channel codes. It also follows that the complicated and difficult to implement physical-layer tomography can be replaced in the logical level by binary error-correction schemes that are easier to implement. According to a critical security requirement of QKD, in the reconciliation phase only uniform distribution can be transmitted over the classical channel, otherwise the information theoretic security of the protocol cannot be proven [\cref{r1}-\cref{r13}]. The raw data itself follows Gaussian random distribution because these arise from a Gaussian random source; however, by applying some trivial operations on the raw data units, the desired uniform distribution can be reached, and the reconciliation can be performed with unconditional security, as we will show in detail in \sref{sec3}.

A relevant difference of DV and CV protocols is that the physical quantum channel that connects the parties is characterized in a different way. For DVQKD the appropriate channel model is the Binary Symmetric Channel (BSC), which allows the use of the well-known channel-coding and error-correction tools in the post-processing phase. It also follows that for DVQKD there is a clear connection between the characteristics of the quantum channel and the world of traditional communication theory. On the other hand, for a CVQKD system the situation is more complicated, because the proper description of a Gaussian quantum channel requires several physical parameters (transmittance, variance, shot noise, excess noise, etc.) which allows no to draw a clear connection. To solve the situation for one-way CVQKD, the multidimensional reconciliation schemes [\cref{r9}-\cref{r12}] have been introduced, which made possible the conversion of the physical AWGN (Additive White Gaussian Noise) quantum channel to a logical binary AWGN (BAWGN) channel, where the Gaussian random noise arises directly from the quantum-level transmission. Precisely, it works only for low dimensions and the resulted logical channel approximates only a binary Gaussian channel. As the accuracy of the physical-logical channel conversion gets closer to perfect the resulting logical channel gets closer to a binary Gaussian channel. At low SNRs (Signal-to-Noise Ratio) the capacities of the Gaussian quantum channel and the binary Gaussian channel coincidence, and this is particularly convenient because for low SNRs the problem of channel conversion can be reduced to the approximation of a binary Gaussian channel. From this follows, that the efficiency of the channel conversion procedure can be described by the relevant parameters of the resulting logical binary channel (such as its variance and capacity). This conversion efficiency has tremendous importance because it also determines the efficiency of the reconciliation process, i.e., the performance of the protocol. In the multidimensional reconciliation the conversion procedure required the use of the spherical space and its sophisticated operations [\cref{r9}-\cref{r11}], which is a complex process. The difficult computational steps of post-processing just cause further slowing down in the very sensitive key rates that are so difficult to establish. These requirements of the reconciliation phase are strongly undesired in a practical CVQKD scenario, so a simpler reconciliation would be desirable---for both one- and two-way systems. The problem of efficient post-processing is more crucial for two-way CVQKD, due to its more complex physical architecture. 

To exploit the real potential of two-way CVQKD systems, efficient post-processing is needed. It is still missing, which makes it not possible to attain the true performance of two-way CVQKD. This is the main reason why the theoretical maximum of key rates and ranges cannot be exceeded in the current practical scenarios; however, the protocol in its `hardware level' is built to be strong,  and would be capable of more performance than is currently available. To boost up the performance of the two-way CVQKD protocols over the current limits, we introduce an efficient reconciliation method that makes it possible to increase the key rates and to extend the currently available distance ranges. The mathematical apparatus that stands behind the multidimensional reconciliation puts a strict upper bound on the available dimensions, and limits its maximum [\cref{r9}-\cref{r11}], [\cref{r42}]. The reason is that in higher dimensions the required spherical division operations do not exist. In our scheme, we also eliminate this serious drawback and extend the reconciliation of Gaussian variables to arbitrary high dimensions. The proposed approach also makes possible to get a closer and more precise approximation of the binary Gaussian channel, in comparison to the multidimensional case. 

Since the post-processing phase uses the binary form of the continuous variables, in fact, we do not have to decode the Gaussian variables in the multidimensional space. As a corollary, arbitrary high-precision approximation of the logical binary Gaussian channel can be made in the non-spherical space by using considerable dimensions. We exploit it in this work to construct a scalar reconciliation that breaks with the traditions of the previously introduced approaches [\cref{r9}-\cref{r11}], [\cref{r42},\cref{r46}], and uses only the space of scalar variables. The proposed scalar reconciliation is also able to transform the physical Gaussian quantum channel into a logical binary Gaussian channel in two-way CVQKD, and the same benefits can be exploited as in the case of multidimensional reconciliation. However since our scheme is not limited to eight dimensions, an arbitrary precision can be reached in the approximation of the logical binary Gaussian channel. As follows, the accuracy of the conversion between the physical Gaussian quantum channel and the logical Gaussian channel can be improved beyond the current limits. Another issue in the current approaches is the requirement of spherical calculations. To make the existing post-processing approaches more efficient, we have to eliminate the multidimensional operations. The reconciliation of Gaussian variables would be much easier, if we found a solution that would make it possible to extract the final key from the noisy data by simple calculations in the level of scalar space. It immediately follows that this would significantly increase the efficiency of the reconciliation process, and would lead to a negligible complexity and computational power in the error-correction procedure. 

In this paper we define \textit{low-dimensional} (\textit{scalar) reconciliation} for CVQKD. It brings significantly higher noise-resistance and information-transmission capability, extended transmission distances, and improved key rates. The proposed method does the reconciliation of Gaussian variables without the need of any physical-layer tomography or multidimensional operations. We demonstrate the results for two-way CVQKD. The scheme is backward compatible it also can be applied to one-way CVQKD. 

The novel contribution of our paper is as follows:
\begin{itemize}
\item  \textit{The reconciliation process of correlated Gaussian variables is a complex problem that requires either tomography in the physical layer that is intractable in a practical scenario, or high-cost calculations in the multidimensional spherical space with strict dimensional limitations.}
\item \textit{To avoid these issues, we propose an efficient logical layer-based reconciliation method for CVQKD to extract binary information from correlated Gaussian variables.}
\item \textit{We demonstrate that by operating on the raw-data level, the noise of the quantum channel can be corrected in the low-dimensional scalar space and the reconciliation can be extended to arbitrary dimensions.}
\item \textit{We prove that the error probability of scalar reconciliation is zero in any practical CVQKD scenario, and provides unconditional security.}
\item \textit{The results allow to significantly improve the currently available key rates and transmission distances of CVQKD.}
\end{itemize}

This paper is organized as follows. In \sref{sec2}, preliminary findings are summarized. In \sref{sec3}, we introduce the reconciliation scheme. \sref{sec4} provides the theorems and proofs. In \sref{sec5}, a numerical evidence is proposed. Finally, in \sref{sec6}, we conclude the paper. Supplemental information is included in the Appendix.
 
\section{System Model}
\label{sec2}
 In comparison to one-way CVQKD protocols, in two-way CVQKD the two uses of the quantum channel lead to superadditive private classical capacity (more precisely, the superadditivity of security threshold leads to a subadditive eavesdropper [\cref{r1}-\cref{r8}], [\cref{r14}]), which makes it possible to decrease the amount of valuable information leaked to Eve. The subadditive eavesdropper is a consequence of the multiple uses of the quantum channel. The superadditivity of the security threshold can also be expressed in terms of tolerable excess noise and the channel transmission [\cref{r1}]. In the two-way scenario, Eve perturbs the quantum channel ${\mathcal{N}}_{\mathrm{1}},$ which causes a noise in the transmission that will have an effect on the success of her second attack. From the two attacks, comparatively lower valuable information will be available to Eve so that she would not have made an attack on ${\mathcal{N}}_{\mathrm{1}}$. The reason for this is that the amount of valuable information transmitted over ${\mathcal{N}}_{\mathrm{2}}$ is already decreased by the attack of ${\mathcal{N}}_{\mathrm{1}}$. More attacks add more noise into the transmission, which also decreases the amount of mutual information between Alice and Bob. With the increased number of channel uses we allow Eve to get as much less valuable information as possible. If Alice encodes her information into the noisy state that is received from ${\mathcal{N}}_{\mathrm{1}}$, and then sends it back to Bob over ${\mathcal{N}}_{\mathrm{2}}$, then the parties can achieve the desired phenomenon of superadditivity [\cref{r1}-\cref{r4}]. The amount of valuable information leaked to Eve is also decreased by the multiple uses of the quantum channel. The errors caused by more channel uses can be corrected in the reconciliation phase by traditional error-correction tools. In fact, by utilizing multiple channel uses, we `set a trap' for Eve, since again and again she will attack the quantum channel. Eve will also simultaneously decrease the amount of eavesdropped information by her actions. The idea works well, because in the post-processing phase the parties can correct the errors caused by Eve, thus, finally, it can be concluded that it was a correct decision to increase the number of channel uses. Of course, if we had perfect amplifiers and ideal devices, then, in theory, it would be possible to completely eliminate Eve from the picture in the asymptotic scenario to make unnecessary the privacy amplification by allowing an infinite amount of channel uses to maximally exploit the superadditivity property (more precisely, the superadditivity of the security-threshold parameter hence the strong subadditivity of Eve). However, in practice it is trivially not possible to circulate over and over the same beam an infinite amount of times, due to the losses and imperfections of the physical devices. 

 Let us review the data components of the protocol that are needed for the appropriate description of the scalar reconciliation for the two-way CVQKD protocol. Our description will be as detailed as desired for further analysis, and will not take into account the particular description of any components of an experimental protocol. The \textit{raw data }is generated by the use of noisy Gaussian channels ${\mathcal{N}}_{\mathrm{1}}$ and ${\mathcal{N}}_{\mathrm{2}}$, and by the parties' internal secrets. The aim of the quantum-level transmission is to generate two nearly identical classical bitstrings between the parties. All quantum-level interactions are closed at this point, and the post-processing phase, which uses the raw data of the parties and a classical authenticated channel, is brought to life. The post-processing phase consists of the processes of reconciliation and privacy amplification. The valuable key will be generated in the reconciliation phase by using the raw data and a random secret. It consists of error-correction phases as well. The privacy amplification is geared toward performing security checks on the elements of the generated key, and it is not part of our description. We will assume reverse reconciliation (RR), which is desirable since the mutual information between Bob and Eve is provably lower than between Alice and Eve  [\cref{r1}-\cref{r6}],  [\cref{r9}-\cref{r14}],  [\cref{r50}]. It is because if Bob starts to run the reconciliation phase using his already noisy raw data, then only lower valuable information can be leaked to Eve during the procedure in comparison to if Alice would have started to run the reconciliation, from her ideal raw data (from the perspective of the raw data-level reconciliation, the noise that arises from the first channel use has no relevance, as will be clarified later, and Alice's raw data can be viewed as ideal). 

 The run of the protocol is sketched as follows. Let us denote Alice's binary raw data by $X$, and Bob's binary raw data by $X\mathrm{'}$, where $\left|X\right|\mathrm{=}\left|X\mathrm{'}\right|\mathrm{=}N$ units. Alice's raw data is generated by a random quadrature measurement of $M_{\mathrm{1}}$. Alice's selects two random variables \textit{x} and \textit{p} each drawn from a Gaussian distribution, that encodes her position and momentum quadratures and obtains a phase space vector $S_{Alice}\mathrm{=}\left|\left.x_A\mathrm{+}ip_A\right\rangle \right.$. Bob also draws a phase space vector $S_{Bob}\mathrm{=}\left|\left.x_B\mathrm{+}ip_B\right\rangle \right.$. The noisy ${{{S}'_{Bob}}}$ is received by Alice in the first phase via channel ${\mathcal{N}}_{\mathrm{1}}$ in the beam $B_{out}$. Alice's raw data is defined as follows:\textbf{}
\begin{equation} \label{1)} 
X\mathrm{\equiv }M_{\mathrm{1}}\left(B_{out}\mathrm{+}S_{Alice}\right)\mathrm{=}{\mathcal{N}}_{\mathrm{1}}\left(S_{Bob}\right)\mathrm{+}S_{Alice}.             
\end{equation} 
The outgoing beam $A_{out}$ will contain the other mode of the coupled beam. Bob's raw data is generated by the $M_{\mathrm{2}}$ random quadrature measurement applied on the beam $A_{out}$, as:
\begin{equation} \label{2)} 
X\mathrm{'}\mathrm{\equiv }M_{\mathrm{2}}\left(A_{out}\right)\mathrm{=}{{{B}'_{out}}}\mathrm{+}{{{S}'_{Alice}}}\mathrm{=}{\mathcal{N}}_{\mathrm{2}}\left({\mathcal{N}}_{\mathrm{1}}\left(S_{Bob}\right)\right)\mathrm{+}{\mathcal{N}}_{\mathrm{2}}\left(S_{Alice}\right), 
\end{equation} 
where $A_{out}$ contains the noisy version of the second mode of the beam. A detailed description will be given in \sref{sec2_1}. 

\begin{center}
\begin{figure}[h!]
\begin{center}
\includegraphics[angle = 0,width=0.7\linewidth]{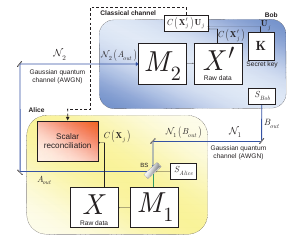}
\caption{Simplified model of a PM-RR two-way CVQKD protocol with the scalar reconciliation. The modulated Gaussian variables are sent through a Gaussian quantum channel (AWGN) depicted by ${\mathcal{N}}_{\mathrm{1}}$ and ${\mathcal{N}}_{\mathrm{2}}$ (same physical link). The classical channel is depicted by the dashed line. Bob sends $S_{Bob}$ to Alice over ${\mathcal{N}}_{\mathrm{1}}$. Alice adds to it her secret $S_{Alice}$ by a BS, and applies measurement $M_{\mathrm{1}}$, which defines her raw data $X\mathrm{=}M_{\mathrm{1}}\left({\mathcal{N}}_{\mathrm{1}}\left(S_{Bob}\right)\mathrm{+}S_{Alice}\right)$. The other mode is sent back to Bob over ${\mathcal{N}}_{\mathrm{2}}$, who applies $M_{\mathrm{2}}$, which results in his $X\mathrm{'}\mathrm{=}M_{\mathrm{2}}\left({\mathcal{N}}_{\mathrm{2}}\left({\mathcal{N}}_{\mathrm{1}}\left(S_{Bob}\right)\mathrm{+}S_{Alice}\right)\right)$.} 
 \label{fig1}
 \end{center}
\end{figure}
\end{center}

 A simplified view of a PM (Prepare-and-Measure: entanglement-free) two-way CVQKD protocol with homodyne measurements $M_{\mathrm{1}}$, $M_{\mathrm{2}}$ at the parties and with RR is shown in \fref{fig1}. Alice and Bob are connected by a noisy quantum channel and a classical authenticated channel. The quantum communication is started by Bob. Alice receives Bob's quantum message and then couples it with her quantum message using a BS (Beam Splitter) to create a correlated signal. The first mode of the beam is measured by Alice, using a random quadrature measurement; the second mode is sent back to Bob, who will also apply a random quadrature measurement on the received beam. After the measurements have been performed, the parties inform each other about the used position and momentum quadratures over the classical channel, and discard the irrelevant data. The resulted raw data is a collection of correlated Gaussian variables. Since these binary strings follow Gaussian random distribution, they cannot be transmitted directly over the classical channel. In reverse reconciliation, Bob has to make the probability distribution of his raw data to uniform. He can do this by applying an appropriate function $C\left(\mathrm{\cdot }\right)$ (will be clarified in \sref{sec3}) on his \textit{j}-th raw data block, denoted by ${{{\boldsymbol{\mathrm{X}}}'_{j}}}$. Bob then generates a random key ${\boldsymbol{\mathrm{U}}}_j$ (the full key vector $\boldsymbol{\mathrm{K}}$ is granulated into several ${\boldsymbol{\mathrm{U}}}_j$-s), and multiplies it with his raw data $C\left({{{\boldsymbol{\mathrm{X}}}'_{j}}}\right)$. Alice receives $C\left({{{\boldsymbol{\mathrm{X}}}'_{j}}}\right){\boldsymbol{\mathrm{U}}}_j$, and using her $C\left({\boldsymbol{\mathrm{X}}}_j\right)$, she computes the noisy ${{{\boldsymbol{\mathrm{U}}}'_{j}}}$. Next, the errors of the secret key that arise from the noise of the quantum channel will be corrected. This phase is modeled by the scalar reconciliation box at Alice's side. The aim of the scalar reconciliation is to share an error-free key $\boldsymbol{\mathrm{K}}$ between Alice and Bob. From Alice, it requires the correction of the noise on ${{{\boldsymbol{\mathrm{U}}}'_{j}}}$ to get back Bob's ${\boldsymbol{\mathrm{U}}}_j$, using only scalar operations without the need of the multidimensional spherical space.
 
\subsection{Coding Scheme}
\label{sec2_1}
 In the following description we give a considerable view of the coding of two-way CVQKD, focusing on the contributions of information theory. Let us denote the quadratures of the \textit{i}-th signal $S_{Alice,i}$ in the phase space ${\mathcal{S}}_A$ by $x_{A,i},p_{A,i}$, and the quadratures of Bob's signal $S_{Bob,i}$ in the phase space ${\mathcal{S}}_B$ by $x_{B,i},p_{B,i}$, where  $x_{A,i},p_{A,i}\mathrm{\in }\mathcal{N}\left(0,{\sigma }^{\mathrm{2}}_{\omega }\right)$ and $x_{B,i},p_{B,i}\mathrm{\in }\mathcal{N}\left(0,{\sigma }^{\mathrm{2}}_{\omega }\right)$ are drawn from a Gaussian random distribution with mean $\mu \mathrm{=0}$, and variance ${\sigma }^{\mathrm{2}}_{\omega }$, where ${\sigma }^{\mathrm{2}}_{\omega }$ is the modulation variance  [\cref{r1}-\cref{r10}]. 

 The coherent states $S_{Alice,i}\mathrm{=}\left|\left.x_{A,i}\mathrm{+}ip_{A,i}\right\rangle \right.\mathrm{\in }{\mathcal{S}}_A$ and $S_{Bob,i}\mathrm{=}\left|\left.x_{B,i}\mathrm{+}ip_{B,i}\right\rangle \right.\mathrm{\in }{\mathcal{S}}_B$ are encoded by Gaussian modulation with dedicated centers $\left(x_{A,i},p_{A,i}\right)\mathrm{\in }{\mathcal{S}}_A$ and $\left(x_{B,i},p_{B,i}\right)\mathrm{\in }{\mathcal{S}}_B$, respectively (\textit{Note}: Each $S_i$ define a zero-mean, circular symmetric complex Gaussian random variable $\mathcal{C}\mathcal{N}\left(0,{\sigma }^{\mathrm{2}}_{S_i}\right)$ with variance ${\sigma }^{\mathrm{2}}_{S_i}\mathrm{=}\mathbb{E}\left[{\left|S_i\right|}^{\mathrm{2}}\right]$ in the phase space $\mathcal{S}$, with i.i.d. real and imaginary components $x_i,p_i\mathrm{\in }\mathcal{N}\left(0,{\sigma }^{\mathrm{2}}_{\omega}\right)$, thus ${\sigma }^{\mathrm{2}}_{S_i}\mathrm{=2}{\sigma }^{\mathrm{2}}_{\omega }$. The squared magnitude ${\left|S_i\right|}^{\mathrm{2}}$, ${\left|S_i\right|}^{\mathrm{2}}\mathrm{\ge }\mathrm{0}$ is exponentially distributed with density $f\left({\left|S_i\right|}^{\mathrm{2}}\right)\mathrm{=}{\mathrm{1}}/{{\sigma }^{\mathrm{2}}_{S_i}}\mathrm{exp}\left({\mathrm{-}{\left|S_i\right|}^{\mathrm{2}}}/{{\sigma }^{\mathrm{2}}_{S_i}}\right)$. The two beams are correlated at Alice's BS, which results in a combined signal in the combined phase space ${\mathcal{S}}_{A\mathrm{\times }B}$. The modulation noise $\mathrm{\partial }\mathrm{\in }\mathcal{C}\mathcal{N}\left(0,{\sigma }^{\mathrm{2}}_{\mathrm{\partial }}\right)$, is precisely centered around $\left(x_{A,i}\mathrm{+}x_{B,i},p_{A,i}\mathrm{+}p_{B,i}\right)\mathrm{\in }{\mathcal{S}}_{A\mathrm{\times }B}$ and $\left(x_{A,i}\mathrm{-}x_{B,i},p_{A,i}\mathrm{-}p_{B,i}\right)\\\mathrm{\in }{\mathcal{S}}_{A\mathrm{\times }B}$ in ${\mathcal{S}}_{A\mathrm{\times }B}$. After the two beams $S_{Alice,i}$ and ${{{S}'_{Bob,i}}}$ are correlated at a BS at Alice's side, where ${{{S}'_{Bob,i}}}$ is the noisy version of $S_{Bob,i}$, Alice applies a random quadrature measurement $M_{\mathrm{1}}$ on the first mode of the beam, while the second mode is transmitted back to Bob over quantum channel ${\mathcal{N}}_{\mathrm{2}}$. Alice's state in the combined phase space ${\mathcal{S}}_{A\mathrm{\times }B}$ is as follows:
\begin{equation} \label{3)} 
\left|\left.{\varphi }_i\right\rangle \right.\mathrm{=}\left|\left.x_{A,i}\mathrm{+}{{{x}'_{B,i}}}\mathrm{+}i\left(p_{A,i}\mathrm{+}{{{p}'_{B,i}}}\right)\right\rangle \right.\mathrm{\in }\mathcal{C}\mathcal{N}\left(0,{\sigma }^{\mathrm{2}}_{{\varphi }_i}\right)\mathrm{\in }{\mathcal{S}}_{A\mathrm{\times }B},                      
\end{equation} 
with Gaussian random quadrature components $\mathcal{N}\left(\mathrm{0,2}{\sigma }^{\mathrm{2}}_{\omega }\mathrm{+}{\sigma }^{\mathrm{2}}_{{\mathcal{N}}_{\mathrm{1}}}\right)$, where $\mathrm{2}{\sigma }^{\mathrm{2}}_{\omega }$ is the cumulated modulation variance, ${\sigma }^{\mathrm{2}}_{{\mathcal{N}}_{\mathrm{1}}}$ is the variance of ${\mathcal{N}}_{\mathrm{1}}$, ${{{x}'_{B,i}}}$, ${{{p}'_{B,i}}}$ are Bob's noisy quadratures modified by ${\mathcal{N}}_{\mathrm{1}}$, while ${\sigma }^{\mathrm{2}}_{{\varphi }_i}\mathrm{=}\mathbb{E}\left[{\left|{\varphi }_i\right|}^{\mathrm{2}}\right]$. Assuming a homodyne measurement $M_{\mathrm{1}}$, Alice gets an $X_i$ \textit{unit} of her \textit{raw data,} which is a binary string. If she measured in the position quadrature basis she obtains:
\begin{equation} \label{4)} 
X_i\mathrm{=}x_{A,i}\mathrm{+}{{{x}'_{B,i}}} 
\end{equation} 
or, if she used the momentum quadrature basis she gets
\begin{equation} \label{5)} 
X_i\mathrm{=}p_{A,i}\mathrm{+}{{{p}'_{B,i}}}.                                                    
\end{equation} 
The second mode of the combined signal in ${\mathcal{S}}_{A\mathrm{\times }B}$ is transmitted directly back to Bob over the noisy channel ${\mathcal{N}}_{\mathrm{2}}$, given as: 
\begin{equation} \label{ZEqnNum542216} 
\left|\left.{\phi }_i\right\rangle \right.\mathrm{=}\left|\left.x_{A,i}\mathrm{-}{{{x}'_{B,i}}}\mathrm{+}i\left(p_{A,i}\mathrm{-}{{{p}'_{B,i}}}\right)\right\rangle \right.\mathrm{\in }\mathcal{C}\mathcal{N}\left(0,{\sigma }^{\mathrm{2}}_{{\phi }_i}\right)\mathrm{\in }{\mathcal{S}}_{A\mathrm{\times }B},                     
\end{equation} 
with $\mathcal{N}\left(\mathrm{0,2}{\sigma }^{\mathrm{2}}_{\omega }\mathrm{+}{\sigma }^{\mathrm{2}}_{{\mathcal{N}}_{\mathrm{1}}}\right)$ Gaussian random quadratures, and ${\sigma }^{\mathrm{2}}_{{\phi }_i}\mathrm{=}\mathbb{E}\left[{\left|{\phi }_i\right|}^{\mathrm{2}}\right]$. The Gaussian noise of the quantum channel ${\mathcal{N}}_{\mathrm{2}}$ defines a noise vector ${\mathrm{\Delta }}_i\mathrm{\in }\mathcal{C}\mathcal{N}\left(0,{\sigma }^{\mathrm{2}}_{{\mathrm{\Delta }}_i}\right)\mathrm{\in }{\mathcal{S}}_{A\mathrm{\times }B}$, with noise components ${\mathrm{\Delta }}_{x_i}\mathrm{\in }\mathcal{N}\left(0,{\sigma }^{\mathrm{2}}_{{\mathcal{N}}_{\mathrm{2}}}\right)$, ${\mathrm{\Delta }}_{p_i}\mathrm{\in }\mathcal{C}\mathcal{N}\left(0,{\sigma }^{\mathrm{2}}_{{\mathcal{N}}_{\mathrm{2}}}\right)$ which results in the noisy state $\left|\left.{\xi }_i\right\rangle \right.\mathrm{\in }{\mathcal{S}}_{A\mathrm{\times }B}$ as follows:
\begin{equation} \label{ZEqnNum576926} 
\left|\left.{\xi }_i\right\rangle \right.\mathrm{=}\left|\left.{\phi }_i\right\rangle \right.\mathrm{+}{\mathrm{\Delta }}_i\mathrm{=}\left|\left.{{{x}'_{A,i}}}\mathrm{-}{{{x}''_{B,i}}}\mathrm{+}i\left({{{p}'_{A,i}}}\mathrm{-}{{{p}''_{B,i}}}\right)\right\rangle \right.\mathrm{\in }\mathcal{C}\mathcal{N}\left(0,{\sigma }^{\mathrm{2}}_{{\xi }_i}\right)\mathrm{\in }{\mathcal{S}}_{A\mathrm{\times }B},              
\end{equation} 
with $\mathcal{N}\left(\mathrm{0,2}{\sigma }^{\mathrm{2}}_{\omega }\mathrm{+}{\sigma }^{\mathrm{2}}_{{\mathcal{N}}_{\mathrm{1}}}\mathrm{+}{\sigma }^{\mathrm{2}}_{{\mathcal{N}}_{\mathrm{2}}}\right)$ distributed Gaussian random quadratures, and ${\sigma }^{\mathrm{2}}_{{\xi }_i}\mathrm{=}\mathbb{E}\left[{\left|{\xi }_i\right|}^{\mathrm{2}}\right]$, where ${{{x}'_{A,i}}}$, ${{{p}'_{A,i}}}$ are Alice's noisy quadratures modified by ${\mathcal{N}}_{\mathrm{2}}$, while ${{{x}''_{B,i}}}$, ${{{p}''_{B,i}}}$ are Bob's noisy quadratures modified by ${\mathcal{N}}_{\mathrm{2}}$. 

 In the next phase, Bob applies a random quadrature measurement $M_{\mathrm{2}}$ (assumed to be homodyne) and gets block $Y_i$. If he used a position quadrature basis, he gets
\begin{equation} \label{8)} 
{{{Y}'_{i}}}\mathrm{=}{{{x}'_{A,i}}}\mathrm{-}{{{x}''_{B,i}}} 
\end{equation} 
and for the momentum quadrature basis he obtains:
\begin{equation} \label{9)} 
{{{Y}'_{i}}}\mathrm{=}{{{p}'_{A,i}}}\mathrm{-}{{{p}''_{B,i}}}.                                                      
\end{equation} 
Bob, calibrating his resulted block ${Y_i}^{\mathrm{'}}$ by $\mathrm{2}{{{x}''_{B,i}}}$ or $\mathrm{2}{{{p}''_{B,i}}}$ (depending on the used quadrature measurement), gets back the noisy version ${{{X}'_{i}}}$ of Alice's raw data unit $X_i$ as:
\begin{equation} \label{ZEqnNum778355} 
{{{X}'_{i}}}\mathrm{=}{{{Y}'_{i}}}\mathrm{+2}{{{x}''_{B,i}}}\mathrm{=}{{{x}'_{A,i}}}\mathrm{-}{{{x}''_{B,i}}}\mathrm{+2}{{{x}''_{B,i}}}\mathrm{=}{{{x}'_{A,i}}}\mathrm{+}{{{x}''_{B,i}}},                         
\end{equation} 
and
\begin{equation} \label{ZEqnNum920181} 
{{{X}'_{i}}}\mathrm{=}{{{Y}'_{i}}}\mathrm{+2}{{{p}''_{B,i}}}\mathrm{=}{{{p}'_{A,i}}}\mathrm{-}{{{p}''_{B,i}}}\mathrm{+2}{{{p}''_{B,i}}}\mathrm{=}{{{p}'_{A,i}}}\mathrm{+}{{{p}''_{B,i}}},                      
\end{equation} 
which is referred as Bob's \textit{raw data unit}. The nature of the of error of the quantum channel will be characterized in detail in \sref{sec4}, however at this point we can surmise that the noise of the quantum channel is analogous to the addition of a non-standard Gaussian random noise vector ${\mathrm{\Delta }}_i$ to Alice's raw data block $X_i$. 

\begin{center}
\begin{figure}[h!]
\begin{center}
\includegraphics[angle = 0,width=1\linewidth]{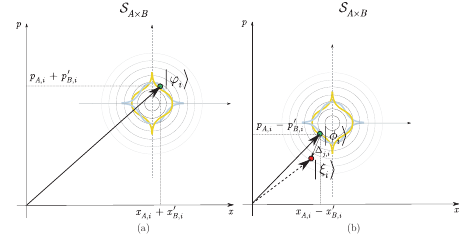}
\caption{The combined signals $\left|\left.{\varphi }_i\right\rangle \right.\mathrm{\in }\mathcal{C}\mathcal{N}\left(0,{\sigma }^{\mathrm{2}}_{{\varphi }_i}\right)$ (\textbf{a}) and $\left|\left.{\phi }_i\right\rangle \right.\mathrm{\in }\mathcal{C}\mathcal{N}\left(0,{\sigma }^{\mathrm{2}}_{{\phi }_i}\right)$ (\textbf{b}) in the combined phase space, ${\mathcal{S}}_{A\mathrm{\times }B}$. The modulation noise $\mathrm{\partial }\mathrm{\in }\mathcal{C}\mathcal{N}\left(0,{\sigma }^{\mathrm{2}}_{\mathrm{\partial }}\right)$ in ${\mathcal{S}}_{A\mathrm{\times }B}$ is illustrated by the Gaussian curves. The noise ${\mathrm{\Delta }}_i\mathrm{\in }\mathcal{N}\left(0,{\sigma }^{\mathrm{2}}_{{\mathcal{N}}_{\mathrm{2}}}\right)$ of quantum channel ${\mathcal{N}}_{\mathrm{2}}$ distorts the distribution of the quadratures from $\mathcal{N}\left(\mathrm{0,2}{\sigma }^{\mathrm{2}}_{\omega }\mathrm{+}{\sigma }^{\mathrm{2}}_{{\mathcal{N}}_{\mathrm{1}}}\right)$ into $\mathcal{N}\left(\mathrm{0,2}{\sigma }^{\mathrm{2}}_{\omega }\mathrm{+}{\sigma }^{\mathrm{2}}_{{\mathcal{N}}_{\mathrm{1}}}\mathrm{+}{\sigma }^{\mathrm{2}}_{{\mathcal{N}}_{\mathrm{2}}}\right)$. Alice's raw data variance is ${\sigma }^{\mathrm{2}}_X\mathrm{=}\left(\mathrm{2}{\sigma }^{\mathrm{2}}_{\omega }\mathrm{+}{\sigma }^{\mathrm{2}}_{{\mathcal{N}}_{\mathrm{1}}}\right)$, while Bob's raw data variance is ${\sigma }^{\mathrm{2}}_{X\mathrm{'}}\mathrm{=}\left(\mathrm{2}{\sigma }^{\mathrm{2}}_{\omega }\mathrm{+}{\sigma }^{\mathrm{2}}_{{\mathcal{N}}_{\mathrm{1}}}\mathrm{+}{\sigma }^{\mathrm{2}}_{{\mathcal{N}}_{\mathrm{2}}}\right)$.} 
 \label{fig2}
 \end{center}
\end{figure}
\end{center}

 Alice's and Bob's modes in the combined phase space ${\mathcal{S}}_{A\mathrm{\times }B}$ right after being outputted from the BS are $\left|\left.{\varphi }_i\right\rangle \right.$ and $\left|\left.{\phi }_i\right\rangle \right.$, as shown in \fref{fig2}. Alice obtains the first mode of the beam, $\left|\left.{\varphi }_i\right\rangle \right.$, the second mode $\left|\left.{\phi }_i\right\rangle \right.$ is sent back to Bob. The noise that exists in ${\mathcal{S}}_{A\mathrm{\times }B}$ arises from the modulation noise $\mathrm{\partial }\mathrm{\in }\mathcal{C}\mathcal{N}\left(0,{\sigma }^{\mathrm{2}}_{\mathrm{\partial }}\right)$ (already included in the quadrature distributions) and the two channel uses, ${\mathcal{N}}_{\mathrm{1}}$ and ${\mathcal{N}}_{\mathrm{2}}$. The measurements performed on $\left|\left.{\varphi }_i\right\rangle \right.$ and $\left|\left.{\xi }_i\right\rangle \right.$ result in raw data units $X_i\mathrm{\in }\mathcal{N}\left(0,{\sigma }^{\mathrm{2}}_X\right)$ and ${{{X}'_{i}}}\mathrm{\in }\mathcal{N}\left(0,{\sigma }^{\mathrm{2}}_{X\mathrm{'}}\right)$. The noise of the first channel changes the Gaussian random distribution of the quadratures from $\mathcal{N}\left(\mathrm{0,2}{\sigma }^{\mathrm{2}}_{\omega }\right)$ to $\mathcal{N}\left(\mathrm{0,2}{\sigma }^{\mathrm{2}}_{\omega }\mathrm{+}{\sigma }^{\mathrm{2}}_{{\mathcal{N}}_{\mathrm{1}}}\right)$ in the combined phase space ${\mathcal{S}}_{A\mathrm{\times }B}$, with mean $\mu \mathrm{=0}$, and results $X$ raw data level variance ${\sigma }^{\mathrm{2}}_X\mathrm{=}\left(\mathrm{2}{\sigma }^{\mathrm{2}}_{\omega }\mathrm{+}{\sigma }^{\mathrm{2}}_{{\mathcal{N}}_{\mathrm{1}}}\right)$, and where noise variance ${\sigma }^{\mathrm{2}}_{{\mathcal{N}}_{\mathrm{1}}}$ arises from the first channel use. The quadratures of the second mode of the coupled beam is also characterized by the same variance, i.e., $\left|\left.{\phi }_i\right\rangle \right.\mathrm{\in }\mathcal{C}\mathcal{N}\left(0,{\sigma }^{\mathrm{2}}_{{\phi }_i}\right)$. The noise of ${\mathcal{N}}_{\mathrm{2}}$ transforms $\left|\left.{\phi }_i\right\rangle \right.\mathrm{\in }{\mathcal{S}}_{A\mathrm{\times }B}$ into $\left|\left.{\xi }_i\right\rangle \right.\mathrm{\in }{\mathcal{S}}_{A\mathrm{\times }B}$ and further modifies the distribution. Finally, Bob's received quadratures will follow a Gaussian distribution $\mathcal{N}\left(\mathrm{0,2}{\sigma }^{\mathrm{2}}_{\omega }\mathrm{+}{\sigma }^{\mathrm{2}}_{{\mathcal{N}}_{\mathrm{1}}}\mathrm{+}{\sigma }^{\mathrm{2}}_{{\mathcal{N}}_{\mathrm{2}}}\right)$. The $X\mathrm{'}$ raw data level variance is evaluated as ${\sigma }^{\mathrm{2}}_{X\mathrm{'}}\mathrm{=}\left(\mathrm{2}{\sigma }^{\mathrm{2}}_{\omega }\mathrm{+}{\sigma }^{\mathrm{2}}_{{\mathcal{N}}_{\mathrm{1}}}\mathrm{+}{\sigma }^{\mathrm{2}}_{{\mathcal{N}}_{\mathrm{2}}}\right)$, which fact arises from the cumulated Gaussian random noise of ${\mathcal{N}}_{\mathrm{1}}$ and ${\mathcal{N}}_{\mathrm{2}}$.

On the raw data level, only the difference of the variance of Alice's and Bob's raw data ${\sigma }^{\mathrm{2}}_X$ and ${\sigma }^{\mathrm{2}}_{X\mathrm{'}}$ has relevance and $\sigma _{{{\mathcal{N}}_{1}}}^{2}$ vanishes from the picture. This difference is, indeed, ${\sigma }^{\mathrm{2}}_{{\mathcal{N}}_{\mathrm{2}}}$. In the level of raw data manipulations Alice's $X_i$ will serve as a reference unit to correct Bob's noisy unit, ${{{X}'_{i}}}$. In other words, the first channel use will have no relevance in the raw data-level calculations, hence the noise of ${\mathcal{N}}_{\mathrm{1}}$ can be excluded from the error-correction process. Precisely, the use of ${\mathcal{N}}_{\mathrm{1}}$ has only one consequence: it increases the initial variance $\mathrm{2}{\sigma }^{\mathrm{2}}_{\omega }$ by ${\sigma }^{\mathrm{2}}_{{\mathcal{N}}_{\mathrm{1}}}$, which finally results in $\mathcal{N}\left(0,{\sigma }^{\mathrm{2}}_X\right)$ on the level of raw data blocks. In particular, only ${\mathcal{N}}_{\mathrm{2}}$ will have significance, and, in fact, only the noise of the second channel use has to be corrected in the reconciliation phase. (Note: Throughout the manuscript, the noise will be modeled on the quadrature-level via a real vector).

 In the reconciliation phase, our task is to share an error-free secret key between the parties. This requires the raw data-level error-correction of the noise that arises from the quantum-level transmission. First we review the background of the multidimensional reconciliation and then we introduce our solution.
 
 \subsection{Uniform Distribution in the Spherical Space}

 In this section we review the background of the multidimensional approaches, and the properties of Gaussian random vectors in the spherical space. The multidimensional reconciliation processes for CVQKD were not implementable without the use of spherical codes and a high-dimensional spherical space.

 First, let us clarify how a \textit{d}-dimensional Gaussian random vector is formulated in the framework of a two-way CVQKD protocol. The outcoming beam from Alice (and Bob) can be regarded as a collection of Gaussian random variables. A standard Gaussian random variable $g\mathrm{\in }\mathcal{N}\left(\mathrm{0,1}\right)\mathrm{\in }\mathbb{R}$ is a real variable selected from a Gaussian distribution. A standard Gaussian variable $g\mathrm{\in }\mathcal{N}\left(\mathrm{0,1}\right)$ has probability density function  [\cref{r15}, \cref{r18}]:
\begin{equation} \label{12)} 
f\left(g\right)\mathrm{=}\frac{\mathrm{1}}{\sqrt{\mathrm{2}\pi }}e^{\frac{\mathrm{-}g^{\mathrm{2}}}{\mathrm{2}}} .                                                  
\end{equation} 
A non-standard Gaussian random variable $g^{\mathrm{*}}\mathrm{\in }\mathcal{N}\left(\mu ,{\sigma }^{\mathrm{2}}\right)\mathrm{\in }\mathbb{R}$ with nonzero mean $\mu \mathrm{\neq }\mathrm{0}$, and variance ${\sigma }^{\mathrm{2}}$, can be expressed from $g\mathrm{\in }\mathcal{N}\left(\mathrm{0,1}\right)$ as $g^{\mathrm{*}}\mathrm{=}g\sigma \mathrm{+}\mu $. A non-standard Gaussian random variable $g^{\mathrm{*}}$ has probability density function:   
\begin{equation} \label{13)} 
f\left(g^{\mathrm{*}}\right)\mathrm{=}\frac{\mathrm{1}}{\sqrt{\mathrm{2}\pi {\sigma }^{\mathrm{2}}}}e^{\frac{\mathrm{-}{\left(g^{\mathrm{*}}\mathrm{-}\mu \right)}^{\mathrm{2}}}{\mathrm{2}{\sigma }^{\mathrm{2}}}}.                                               
\end{equation} 
In Alice's raw data, a \textit{d}-dimensional Gaussian vector
\begin{equation} \label{14)} 
{\boldsymbol{\mathrm{X}}}_j\mathrm{=}{\left(X_{j,0}\mathrm{,\dots ,}X_{j,d\mathrm{-}\mathrm{1}}\right)}^T\mathrm{\in }\mathcal{N}{\left(0,{\sigma }^{\mathrm{2}}_X\right)}_d\mathrm{\in }{\mathbb{R}}^d 
\end{equation} 
is a collection of \textit{d} independent Gaussian random variables $X_{j,0}\mathrm{,\dots ,}X_{j,d\mathrm{-}\mathrm{1}}$, where each $X_{j,i}$ is a real variable $\mathbb{R}$ drawn from a Gaussian random distribution $\mathcal{N}\left(0,{\sigma }^{\mathrm{2}}_X\right)$. Alice's Gaussian vector is referred by ${\boldsymbol{\mathrm{X}}}_j\mathrm{\in }\mathcal{N}{\left(0,{\sigma }^{\mathrm{2}}_X\right)}_d\mathrm{\in }{\mathbb{R}}^d$, and its noisy version at Bob's side is denoted by ${{{\boldsymbol{\mathrm{X}}}'_{j}}}\mathrm{\in }\mathcal{N}{\left(0,{\sigma }^{\mathrm{2}}_{X\mathrm{'}}\right)}_d\mathrm{\in }{\mathbb{R}}^d$. The values of Bob's units are affected by the Gaussian noise that arises from the quantum channel. 

 First, let us evaluate why the \textit{normalized} vector structure has importance in the multidimensional scenario. The reason: the normalized \textit{d}-dimensional Gaussian vectors change the probability distribution from Gaussian random to uniform on the \textit{d}-dimensional unit sphere, ${\mathrm{\Gamma }}^{d\mathrm{-}\mathrm{1}}$. It has a relevance, since only uniform distribution is allowed in the reconciliation phase. The result clearly follows from the Rayleigh law  [\cref{r18}], the application of Stirling's formula [\cref{r19}], Gersho's conjecture [\cref{r22}], and Sakrison's result [\cref{r23}], which are connected to the contributions of spherical coding [\cref{r24}]. 

 We formulate \textit{d-}length blocks ${{{\boldsymbol{\mathrm{X}}}'_{j}}}\mathrm{=}{\left({{{X}'_{j,0}}}\mathrm{,\dots ,}{{{X}'_{j,d-1}}}\right)}^T\mathrm{\in }\mathcal{N}{\left(0,{\sigma }^{\mathrm{2}}_{X\mathrm{'}}\right)}_d\mathrm{\in }{\mathbb{R}}^d$, where ${{{X}'_{j,i}}}\mathrm{\in }\mathcal{N}\left(0,{\sigma }^{\mathrm{2}}_{X\mathrm{'}}\right)\mathrm{\in }\mathbb{R}$, for $i\mathrm{\in }\left[d\right]$. The \textit{d}-length Gaussian random vector ${{{\boldsymbol{\mathrm{X}}}'_{j}}}$ has norm $\left\|{{{\boldsymbol{\mathrm{X}}}'_{j}}}\right\|$, mean 
\begin{equation} \label{15)} 
\mathbb{E}\left[\left\|{{{\boldsymbol{\mathrm{X}}}'_{j}}}\right\|\right]\mathrm{=}{\sigma }_{X\mathrm{'}}\sqrt{d\mathrm{-}\frac{\mathrm{1}}{\mathrm{2}}} 
\end{equation} 
and \textit{variance} 
\begin{equation} \label{16)} 
\mathrm{var}\left[\left\|{{{\boldsymbol{\mathrm{X}}}'_{j}}}\right\|\right]\mathrm{\le }\frac{{\sigma }^{\mathrm{2}}_{X\mathrm{'}}}{\mathrm{2}}.                                                   
\end{equation} 
We step further from this point. Since the variance of ${{{\boldsymbol{\mathrm{X}}}'_{j}}}$ is not unit, the covariance matrix $\mathfrak{C}\left({{{\boldsymbol{\mathrm{X}}}'_{j}}}\right)$ is not equal to identity, but the random units ${{{X}'_{j,i}}}$ are uncorrelated, thus $\mathfrak{C}\left({{{\boldsymbol{\mathrm{X}}}'_{j}}}\right)$ is diagonal. 

 The normalized vector ${{{{{\boldsymbol{\mathrm{X}}}'_{j}}}}}/{\sqrt{d{\sigma }^{\mathrm{2}}_{X\mathrm{'}}}}$ with norm $\left\|{{{{\boldsymbol{\mathrm{X}}}'_{j}}}}/{\sqrt{d{\sigma }^{\mathrm{2}}_{X\mathrm{'}}}}\right\|$, can be identified on the unit sphere ${\mathrm{\Gamma }}^{d\mathrm{-}\mathrm{1}}$ [\cref{r18}, \cref{r24}], with radius $r\mathrm{=}\left\|{{{{\boldsymbol{\mathrm{X}}}'_{j}}}}/{\sqrt{d{\sigma }^{\mathrm{2}}_{X\mathrm{'}}}}\right\|$. The mean of ${\left\|{{{\boldsymbol{\mathrm{X}}}'_{j}}}\right\|}/{\sqrt{d{\sigma }^{\mathrm{2}}_{X\mathrm{'}}}}$ is 
\begin{equation} \label{17)} 
\mathbb{E}\left[{\left\|{{{\boldsymbol{\mathrm{X}}}'_{j}}}\right\|}/{\sqrt{d{\sigma }^{\mathrm{2}}_{X\mathrm{'}}}}\right]\mathrm{=}{{\sigma }_{X\mathrm{'}}\sqrt{d\mathrm{-}\frac{\mathrm{1}}{\mathrm{2}}}}/{\sqrt{d{\sigma }^{\mathrm{2}}_{X\mathrm{'}}}}.                               
\end{equation} 
The vector ${{{{\boldsymbol{\mathrm{X}}}'_{j}}}}/{\sqrt{d{\sigma }^{\mathrm{2}}_{X\mathrm{'}}}}$ on the unit sphere ${\mathrm{\Gamma }}^{d\mathrm{-}\mathrm{1}}$ is identified as
\begin{equation} \label{18)} 
{{{{\boldsymbol{\mathrm{X}}}'_{j}}}}/{\sqrt{d{\sigma }^{\mathrm{2}}_{X\mathrm{'}}}}\mathrm{=}r\frac{{{{\boldsymbol{\mathrm{X}}}'_{j}}}}{\left\|{{{\boldsymbol{\mathrm{X}}}'_{j}}}\right\|}\mathrm{=}\frac{\left\|{{{{\boldsymbol{\mathrm{X}}}'_{j}}}}/{\sqrt{d{\sigma }^{\mathrm{2}}_{X\mathrm{'}}}}\right\|{{{{\boldsymbol{\mathrm{X}}}'_{j}}}}}{\left\|{{{\boldsymbol{\mathrm{X}}}'_{j}}}\right\|}.                                    
\end{equation} 
Precisely, the normalized quantity ${\left\|{{{\boldsymbol{\mathrm{X}}}'_{j}}}\right\|}/{\sqrt{d\sigma _{{{X}'}}^{2}}}$ has variance $\mathrm{var}\left[{\left\|{{{\boldsymbol{\mathrm{X}}}'_{j}}}\right\|}/{\sqrt{d{\sigma }^{\mathrm{2}}_{X\mathrm{'}}}}\right]\mathrm{\le }{\frac{{\sigma }^{\mathrm{2}}_{X\mathrm{'}}}{\mathrm{2}}}/{d{\sigma }^{\mathrm{2}}_{X\mathrm{'}}}$. 

 From the spherical symmetry, it follows that if $d\mathrm{\to }\mathrm{\infty }$, the normalized random vector ${{{{\boldsymbol{\mathrm{X}}}'_{j}}}}/{\sqrt{d{\sigma }^{\mathrm{2}}_{X\mathrm{'}}}}$ will be equipped with uniform distribution on ${\mathrm{\Gamma }}^{d\mathrm{-}\mathrm{1}}$. The background of this phenomenon is as follows. 

 First, for $d\mathrm{\to }\mathrm{\infty }$, the mean $\mathbb{E}\left[\mathrm{\cdot }\right]$ of the normalized quantity ${\left\|{{{\boldsymbol{\mathrm{X}}}'_{j}}}\right\|}/{\sqrt{d{\sigma }^{\mathrm{2}}_{X\mathrm{'}}}}$ will tend to one, i.e., 
\begin{equation} \label{19)} 
\mathop{\mathrm{lim}}_{d\mathrm{\to }\mathrm{\infty }}\mathbb{E}\left[\frac{\left\|{{{\boldsymbol{\mathrm{X}}}'_{j}}}\right\|}{\sqrt{d{\sigma }^{\mathrm{2}}_{X\mathrm{'}}}}\right]\mathrm{=}\mathop{\mathrm{lim}}_{d\mathrm{\to }\mathrm{\infty }}\frac{{\sigma }_{X\mathrm{'}}\sqrt{d\mathrm{-}\frac{\mathrm{1}}{\mathrm{2}}}}{\sqrt{d{\sigma }^{\mathrm{2}}_{X\mathrm{'}}}}\mathrm{=1}.                                     
\end{equation} 
Second, the variance $\mathrm{var}\left[\mathrm{\cdot }\right]$ of ${\left\|{{{\boldsymbol{\mathrm{X}}}'_{j}}}\right\|}/{\sqrt{d{\sigma }^{\mathrm{2}}_{X\mathrm{'}}}}$ will tend to zero, 
\begin{equation} \label{20)} 
\mathop{\mathrm{lim}}_{d\mathrm{\to }\mathrm{\infty }}\mathrm{var}\left[\frac{\left\|{{{\boldsymbol{\mathrm{X}}}'_{j}}}\right\|}{\sqrt{d{\sigma }^{\mathrm{2}}_{X\mathrm{'}}}}\right]\mathrm{=}\mathop{\mathrm{lim}}_{d\mathrm{\to }\mathrm{\infty }}\frac{\frac{\mathrm{1}}{\mathrm{2}}{\sigma }^{\mathrm{2}}_{X\mathrm{'}}}{d{\sigma }^{\mathrm{2}}_{X\mathrm{'}}}\mathrm{=0}.                                        
\end{equation} 
These implies that for $d\mathrm{\to }\mathrm{\infty }$, the normalized Gaussian random vector ${{{{\boldsymbol{\mathrm{X}}}'_{j}}}}/{\sqrt{d{\sigma }^{\mathrm{2}}_{X\mathrm{'}}}}$ becomes uniformly distributed on the unit sphere ${\mathrm{\Gamma }}^{d\mathrm{-}\mathrm{1}}$. Third, as the dimension increases the distribution of the norm of ${{{{\boldsymbol{\mathrm{X}}}'_{j}}}}/{\sqrt{d{\sigma }^{\mathrm{2}}_{X\mathrm{'}}}}$ (i.e., the radius on ${\mathrm{\Gamma }}^{d\mathrm{-}\mathrm{1}}$) will approximate the Dirac distribution $\mathcal{D}\left(d\right)$ [\cref{r9}-\cref{r11}], [\cref{r18}], and it will also converge to one, $r\mathrm{=}\mathop{\mathrm{lim}}_{d\mathrm{\to }\mathrm{\infty }}\left\|{{{{\boldsymbol{\mathrm{X}}}'_{j}}}}/{\sqrt{d{\sigma }^{\mathrm{2}}_{X\mathrm{'}}}}\right\|\mathrm{=1}$. The unit norms of ${{{{\boldsymbol{\mathrm{X}}}'_{j}}}}/{\sqrt{d{\sigma }^{\mathrm{2}}_{X\mathrm{'}}}}$ play exactly the role of unit fading-coefficients for a logical binary Gaussian channel, since during the transmissions of the messages generated from ${{{{\boldsymbol{\mathrm{X}}}'_{j}}}}/{\sqrt{d{\sigma }^{\mathrm{2}}_{X\mathrm{'}}}}$ the unit norms $r\mathrm{=}\left\|{{{{\boldsymbol{\mathrm{X}}}'_{j}}}}/{\sqrt{d{\sigma }^{\mathrm{2}}_{X\mathrm{'}}}}\right\|\mathrm{=1}$ are also transmitted [\cref{r11}, \cref{r21}]. 

 To be more exact, the unit norms are only approximated and the distribution of the unit norms also depends on \textit{d}, and as $d\mathrm{\to }\mathrm{\infty }$, it precisely can be described by the Dirac distribution
\begin{equation} \label{21)} 
{\mathcal{D}}_d\left(x\right)\mathrm{=}\left({\mathrm{1}}/{a\sqrt{\pi }}\right)e^{\mathrm{-}{{\left(x\mathrm{-}r\right)}^{\mathrm{2}}}/{a^{\mathrm{2}}}},                                        
\end{equation} 
where $a\mathrm{=}{\mathrm{1}}/{\sqrt{d}}$ and  
\begin{equation} \label{22)} 
r\mathrm{=}\mathop{\mathrm{lim}}_{d\mathrm{\to }\mathrm{\infty }}\frac{\left\|{{{\boldsymbol{\mathrm{X}}}'_{j}}}\right\|}{\sqrt{d{\sigma }^{\mathrm{2}}_{X\mathrm{'}}}}\mathrm{=1}.                                               
\end{equation} 
From ${\mathcal{D}}_d\left(x\right)$ it immediately follows, that the unit norms of the normalized random Gaussian vectors gets closer to 1, as \textit{d} goes to infinity [\cref{r18}]. As follows from these, for low values of $d$ the uniform distribution of ${{{{\boldsymbol{\mathrm{X}}}'_{j}}}}/{\sqrt{d{\sigma }^{\mathrm{2}}_{X\mathrm{'}}}}$ cannot be achieved. 

 In comparison to the multidimensional reconciliation where the required mathematical operations (the spherical division operator at Alice's side) exist only in $d\mathrm{=1,}$ 2, 4 or 8 dimensions [\cref{r9}-\cref{r11}], [\cref{r18}], the scalar reconciliation process are also existent for arbitrary high dimensions, which makes possible to give a more closer approximation, however it will not refer to the Dirac distribution. Analyzing the situation if the noisy raw data follows Gaussian random distribution with ${\sigma }^{\mathrm{2}}_{X\mathrm{'}}\mathrm{>1}$, the speed of convergence of the mean $\mathbb{E}\left[{{{{\boldsymbol{\mathrm{X}}}'_{j}}}}/{\sqrt{d{\sigma }^{\mathrm{2}}_{X\mathrm{'}}}}\right]$ and variance $\mathrm{var}\left[{{{{\boldsymbol{\mathrm{X}}}'_{j}}}}/{\sqrt{d{\sigma }^{\mathrm{2}}_{X\mathrm{'}}}}\right]$ will be lower for any \textit{d}, in comparison if ${\sigma }^{\mathrm{2}}_{X\mathrm{'}}\mathrm{=1}$ would have hold. 

\begin{center}
\begin{figure}[h!]
\begin{center}
\includegraphics[angle = 0,width=0.7\linewidth]{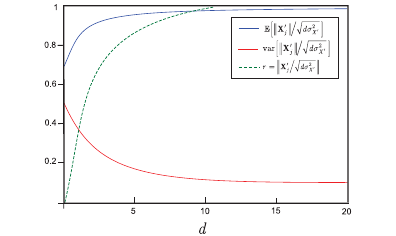}
\caption{The mean $\mathbb{E}\left[\mathrm{\cdot }\right]$, variance $\mathrm{var}\left[\mathrm{\cdot }\right]$ of the normalized quantity ${\left\|{{{\boldsymbol{\mathrm{X}}}'_{j}}}\right\|}/{\sqrt{d{\sigma }^{\mathrm{2}}_{X\mathrm{'}}}}$ and the norm $\left\|{{{{\boldsymbol{\mathrm{X}}}'_{j}}}}/{\sqrt{d{\sigma }^{\mathrm{2}}_{X\mathrm{'}}}}\right\|$ of the normalized Gaussian random vector ${{{{\boldsymbol{\mathrm{X}}}'_{j}}}}/{\sqrt{d{\sigma }^{\mathrm{2}}_{X\mathrm{'}}}}$. Vector ${{{\boldsymbol{\mathrm{X}}}'_{j}}}$ is formulated from \textit{d }number of ${{{X}'_{j,i}}}$ elements of Bob's noisy raw data $X\mathrm{'}$. The approximation of the logical binary Gaussian gets more precise as the norm approaches to one, which requires the use of higher dimensions.} 
 \label{fig3}
 \end{center}
\end{figure}
\end{center}

 For ${\sigma }^{\mathrm{2}}_{X\mathrm{'}}\mathrm{=1}$, the situation for various dimensions of ${{{\boldsymbol{\mathrm{X}}}'_{j}}}$ is summarized in \fref{fig3}.
 
As we have mentioned, the multidimensional approaches are limited in the dimension, specifically, $d\mathrm{=8}$ in [\cref{r9}-\cref{r11}]. In this case, the Gaussian random vectors form the so-called \textit{octonions } [\cref{r20}]. In the level of Gaussian random raw data, an octonion ${\mathrm{O}}_j\mathrm{\in }{\mathbb{R}}^{\mathrm{8}}$ is built up from eight units $X_{j\mathrm{,0\dots }j\mathrm{,7}}\mathrm{\in }\mathcal{N}\left(0,{\sigma }^{\mathrm{2}}_X\right)$, as: 
\begin{equation} \label{23)} 
{\mathrm{O}}_j\mathrm{=}X_{j,0}\mathrm{Re+}X_{j\mathrm{,1}}\mathrm{I}{\mathrm{m}}_{\mathrm{1}}\mathrm{+\dots +}X_{j\mathrm{,7}}\mathrm{I}{\mathrm{m}}_{\mathrm{7}}, 
\end{equation} 
where $\mathrm{Re}\mathrm{\in }\mathbb{R}$ stands for the real part, while $\mathrm{I}{\mathrm{m}}_i\mathrm{\in }\mathbb{C}$, $for\ i\mathrm{=1,}i\mathrm{\le }\mathrm{7}$ indentifies the \textit{i}-th imaginary units, respectively. Bob's noisy ${\mathrm{O}\mathrm{'}}_j$ is ${\mathrm{O}\mathrm{'}}_j\mathrm{=}{{{X}'_{j,0}}}\mathrm{Re+}{{{X}'_{j\mathrm{,1}}}}\mathrm{I}{\mathrm{m}}_{\mathrm{1}}\mathrm{+\dots +}{X\mathrm{'}}_{j\mathrm{,7}}\mathrm{I}{\mathrm{m}}_{\mathrm{7}},$ where ${X\mathrm{'}}_{j\mathrm{,0\dots 7}}\mathrm{\in }\mathcal{N}\left(0,{\sigma }^{\mathrm{2}}_{X\mathrm{'}}\right)$. In the multidimensional case the uniformity of the \textit{d}-dimensional Gaussian random raw data vectors ${\boldsymbol{\mathrm{X}}}_j\mathrm{\in }{\mathbb{R}}^d$, $d\mathrm{\le }\mathrm{8}$, can be achieved only in the multidimensional spherical space, over the unit sphere ${\mathrm{\Gamma }}^{d\mathrm{-}\mathrm{1}}$. The process requires complex operations and transformations [\cref{r9}-\cref{r11}] that are undesirable in a practical CVQKD scenario. In comparison to these approaches, our proposed scalar reconciliation uses only simple scalar operations on the raw data, which makes it possible to eliminate the spherical calculations from the reconciliation phase. 

 \section{Low-Dimensional Reconciliation}
\label{sec3}
 We start our description from the point at which the quantum states are completely transmitted through the quantum channel from Alice to Bob. At this point all interactions with the quantum channel are closed, and the post-processing phase is being started. First, Alice and Bob exclude from the raw data those measurements that have been performed in different quadratures that results in the \textit{N-}unit length raw data vectors. Then formulate ${N}/{d}$ number of \textit{d}-dimensional vectors ${\boldsymbol{\mathrm{X}}}_j\mathrm{\in }{\mathbb{R}}^d$, ${{{\boldsymbol{\mathrm{X}}}'_{j}}}\mathrm{\in }{\mathbb{R}}^d$. These quantities are introduced as follows.
 
\subsection{Notations}
\label{sec3_1}
 Let $X\mathrm{\in }{\mathbb{R}}^N$ and $X\mathrm{'}\mathrm{\in }{\mathbb{R}}^N$ the \textit{N}-unit length raw data of Alice and Bob. The \textit{d}-dimensional vectors ${\boldsymbol{\mathrm{X}}}_j\mathrm{\in }{\mathbb{R}}^d$ and ${{{\boldsymbol{\mathrm{X}}}'_{j}}}\mathrm{\in }{\mathbb{R}}^d$, $for\ j\mathrm{=}0,j\mathrm{\le }\left({N}/{d}\right)\mathrm{-}\mathrm{1}$, of Alice and Bob are defined as: 
\begin{equation} \label{24)} 
{\boldsymbol{\mathrm{X}}}_j\mathrm{=}{\left(X_{j,0}\mathrm{,\dots ,}X_{j,d\mathrm{-}\mathrm{1}}\right)}^T\mathrm{\in }\mathcal{N}{\left(0,{\sigma }^{\mathrm{2}}_X\right)}_d 
\end{equation} 
and
\begin{equation} \label{25)} 
{{{\boldsymbol{\mathrm{X}}}'_{j}}}\mathrm{=}{\left({{{X}'_{j,0}}}\mathrm{,\dots ,}{{{X}'_{j,d-1}}}\right)}^T\mathrm{\in }\mathcal{N}{\left(0,{\sigma }^{\mathrm{2}}_{X\mathrm{'}}\right)}_d, 
\end{equation} 
where 
\begin{equation} \label{26)} 
X_{j,i}\mathrm{\in }\mathcal{N}\left(0,{\sigma }^{\mathrm{2}}_X\right)\mathrm{\in }\mathbb{R} 
\end{equation} 
and 
\begin{equation} \label{27)} 
{{{X}'_{j,i}}}\mathrm{\in }\mathcal{N}\left(0,{\sigma }^{\mathrm{2}}_{X\mathrm{'}}\right)\mathrm{\in }\mathbb{R} 
\end{equation} 
refer to the \textit{i}-th unit of the \textit{j}-th vector, respectively. Alice and Bob have to share a common secret \textit{b}y using their correlated raw data. For this purpose, they establish a proper code-alphabet $\mathcal{A}\mathrm{=}\left\{a,b\right\}$, where $a\mathrm{\in }\mathbb{R}$ and $b\mathrm{\in }\mathbb{R}$ are two public variables (i.e., Eve also has access to it). In the reverse reconciliation these will be selected uniformly at random in the form of several $U_j\mathrm{\in }\left\{a,b\right\}$-s at Bob's side, with $\mathrm{Pr}\left(a\right)\mathrm{=Pr}\left(b\right)\mathrm{=0.5}$. 

 A secret\textit{ d}-dimensional\textit{ key vector} ${\boldsymbol{\mathrm{U}}}_j$ is drawn from a uniform distribution $\mathcal{U}$ and built up from \textit{d} units, $U_{j,i}\mathrm{\in }\mathbb{R}$, as:
\begin{equation} \label{ZEqnNum650790} 
{\boldsymbol{\mathrm{U}}}_j\mathrm{\in }{\mathbb{R}}^d\mathrm{:}{\left(U_{j,0}\mathrm{,\dots ,}U_{j,d\mathrm{-}\mathrm{1}}\right)}^T, for\text{ }j\mathrm{=0,}\text{ }j\mathrm{\le }\left({N}/{d}\right)\mathrm{-}\mathrm{1}.               
\end{equation} 
The \textit{d} units $U_{j,i}\mathrm{\in }\mathcal{U}$ of ${\boldsymbol{\mathrm{U}}}_j$ are uniform random variables, and define $U_j\mathrm{\in }\mathbb{R}$ as follows: 
\begin{equation} \label{ZEqnNum462129} 
U_j\mathrm{=}\sum^{d\mathrm{-}\mathrm{1}}_{i\mathrm{=0}}{U_{j,i}}\mathrm{\in }\mathcal{U}.                                         
\end{equation} 
The noisy version of \eqref{ZEqnNum462129}, ${{{U}'_{j}}}$, is defined as
\begin{equation} \label{30)} 
{{{U}'_{j}}}\mathrm{=}\sum^{d\mathrm{-}\mathrm{1}}_{i\mathrm{=0}}{{{{U}'_{j,i}}}}.                                              
\end{equation} 
From \eqref{ZEqnNum462129} follows, that \eqref{ZEqnNum650790} can be rewritten as ${\boldsymbol{\mathrm{U}}}_j\mathrm{\in }\left\{\boldsymbol{\mathrm{A}},\boldsymbol{\mathrm{B}}\right\}\mathrm{\in }{\mathbb{R}}^d$, with vectors $\boldsymbol{\mathrm{A}},\boldsymbol{\mathrm{B}}$ as:
\begin{equation} \label{31)} 
\boldsymbol{\mathrm{A}}\mathrm{:}{\left(a_{j,0}\mathrm{,\dots ,}a_{j,d\mathrm{-}\mathrm{1}}\right)}^T,\left\{\sum^{d\mathrm{-}\mathrm{1}}_{i\mathrm{=0}}{a_{j,i}}\mathrm{=}a\right\}, \boldsymbol{\mathrm{B}}\mathrm{:}{\left(b_{j,0}\mathrm{,\dots ,}b_{j,d\mathrm{-}\mathrm{1}}\right)}^T,\left\{\sum^{d\mathrm{-}\mathrm{1}}_{i\mathrm{=0}}{b_{j,i}}\mathrm{=}b\right\}.         
\end{equation} 
As follows, Bob granulates the selected \textit{a} or \textit{b} into \textit{d} number of uniformly random variables $U_{j,i}$, so that the sum of the units will be equal to the selected value. 

 The \textit{full} \textit{key} $\boldsymbol{\mathrm{K}}$ is built up as: 
\begin{equation} \label{32)} 
\boldsymbol{\mathrm{K}}\mathrm{\in }{\mathbb{R}}^{{N}/{d}}\mathrm{:}{\left(U_0\mathrm{,\dots ,}U_{\left({N}/{d}\right)\mathrm{-}\mathrm{1}}\right)}^T.                                        
\end{equation} 
Alice and Bob first agree on \textit{d}. Bob then sends the \textit{d} blocks of
\begin{equation} \label{33)} 
C\left({{{\boldsymbol{\mathrm{X}}}'_{j}}}\right){\boldsymbol{\mathrm{U}}}_j\mathrm{=}{\left(C\left({{{X}'_{j,0}}}\right)U_{j,0}\mathrm{,\dots ,}C\left({{{X}'_{j,d-1}}}\right)U_{j,d\mathrm{-}\mathrm{1}}\right)}^T\mathrm{\in }{\mathbb{R}}^d,                    
\end{equation} 
$for\ j\mathrm{=0,}\text{ }j\mathrm{\le }\left({N}/{d}\right)\mathrm{-}\mathrm{1}$, over a classical channel. The scalar quantities  $C\left(X_j\right)$, $C\left({{{X}'_{j}}}\right)$, and $C\left({{{X}'_{j}}}\right)U_j$ are evaluated as
\begin{equation} \label{ZEqnNum391203} 
C\left(X_j\right)\mathrm{=}\sum^{d\mathrm{-}\mathrm{1}}_{i\mathrm{=0}}{C\left(X_{j,i}\right)}\mathrm{\in }\mathbb{R}, C\left({{{X}'_{j}}}\right)\mathrm{=}\sum^{d\mathrm{-}\mathrm{1}}_{i\mathrm{=0}}{C\left({{{X}'_{j,i}}}\right)}\mathrm{\in }\mathbb{R},                   
\end{equation} 
and
\begin{equation} \label{ZEqnNum766670} 
C\left({{{X}'_{j}}}\right)U_j\mathrm{=}\sum^{d\mathrm{-}\mathrm{1}}_{i\mathrm{=0}}{C\left({{{X}'_{j,i}}}\right)U_{j,i}}\mathrm{\in }\mathbb{R},                                 
\end{equation} 
respectively. 

 Alice receives the \textit{d} noisy ${{{U}'_{j,i}}}$ units, and by the addition of the \textit{d} units, and via the application of $C\left(X_j\right)$ she computes ${{{U}'_{j}}}$ as
\begin{equation} \label{36)} 
 \begin{array}{l}
\begin{split}
{{{U}'_{j}}}\mathrm{=}\sum^{d\mathrm{-}\mathrm{1}}_{i\mathrm{=0}}{{{{U}'_{j,i}}}}&\mathrm{=}C\left({{{X}'_{j}}}\right)U_j\frac{\mathrm{1}}{C\left(X_j\right)} \\ 
&\mathrm{=}\left(\sum^{d\mathrm{-}\mathrm{1}}_{i\mathrm{=0}}{{C\left({{{X}'_{j,i}}}\right)}/{\sum^{d\mathrm{-}\mathrm{1}}_{i\mathrm{=0}}{C\left(X_{j,i}\right)}}}\right)\sum^{d\mathrm{-}\mathrm{1}}_{i\mathrm{=0}}{U_{j,i}}. 
\end{split}
\end{array}
\end{equation} 
Thus, Alice has to make an error-correction to remove the noise from ${{{U}'_{j}}}$ to get achieve $U_j$. 

 \subsection{Achieving the Uniform Distribution}

 In comparison to the multidimensional reconciliation, the scalar reconciliation uses a fundamentally different solution to achieve the uniform distribution of the raw data. While the former is based on sophisticated multidimensional spherical operations, our solution requires only the use of a simple function in the scalar space. In our scheme, the \textit{uniform distribution }of the correlated raw data units is achieved by the Gaussian\textit{ Cumulative Distribution Function} (CDF) [\cref{r26}], [\cref{r43}-\cref{r45}]. Another important difference is that the approximation of the logical binary Gaussian channel can be achieved by arbitrary dimension with arbitrary accuracy, which is justified by the \textit{Central Limit Theorem} (CLT) [\cref{r26}], [\cref{r43}-\cref{r45}].

\subsubsection{Gaussian Cumulative Distribution Function}
On Alice's and Bob's side, the Gaussian CDF function can be used to reach the uniform distribution of the correlated raw data. Since we assumed reverse reconciliation let us to start the description from Bob's perspective. Let Bob's raw data unit ${{{X}'_{j,i}}}$ with Gaussian random distribution $\mathcal{N}\left(0,{\sigma }^{\mathrm{2}}_{X\mathrm{'}}\right)$. The Gaussian CDF-transformation $C\left(\mathrm{\cdot }\right)\mathrm{:}\mathbb{R}\mathrm{\to }\mathbb{R}$ for a unit ${{{X}'_{j,i}}}$ is as follows: 
\begin{equation} \label{} 
C\left({{{X}'_{j,i}}}\right)\mathrm{=}\frac{\mathrm{1}}{\mathrm{2}}\left(\mathrm{1+}erf\left(\frac{{{{X}'_{j,i}}}}{\sqrt{\mathrm{2}{\sigma }^{\mathrm{2}}_{X\mathrm{'}}}}\right)\right), for\text{ }i\mathrm{\in }\left[d\right],                             \label{ZEqnNum407819}
\end{equation}
 where 
\begin{equation} \label{38)} 
erf\left(\frac{{{{X}'_{j,i}}}}{\sqrt{\mathrm{2}{\sigma }^{\mathrm{2}}_{X\mathrm{'}}}}\right)\mathrm{=}\frac{\mathrm{2}}{\sqrt{\pi }}\int^{{{{{X}'_{j,i}}}}/{\sqrt{\mathrm{2}{\sigma }^{\mathrm{2}}_{X\mathrm{'}}}}}_0{e^{\mathrm{-}t^{\mathrm{2}}}dt} 
\end{equation} 
is the Gauss error function, and $C\left({{{X}'_{j,i}}}\right)\mathrm{\in }\mathbb{R}$ is a real variable from the range of $\left[\mathrm{0,1}\right]$, with $\mathcal{U}$ \textit{uniform} distribution (for a plausible example \sref{sec5}). The quantity $C\left({{{X}'_{j,i}}}\right)$ will be referred as the \textit{CDF-transformed} unit. 

 Alice also applies the CDF transformation, and takes into account her raw data variance ${\sigma }^{\mathrm{2}}_X$ for the units of $X_{j,i}$ to get $C\left(X_{j,i}\right)$:
\begin{equation} \label{} 
C\left(X_{j,i}\right)\mathrm{=}\frac{\mathrm{1}}{\mathrm{2}}\left(\mathrm{1+}erf\left(\frac{X_{j,i}}{\sqrt{\mathrm{2}{\sigma }^{\mathrm{2}}_X}}\right)\right), for\text{ }i\mathrm{\in }\left[d\right],                             \label{ZEqnNum847571}
\end{equation} 
 and the result of \eqref{ZEqnNum407819} and \eqref{ZEqnNum847571} is the correlated uniform raw data $C\left(X_{j,i}\right)\mathrm{\approx }C\left({{{X}'_{j,i}}}\right)$. In the reconciliation process, only Alice can correct ${{{U}'_{j}}}$ into $U_j$, because nobody knows the CDF-transformed raw data units $C\left(X_{j,i}\right)$, except Alice. 

 For a given ${\boldsymbol{\mathrm{X}}}_j\mathrm{\in }{\mathbb{R}}^d$, the CDF function $C\left(\mathrm{\cdot }\right)\mathrm{:}\mathbb{R}\mathrm{\to }\mathbb{R}$ reads as
\begin{equation} \label{} 
C\left({\boldsymbol{\mathrm{X}}}_j\right)\mathrm{=}C\left(X_{j,0}\right)\mathrm{,\dots ,}C\left(X_{j,d\mathrm{-}\mathrm{1}}\right)\mathrm{=}\frac{\mathrm{1}}{\mathrm{2}}\left(\mathrm{1+}erf\left(\frac{X_{j,i}}{\sqrt{\mathrm{2}{\sigma }^{\mathrm{2}}}}\right)\right)\mathrm{\in }\mathbb{R}, for\text{ }i\mathrm{\in }\left[d\right],         \label{40)}
\end{equation}
 Applying the results for Bob's raw data the CDF-transformed vector is: 
\begin{equation} \label{} 
C\left({{{\boldsymbol{\mathrm{X}}}'_{j}}}\right)\mathrm{=}C\left({{{X}'_{j,0}}}\right)\mathrm{,\dots ,}C\left({{{X}'_{j,d-1}}}\right)\mathrm{=}\frac{\mathrm{1}}{\mathrm{2}}\left(\mathrm{1+}erf\left(\frac{{{{X}'_{j,i}}}}{\sqrt{\mathrm{2}{\sigma }^{\mathrm{2}}}}\right)\right)\mathrm{\in }\mathbb{R}, for\text{ }i\mathrm{\in }\left[d\right].          \label{41)}
\end{equation}
 The CDF-transformed $C\left({\boldsymbol{\mathrm{X}}}_j\right)$, $C\left({{{\boldsymbol{\mathrm{X}}}'_{j}}}\right)$ raw data vectors each consist of \textit{d} real $\mathbb{R}$ variables as: 
\begin{equation} \label{42)} 
C\left({\boldsymbol{\mathrm{X}}}_j\right)\mathrm{=}{\left(C\left(X_{j,0}\right)\mathrm{,\dots ,}C\left(X_{j,d\mathrm{-}\mathrm{1}}\right)\right)}^T, C\left({{{\boldsymbol{\mathrm{X}}}'_{j}}}\right)\mathrm{=}{\left(C\left({{{X}'_{j,0}}}\right)\mathrm{,\dots ,}C\left({{{X}'_{j,d-1}}}\right)\right)}^T. 
\end{equation} 
 
\subsubsection{Central Limit Theorem}
In the multidimensional case, the precision of the approximation of the logical binary Gaussian channel (i.e., the quality of the physical-logical channel conversion) was quantified by the Dirac distribution [\cref{r9}-\cref{r11}]. Since in the scalar reconciliation the spherical space is eliminated, a different solution was needed to analyze the accuracy of the conversion between the physical-logical Gaussian channels. Our answer for the problem is the \textit{Central Limit Theorem} [\cref{r26}], [\cref{r43}-\cref{r45}] and a mathematical result from the 19th century -- the so-called \textit{Lyapunov-condition} [\cref{r26},\cref{r45}]. The accuracy of the physical-logical conversion of scalar reconciliation can be maximized and it can be made in arbitrary high dimensions as it is being stated in Lemma 1.

\begin{lemma}The noise variance of the converted logical binary Gaussian channel asymptotically coincidences with the noise variance of the physical quantum channel, which allows to reach the theoretical maximum of the capacity of the converted logical binary channel.
\end{lemma}
\begin{proof}
 Let $X_{j,i}\mathrm{\in }\mathbb{R}$ and ${{{X}'_{j,i}}}\mathrm{\in }\mathbb{R}$ the \textit{j}-th units of Alice's and Bob's raw data, respectively. For a \textit{d}-dimensional vector ${\boldsymbol{\mathrm{U}}}_j\mathrm{=}{\left({{{U}'_{j,0}}},\mathrm{\dots ,}{{{U}'_{j,d\mathrm{-}\mathrm{1}}}}\right)}^T$, the sum of the independent noise $\left\{{\delta }_{j,0}\mathrm{,\dots ,}{\delta }_{j,d\mathrm{-}\mathrm{1}}\right\}$ units on the secret noisy key units ${{{U}'_{j,i}}}\mathrm{=}U_{j,i}\mathrm{+}{\delta }_{j,i}$ will approximate a zero-mean Gaussian random variable with mean $\mathbb{E}\left[{\delta }_{j,i}\right]\mathrm{=}{\mu }_{{\delta }_{j,i}}\mathrm{=0}$, noise variance $\mathrm{var}\left[{\delta }_{j,i}\right]\mathrm{=}{\sigma }^{\mathrm{2}}_{{\delta }_{j,i}}$ (see \sref{sec3_1} and \sref{sec3_3} for a detailed derivation) as follows:
\begin{equation} \label{ZEqnNum298267} 
 \begin{array}{c}
\begin{split}
\boldsymbol{\mathrm{CLT}}&\mathrm{:}\frac{\mathrm{1}}{\sqrt{\sum^{d\mathrm{-}\mathrm{1}}_{i\mathrm{=0}}{{\sigma }^{\mathrm{2}}_{{\delta }_{j,i}}}}}{\delta }_j\mathrm{=}\frac{\mathrm{1}}{\sqrt{\sum^{d\mathrm{-}\mathrm{1}}_{i\mathrm{=0}}{{\sigma }^{\mathrm{2}}_{{\delta }_{j,i}}}}}\left(\sum^{d\mathrm{-}\mathrm{1}}_{i\mathrm{=0}}{{\delta }_{j,i}}\right)\mathrm{\to }\mathcal{N}{\left(\mathrm{0,1}\right)}_d \\ 
&{\delta }_j\mathrm{=}\left(\sum^{d\mathrm{-}\mathrm{1}}_{i\mathrm{=0}}{{\delta }_{j,i}}\right)\mathrm{\to }\mathcal{N}\left(0,\sum^{d\mathrm{-}\mathrm{1}}_{i\mathrm{=0}}{{\sigma }^{\mathrm{2}}_{{\delta }_{j,i}}}\right)\mathrm{=}\mathcal{N}{\left(0,{\sigma }^{\mathrm{2}}_{{\delta }_{j,i}}\right)}_d. 
\end{split}
\end{array}
\end{equation} 
To show that \eqref{ZEqnNum298267} holds for the \textit{d}-dimensional noise parameter ${\delta }_j$, we exploit the Lyapunov-condition [\cref{r26}]. Applying the standard mathematical description of the Lyapunov condition [\cref{r45}], let $\mathfrak{L}\mathrm{>0}$, then
\begin{equation} \label{ZEqnNum257966} 
\mathop{\mathrm{lim}}_{d\mathrm{\to }\mathrm{\infty }}\frac{\mathrm{1}}{{\left(\sqrt{\sum^{d\mathrm{-}\mathrm{1}}_{i\mathrm{=0}}{{\sigma }^{\mathrm{2}}_{{\delta }_{j,i}}}}\right)}^{\mathrm{2+}\mathfrak{L}}}\sum^{d\mathrm{-}\mathrm{1}}_{i\mathrm{=0}}{\mathbb{E}\left[{\left|{\delta }_{j,i}\right|}^{\mathrm{2+}\mathfrak{L}}\right]}\mathrm{=0} 
\end{equation} 
is satisfied for any $d\mathrm{\to }\mathrm{\infty }$, by theory. As follows, the noise on ${\boldsymbol{\mathrm{U}}}_j\mathrm{\in }{\mathbb{R}}^d$ will converge to
\begin{equation} \label{45)} 
{\delta }_j\mathrm{=}\left(\sum^{d\mathrm{-}\mathrm{1}}_{i\mathrm{=0}}{{\delta }_{j,i}}\right)\mathrm{\in }\mathcal{N}{\left(0,{\sigma }^{\mathrm{2}}_{{\delta }_j}\right)}_d,                                       
\end{equation} 
and the resulting logical channel will be equivalent to a logical binary Gaussian channel with noise variance ${\sigma }^{\mathrm{2}}_{{\delta }_j}$. By the same argumentation, the variance of the resulting logical binary Gaussian channel will converge to the variance of the physical Gaussian quantum channel ${\sigma }^{\mathrm{2}}_{{\mathcal{N}}_{\mathrm{2}}}$ for $N\mathrm{\to }\mathrm{\infty }$.

 Let again $\mathfrak{L}\mathrm{>0}$, and \textit{d} is an appropriate dimension for which \eqref{ZEqnNum257966} is satisfied, and let the expected variance of ${\delta }_j$ is $\mathrm{var}\left[{\delta }_j\right]\mathrm{=}{\sigma }^{\mathrm{2}}_{{\mathcal{N}}_{\mathrm{2}}}$. Then
\begin{equation} \label{46)} 
\mathop{\mathrm{lim}}_{N\mathrm{\to }\mathrm{\infty }}\frac{\mathrm{1}}{{\left(\sqrt{\sum^{\left({N}/{d}\right)\mathrm{-}\mathrm{1}}_{j\mathrm{=0}}{{\sigma }^{\mathrm{2}}_{{\mathcal{N}}_{\mathrm{2}}}}}\right)}^{\mathrm{2+}\mathfrak{L}}}\sum^{\left({N}/{d}\right)\mathrm{-}\mathrm{1}}_{j\mathrm{=0}}{\mathbb{E}\left[{\left|{\delta }_j\right|}^{\mathrm{2+}\mathfrak{L}}\right]}\mathrm{=0},                          
\end{equation} 
is satisfied by theory, from which 
\begin{equation} \label{ZEqnNum168849} 
 \begin{array}{c}
\begin{split}
\boldsymbol{\mathrm{CLT}}&\mathrm{:}\frac{\mathrm{1}}{\sqrt{\sum^{\left({N}/{d}\right)\mathrm{-}\mathrm{1}}_{j\mathrm{=0}}{{\sigma }^{\mathrm{2}}_{{\mathcal{N}}_{\mathrm{2}}}}}}\left(\sum^{\left({N}/{d}\right)\mathrm{-}\mathrm{1}}_{j\mathrm{=0}}{{\delta }_j}\right)\mathrm{\to }\mathcal{N}{\left(\mathrm{0,1}\right)}_{{N}/{d}} \\ 
&\left(\sum^{\left({N}/{d}\right)\mathrm{-}\mathrm{1}}_{j\mathrm{=0}}{{\delta }_j}\right)\mathrm{\to }\mathcal{N}\left(0,\sum^{\left({N}/{d}\right)\mathrm{-}\mathrm{1}}_{j\mathrm{=0}}{{\sigma }^{\mathrm{2}}_{{\mathcal{N}}_{\mathrm{2}}}}\right)\mathrm{=}\mathcal{N}{\left(0,{\sigma }^{\mathrm{2}}_{{\mathcal{N}}_{\mathrm{2}}}\right)}_{{N}/{d}}, 
\end{split}
\end{array}
\end{equation} 
follows, which proves the statement. Hence one can readily recognize that 
\begin{equation} \label{48)} 
\mathop{\mathrm{lim}}_{N\mathrm{\to }\mathrm{\infty }}\mathrm{var}\left[{\delta }_{\mathrm{0\dots }\left({N}/{d}\right)\mathrm{-}\mathrm{1}}\right]\mathrm{=}{\left({\sigma }^{\mathrm{2}}_{{\mathcal{N}}_{\mathrm{2}}}\right)}_{{N}/{d}}.                                     
\end{equation} 
To conclude the situation, in \eqref{ZEqnNum298267} and \eqref{ZEqnNum168849} the variances of ${\delta }_j$ and $\sum^{\left({N}/{d}\right)\mathrm{-}\mathrm{1}}_{j\mathrm{=0}}{{\delta}_j}$, indeed, are not scaled up by \textit{d} and ${N}/{d}$, which makes possible to convert the physical Gaussian quantum channel to a logical binary Gaussian channel with noise variance $d{\sigma }^{\mathrm{2}}_{{\delta }_j}\mathrm{\approx }{\sigma }^{\mathrm{2}}_{{\mathcal{N}}_{\mathrm{2}}}$ for arbitrary \textit{d}. 

 These results allow for one to obtain the lowest noise variance and hence, the highest SNR of the logical channel that is possible by theory. At the resulting SNR, the capacity of the logical binary Gaussian channel also picks up its maximum. From this one can immediately conclude, that, in fact, it is a favorable result because the logical channel is indeed a binary Gaussian channel which is equipped with the same capacity at low SNRs (which is the situation in an experimental long-distance scenario) than the physical Gaussian quantum channel. In our solution, the lower bound ${\sigma }^{\mathrm{2}}_{{\delta }_j}\mathrm{=}{\sigma }^{\mathrm{2}}_{{\mathcal{N}}_{\mathrm{2}}}$ is precisely reached and is justified by the Lyapunov-condition, which means that our conversion provides the best approximation that is possible.
\end{proof}

\subsubsection{Application}
 In comparison to the multidimensional approaches, here, one can recognize that these results make no necessary the use of the multidimensional spherical space. The key idea is as follows: do the reconciliation in the scalar space to reduce the problem from ${\mathrm{\Gamma }}^{d\mathrm{-}\mathrm{1}}$ of ${\mathbb{R}}^d$ into $\mathbb{R}$. The main drawback of the multidimensional reconciliation approaches is the use of spherical space ${\mathrm{\Gamma }}^{d\mathrm{-}\mathrm{1}}$ of ${\mathbb{R}}^n$ to achieve the uniform distribution. As we have found in a CVQKD scenario it is not a required condition, and completely can be eliminated. The uniformly distributed elements of ${\mathbb{R}}^d$ have to be transmitted over the classical authenticated channel, but it \textit{per se}, does not imply that the reconciliation has to be executed in the spherical space\textit{. }The spherical correction of the errors of the raw data is a completely undesirable and unwanted event in a practical CVQKD, because it would just cause a further decrease in the very fragile\textit{,} sensitive, and so strenuously established secret key rates. The use of ${\mathrm{\Gamma }}^{d\mathrm{-}\mathrm{1}}$ of ${\mathbb{R}}^d$ served only one purpose in the multidimensional reconciliation: to guarantee the security requirements of the QKD post-processing phase. From this it immediately can be concluded that the use of spherical space is, in fact, unnecessary, and a mathematically equivalent and more efficient solution exists in the scalar space of $\mathbb{R}$. 

 One can recognize two improvements in our proposed scheme in comparison to the existing approaches. First, the uniform distribution will be reached by a simple operation, the Gaussian-CDF function applied separately on each unit of the raw data. Second, the approximation of the Gaussian channel will be justified by the CLT, using arbitrary dimensional vectors. As follows, the physical-logical channel conversion can be established with arbitrary high precision, since the $d\mathrm{\le }\mathrm{8}$ limitation has also been eliminated from the picture. To conclude, the spherical space can be replaced by the CDF transformation on the raw data units, and the Dirac distribution can be replaced by the CLT. It is clear now that the existing reconciliation methods require a revision since its application just leads to further slow-down in a practical CVQKD scenario. By these reasons, we drop away the spherical space, and instead of it, use the CDF-transformed units. These improvements allow very efficient decoding and error-correction, however, this step does not modify any property of the code: in other words, it keeps the desired uniform distribution and guarantees the arbitrary high-precision in the approximation of the logical binary Gaussian channel. Finally, we have to emphasize again that the whole reconciliation procedure is implemented through the logical layer only, without any need of physical-layer tomography. 

\subsection{Run of Scalar Reconciliation}
\label{sec3_3}
The run of scalar reconciliation (assuming reverse reconciliation) is sketched as follows. Bob divides his \textit{N}-unit length raw data $X\mathrm{'}$ into $n\mathrm{=}{N}/{d}$ number of \textit{d}-dimensional vectors ${{{\boldsymbol{\mathrm{X}}}'_{j}}}\mathrm{=}{\left({{{X}'_{j,0}}}\mathrm{,\dots ,}{{{X}'_{j,d-1}}}\right)}^T\\\mathrm{\in }{\mathbb{R}}^d$, where $d$ is the length of the vectors measured in units ${{{X}'_{j,i}}}$ in the raw data. 

 Then for each ${{{\boldsymbol{\mathrm{X}}}'_{j}}}$, applies CDF transformation \textit{C} on the units ${{{X}'_{j,i}}}\mathrm{\in }\mathbb{R}$ of ${{{\boldsymbol{\mathrm{X}}}'_{j}}}$, $for\ i\mathrm{=0,}\text{ }i\mathrm{\le }d\mathrm{-}\mathrm{1}$, $for\ j\mathrm{=0,}\text{ }j\mathrm{\le }\left({N}/{d}\right)\mathrm{-}\mathrm{1}$. Bob generates ${\boldsymbol{\mathrm{U}}}_j\mathrm{=}{\left(U_{j,0}\mathrm{\dots }U_{j,d\mathrm{-}\mathrm{1}}\right)}^T\mathrm{\in }{\mathbb{R}}^d,$ $U_{j,i}\mathrm{\in }\mathbb{R}$, computes $C\left({{{\boldsymbol{\mathrm{X}}}'_{j}}}\right){\boldsymbol{\mathrm{U}}}_j\\\mathrm{=}{\left(C\left({{{X}'_{j,0}}}\right)U_{j,0}\mathrm{,\dots ,}C\left({{{X}'_{j,d-1}}}\right)U_{j,d\mathrm{-}\mathrm{1}}\right)}^T$, and sends it to Alice over the classical authenticated channel. 

 Alice also divides her \textit{N}-unit length raw data $X$, into $n\mathrm{=}{N}/{d}$ number of \textit{d}-dimensional vectors ${\boldsymbol{\mathrm{X}}}_j\mathrm{=}{\left(X_{j,0}\mathrm{,\dots ,}X_{j,d\mathrm{-}\mathrm{1}}\right)}^T\mathrm{\in }{\mathbb{R}}^d$, computes the CDF-transformed $C\left({\boldsymbol{\mathrm{X}}}_j\right)\mathrm{=}{\left(C\left(X_{j,0}\right)\mathrm{,\dots ,}C\left(X_{j,d\mathrm{-}\mathrm{1}}\right)\right)}^T\mathrm{\in }{\mathbb{R}}^d$ and using \eqref{ZEqnNum462129}, \eqref{ZEqnNum391203} and \eqref{ZEqnNum766670} computes as
\begin{equation} \label{49)} 
 \begin{array}{l}
\begin{split}
{{{U}'_{j}}}&\mathrm{=}C\left({{{X}'_{j}}}\right)U_j\frac{\mathrm{1}}{C\left(X_j\right)} \\ 
&\mathrm{=}\sum^{d\mathrm{-}\mathrm{1}}_{i\mathrm{=0}}{{{{X}'_{j,i}}}U_{j,i}}\sum^{d\mathrm{-}\mathrm{1}}_{i\mathrm{=0}}{{{{U}'_{j,i}}}} \\ 
&\mathrm{=}\frac{\sum^{d\mathrm{-}\mathrm{1}}_{i\mathrm{=0}}{C\left({{{X}'_{j,i}}}\right)U_{j,i}}}{\sum^{d\mathrm{-}\mathrm{1}}_{i\mathrm{=0}}{C\left(X_{j,i}\right)}}. 
\end{split}
\end{array}
\end{equation} 
Next, she corrects the Gaussian noise on ${{{U}'_{j}}}$ to get $U_j$. From these she rebuilds the error-free full key 
\begin{equation} \label{50)} 
\boldsymbol{\mathrm{K}}\mathrm{\in }{\mathbb{R}}^{{N}/{d}}\mathrm{:}{\left(U_0\mathrm{,\dots ,}U_{\left({N}/{d}\right)\mathrm{-}\mathrm{1}}\right)}^T.                                        
\end{equation} 
 
\subsection{Security}
The scalar reconciliation provides unconditional security. It will be demonstrated for reverse reconciliation. The security of scalar reconciliation is guaranteed by the fact that the transmitted $C\left({{{\boldsymbol{\mathrm{X}}}'_{j}}}\right){\boldsymbol{\mathrm{U}}}_j$ messages follow \textit{uniform} distribution, and the multiplied ${\boldsymbol{\mathrm{U}}}_j$ and ${{{\boldsymbol{\mathrm{X}}}'_{j}}}$ vectors are also uniform and independent. 

 The following conditional probability holds for each $U_j$, $U_j\mathrm{=}U_{\mathrm{0...1}}$ (see also \eqref{ZEqnNum462129},\eqref{ZEqnNum391203} and \eqref{ZEqnNum766670}):
\begin{equation} \label{51)} 
\mathrm{Pr}\left(\left.U_j\mathrm{=}U_{0...\mathrm{1}}\right|C\left({{{X}'_{j}}}\right)U_j\right)\mathrm{=}\frac{\mathrm{1}}{\mathrm{2}}.                                    
\end{equation} 
Since $C\left({{{\boldsymbol{\mathrm{X}}}'_{j}}}\right){\boldsymbol{\mathrm{U}}}_j$ are uniformly distributed, and also independent [\cref{r11}], it follows that:
\begin{equation} \label{52)} 
\mathrm{Pr}\left(C\left({{{X}'_{j,i}}}\right)\mathrm{=}C\left({{{X}'_{j,0}}}\right)\mathrm{\dots }C\left({X\mathrm{'}}_{j,N\mathrm{-}\mathrm{1}}\right)\right)\mathrm{=}\frac{\mathrm{1}}{N} 
\end{equation} 
and
\begin{equation} \label{53)} 
\mathrm{Pr}\left(U_j\mathrm{=}U_{\mathrm{0...1}}\right)\mathrm{=}\frac{\mathrm{1}}{\mathrm{2}}.                                           
\end{equation} 
Since the overall number of \textit{d}-dimensional ${\boldsymbol{\mathrm{U}}}_j\mathrm{\in }{\mathbb{R}}^d$ vectors is ${N}/{d}$, the probability that Eve obtains the full key $\boldsymbol{\mathrm{K}}$ is
\begin{equation} \label{54)} 
\mathrm{P}{\mathrm{r}}_{Eve}\left(\boldsymbol{\mathrm{K}}\mathrm{=}{\left(U_0\mathrm{,\dots ,}U_{\left({N}/{d}\right)\mathrm{-}\mathrm{1}}\right)}^T\right)\mathrm{=}\frac{\mathrm{1}}{{\mathrm{2}}^{{N}/{d}}}. 
\end{equation} 

\subsection{Noise on the Data}
This section reveals the mathematical description of the noise vector of the Gaussian quantum channel ${\mathcal{N}}_{\mathrm{2}}$ and its impacts on Bob's raw data and Alice's received secret key. We also can exploit that in the evaluation of the noise vector only the second channel use ${\mathcal{N}}_{\mathrm{2}}$ has to be taken in to consideration in the error correction.

 The \textit{d}-dimensional \textit{noise vector }${\Delta}_j\in \mathcal{N}{\left(0,{\sigma }^2_{{\mathcal{N}}_2}\right)}_d\in {\mathbb{R}}^d$ of the Gaussian channel ${\mathcal{N}}_{\mathrm{2}}$ on the \textit{j}-th ${{{\boldsymbol{\mathrm{X}}}'_{j}}}$ is a Gaussian random vector defined as:
\begin{equation} \label{ZEqnNum190671} 
{\mathrm{\Delta }}_j\mathrm{=}{{{\boldsymbol{\mathrm{X}}}'_{j}}}\mathrm{-}{\boldsymbol{\mathrm{X}}}_j\mathrm{=}\left\{{\mathrm{\Delta }}_{j,0}\mathrm{,\dots ,}{\mathrm{\Delta }}_{j,d\mathrm{-}\mathrm{1}}\right\}\mathrm{\in }\mathcal{N}{\left(0,{\sigma }^{\mathrm{2}}_{{\mathcal{N}}_{\mathrm{2}}}\right)}_d\mathrm{\in }{\mathbb{R}}^d,                  
\end{equation} 
where ${\mathrm{\Delta }}_{j,i}\mathrm{\in }$$\mathcal{N}\left(0,{\sigma }^{\mathrm{2}}_{{\mathcal{N}}_{\mathrm{2}}}\right)\mathrm{\in }\mathbb{R}$ identifies the Gaussian noise on the \textit{i}-th unit $X_i$ of ${{{\boldsymbol{\mathrm{X}}}'_{j}}}$ as:
\begin{equation} \label{56)} 
{\mathrm{\Delta }}_{j,i}\mathrm{=}{{{X}'_{j,i}}}\mathrm{-}X_{j,i}\mathrm{\in }\mathcal{N}\left(0,{\sigma }^{\mathrm{2}}_{{\mathcal{N}}_{\mathrm{2}}}\right)\mathrm{\in }\mathbb{R}.                                  
\end{equation} 
The noise vector ${{\Delta}}_j$ is added to Alice's ${\boldsymbol{\mathrm{X}}}_j$, hence Bob's noisy ${{{\boldsymbol{\mathrm{X}}}'_{j}}}$ is: 
\begin{equation} \label{ZEqnNum938857} 
{{{{\boldsymbol{\mathrm{X}}}'_{j}}}}_j\mathrm{=}{\boldsymbol{\mathrm{X}}}_j\mathrm{+}{\mathrm{\Delta }}_j\mathrm{\in }{\mathbb{R}}^d.                                           
\end{equation} 
In terms of raw-data vector units, the Gaussian noise vector ${\mathrm{\Delta }}_{j,i}$ is described as follows:
\begin{equation} \label{58)} 
{{{X}'_{j,i}}}\mathrm{=}X_{j,i}\mathrm{+}{\mathrm{\Delta }}_{j,i}\mathrm{\in }\mathbb{R},                                           
\end{equation} 
and \eqref{ZEqnNum938857} can be rewritten as:
\begin{equation} \label{59)} 
 \begin{array}{l}
\begin{split}
{{{\boldsymbol{\mathrm{X}}}'_{j}}}&\mathrm{=}\left\{{{{X}'_{j,0}}}\mathrm{,\dots ,}{{{X}'_{j,d-1}}}\right\} \\ 
&\mathrm{=}\left\{X_{j,0}\mathrm{+}{\mathrm{\Delta }}_{j,0}\mathrm{,\dots ,}X_{j,d\mathrm{-}\mathrm{1}}\mathrm{+}{\mathrm{\Delta }}_{j,d\mathrm{-}\mathrm{1}}\right\}. 
\end{split}
\end{array}
\end{equation} 
In the scalar reconciliation, the error-correction is performed on the level of unit sums ${{{U}'_{j}}}\mathrm{=}\sum^{d\mathrm{-}\mathrm{1}}_{i\mathrm{=0}}{{{{U}'_{j,i}}}}$ in $\mathbb{R}$ as follows. Alice receives the \textit{d}-dimensional $C\left({{{\boldsymbol{\mathrm{X}}}'_{j}}}\right){\boldsymbol{\mathrm{U}}}_j$ from Bob, from which she obtains $C\left({{{X}'_{j}}}\right)U_j$ (see \eqref{ZEqnNum766670}) and divides it by her $C\left(X_j\right)$ (see \eqref{ZEqnNum391203}). The effect of Gaussian noise [\cref{r9}] results in a distorted secret ${{{U}'_{j}}}\mathrm{\in }\mathbb{R}$ as:
\begin{equation} \label{ZEqnNum310092} 
{{{U}'_{j}}}\mathrm{=}\sum^{d\mathrm{-}\mathrm{1}}_{i\mathrm{=0}}{{{{U}'_{j,i}}}}\mathrm{=}\frac{\sum^{d\mathrm{-}\mathrm{1}}_{i\mathrm{=0}}{C\left({{{X}'_{j,i}}}\right)U_{j,i}}}{\sum^{d\mathrm{-}\mathrm{1}}_{i\mathrm{=0}}{C\left(X_{j,i}\right)}}\mathrm{=}\sum^{d\mathrm{-}\mathrm{1}}_{i\mathrm{=0}}{U_{j,i}}\mathrm{+}\sum^{d\mathrm{-}\mathrm{1}}_{i\mathrm{=0}}{{\delta }_{j,i}}\mathrm{=}U_j\mathrm{+}{\delta }_j\mathrm{\in }\mathbb{R},       
\end{equation} 
where ${\delta }_{j,i}$ is the noise on $U_{j,i}$ (for a plausible example, see \sref{sec5}):
\begin{equation} \label{ZEqnNum549840} 
{\delta }_{j,i}\mathrm{=}\frac{U_{j,i}}{C\left(X_{j,i}\right)}C\left({\mathrm{\Delta }}_{j,i}\right)\mathrm{\in }\mathcal{N}\left(0,{\sigma }^{\mathrm{2}}_{{\delta }_{j,i}}\right), 
\end{equation} 
where ${\sigma }^{\mathrm{2}}_{{\delta }_{j,i}}$ is the variance of the distribution of ${\delta }_{j,i}$, while $C\left({\mathrm{\Delta }}_{j,i}\right)$ is the noise of the CDF-transformed raw data units:
\begin{equation} \label{ZEqnNum903756} 
C\left({\mathrm{\Delta }}_{j,i}\right)\mathrm{=}C\left({{{X}'_{j,i}}}\right)\mathrm{-}C\left(X_{j,i}\right)\mathrm{\in }\mathbb{R},                                
\end{equation} 
where $C\left({\mathrm{\Delta }}_{j,i}\right)\mathrm{\in }\mathcal{N}\left(0,{\sigma }^{\mathrm{2}}_{C\left({\mathrm{\Delta }}_{j,i}\right)}\right)$, and $C\left({\mathrm{\Delta }}_j\right)\mathrm{=}C\left({{{\boldsymbol{\mathrm{X}}}'_{j}}}\right)\mathrm{-}C\left({\boldsymbol{\mathrm{X}}}_j\right)\mathrm{\in }{\mathbb{R}}^d$, with a $\mathcal{N}{\left(0,{\sigma }^{\mathrm{2}}_{C\left({\mathrm{\Delta }}_j\right)}\right)}_d$ distribution. The error-corrected $U_j$ can be expressed from the noisy ${{{U}'_{j,i}}}$ as follows:
\begin{equation} \label{ZEqnNum652328} 
U_j\mathrm{=}\sum^{d\mathrm{-}\mathrm{1}}_{i\mathrm{=0}}{{{{U}'_{j,i}}}}\mathrm{-}\sum^{d\mathrm{-}\mathrm{1}}_{i\mathrm{=0}}{{\varsigma }_{j,i}}\mathrm{=}U_j\mathrm{-}{\varsigma }_j\mathrm{\in }\mathbb{R},                            
\end{equation} 
where ${\varsigma }_{j,i}\mathrm{\in }\mathcal{N}\left(0,{\sigma }^{\mathrm{2}}_{{\varsigma }_{j,i}}\right)$ characterizes the same amount of noise as \eqref{ZEqnNum549840}, i.e., and ${\varsigma }_{j,i}\mathrm{=}{\delta }_{j,i}$, however it is evaluated from the noisy raw-data units ${{{U}'_{j,i}}},$ $C\left({{{X}'_{j,i}}}\right)$ as: 
\begin{equation} \label{64)} 
{\varsigma }_{j,i}\mathrm{=}\frac{{{{U}'_{j,i}}}}{C\left({{{X}'_{j,i}}}\right)}C\left({\mathrm{\Delta }}_{j,i}\right)\mathrm{\in }\mathbb{R},                                          
\end{equation} 
with ${\varsigma }_{j,i}\mathrm{\in }\mathcal{N}\left(0,{\sigma }^{\mathrm{2}}_{{\varsigma }_{j,i}}\right)$. The \textit{d}-dimensional vector ${{{\boldsymbol{\mathrm{U}}}'_{j}}}\mathrm{\in }{\mathbb{R}}^d$ can be expressed as:
\begin{equation} \label{ZEqnNum718760} 
{{{\boldsymbol{\mathrm{U}}}'_{j}}}\mathrm{=}{\boldsymbol{\mathrm{U}}}_j\mathrm{+}{\overrightarrow{{{{\delta }_{j}}}}}\mathrm{\in }{\mathbb{R}}^d,                                             
\end{equation} 
where the noise vector ${\overrightarrow{{{{\delta }_{j}}}}}\mathrm{=}\left\{{\delta }_{j,0}\mathrm{,\dots ,}{\delta }_{j,d\mathrm{-}\mathrm{1}}\right\}\mathrm{\in }{\mathbb{R}}^d$ is as follows:
\begin{equation} \label{66)} 
{\overrightarrow{{{{\delta }_{j}}}}}\mathrm{=}\frac{{\boldsymbol{\mathrm{U}}}_j}{C\left({\boldsymbol{\mathrm{X}}}_j\right)}C\left({\mathrm{\Delta }}_j\right)\mathrm{\in }\mathcal{N}{\left(0,{\sigma }^{\mathrm{2}}_{{\delta }_j}\right)}_d\mathrm{=}\mathcal{N}\left(0,{\sigma }^{\mathrm{2}}_{{\delta }_{j,0}\mathrm{,\dots ,}{\delta }_{j,d\mathrm{-}\mathrm{1}}}\right). 
\end{equation} 
According to the CLT, the sum of independent noise on units ${{{U}'_{j,i}}}$ in ${{{\boldsymbol{\mathrm{U}}}'_{j}}}\mathrm{\in }{\mathbb{R}}^d$ is evaluated by a Gaussian random variable as: 
\begin{equation} \label{ZEqnNum820535} 
{\delta }_j\mathrm{=}\sum^{d\mathrm{-}\mathrm{1}}_{i\mathrm{=0}}{{\delta }_{j,i}\mathrm{=}\frac{\sum^{d\mathrm{-}\mathrm{1}}_{i\mathrm{=0}}{C\left({\Delta}_{j,i}\right)U_{j,i}}}{\sum^{d\mathrm{-}\mathrm{1}}_{i\mathrm{=0}}{C\left(X_{j,i}\right)}}}\mathrm{\in }\mathcal{N}\left(0,{\sigma }^{\mathrm{2}}_{{\delta }_j}\mathrm{=}\sum^{d\mathrm{-}\mathrm{1}}_{i\mathrm{=0}}{{\sigma }^{\mathrm{2}}_{{\delta }_{j,i}}}\right). 
\end{equation} 
The \textit{d}-dimensional vector ${\boldsymbol{\mathrm{U}}}_j\mathrm{\in }{\mathbb{R}}^d$ can be expressed as
\begin{equation} \label{68)} 
{\boldsymbol{\mathrm{U}}}_j\mathrm{=}{{{\boldsymbol{\mathrm{U}}}'_{j}}}\mathrm{-}{\overrightarrow{{{{\varsigma }_{j}}}}}\mathrm{\in }{\mathbb{R}}^{d},                                             
\end{equation} 
and the noise vector ${\overrightarrow{{{{\varsigma }_{j}}}}}\mathrm{=}\left\{{\varsigma }_{j,0}\mathrm{,\dots ,}{\varsigma }_{j,d\mathrm{-}\mathrm{1}}\right\}\mathrm{\in }{\mathbb{R}}^d$ is as follows: 
\begin{equation} \label{69)} 
{\overrightarrow{{{{\varsigma }_{j}}}}}\mathrm{=}\frac{{{{\boldsymbol{\mathrm{U}}}'_{j}}}}{C\left({\boldsymbol{\mathrm{X}}}_j\right)\mathrm{+}C\left({\mathrm{\Delta }}_j\right)}C\left({\mathrm{\Delta }}_j\right)\mathrm{\in }\mathcal{N}{\left(0,{\sigma }^{\mathrm{2}}_{{\overrightarrow{{{{\varsigma }_{j}}}}}}\right)}_d. 
\end{equation} 
The sum of independent noise on units ${{{U}'_{j,i}}}$ of ${{{\boldsymbol{\mathrm{U}}}'_{j}}}\mathrm{\in }{\mathbb{R}}^d$ can also be identified as:
\begin{equation} \label{70)} 
{\varsigma }_j\mathrm{=}\sum^{d\mathrm{-}\mathrm{1}}_{i\mathrm{=0}}{{\varsigma }_{j,i}}\mathrm{=}\frac{\sum^{d\mathrm{-}\mathrm{1}}_{i\mathrm{=0}}{C\left({\Delta}_{j,i}\right){{{U}'_{j,i}}}}}{\sum^{d\mathrm{-}\mathrm{1}}_{i\mathrm{=0}}{C\left({{{X}'_{j,i}}}\right)}}\mathrm{=}\mathcal{N}\left(0,{\sigma }^{\mathrm{2}}_{{\varsigma }_j}\mathrm{=}\sum^{d\mathrm{-}\mathrm{1}}_{i\mathrm{=0}}{{\sigma }^{\mathrm{2}}_{{\varsigma }_{j,i}}}\right). 
\end{equation} 
From the physical properties of a Gaussian quantum channel [\cref{r1}-\cref{r11}], we know exactly what happens during the transmission of the coherent combined signal from Alice to Bob. The noise on ${{{X}'_{j,i}}}$ has a non-standard Gaussian random distribution ${\mathrm{\Delta }}_{j,i}\mathrm{\in }\mathcal{N}\left(0,{\sigma }^{\mathrm{2}}_{{\mathcal{N}}_{\mathrm{2}}}\right)$. 

We have to analyze in detail the properties of the noise vector. The vector ${\mathrm{\Delta }}_j\mathrm{\in }\mathcal{N}{\left(0,{\sigma }^{\mathrm{2}}_{{\mathcal{N}}_{\mathrm{2}}}\right)}_d\mathrm{\in }{\mathbb{R}}^d$ of ${{\mathcal{N}}_{2}}$ that generates the noisy ${{{\boldsymbol{\mathrm{X}}}'_{j}}}$ from ${\boldsymbol{\mathrm{X}}}_j$ is characterized as follows. First we decompose the noise vector ${\mathrm{\Delta }}_j$ into its components: 
\begin{equation} \label{71)} 
{\mathrm{\Delta }}_j\mathrm{=}{\boldsymbol{\mathrm{A}}}_j{\mathrm{\Lambda }}_j,                                                    
\end{equation} 
where matrix ${\boldsymbol{\mathrm{A}}}_j$ represents a linear transformation in ${\mathbb{R}}^d$, while ${\mathrm{\Lambda }}_j$ is a the standard Gaussian noise vector ${\mathrm{\Lambda }}_j\mathrm{\in }\mathcal{N}{\left(\mathrm{0,1}\right)}_d\mathrm{\in }{\mathbb{R}}^d$. The probability density function of ${\mathrm{\Lambda }}_j$ is: 
\begin{equation} \label{72)} 
f\left({\mathrm{\Lambda }}_j\right)\mathrm{=}\frac{\mathrm{1}}{{\left(\sqrt{\mathrm{2}\pi }\right)}^d}e^{\mathrm{-}\frac{{\left\|{\mathrm{\Lambda }}_j\right\|}^{\mathrm{2}}}{\mathrm{2}}},                                          
\end{equation} 
where $\left\|{\mathrm{\Lambda }}_j\right\|\mathrm{=}\sqrt{{\mathrm{\Lambda }}^{\mathrm{2}}_{j,0}\mathrm{+\dots +}{\mathrm{\Lambda }}^{\mathrm{2}}_{j,d\mathrm{-}\mathrm{1}}}$ is magnitude, in other words, the Euclidean distance from the origin to ${\mathrm{\Lambda }}_j\mathrm{\in }{\mathbb{R}}^d$. This type of noise exhibits different behavior than the real Gaussian noise of a quantum channel, and it is characterized by the same magnitude $\left\|{\mathrm{\Lambda }}_{j}\right\|$ in every direction. This property is connected to the standard Gaussian random noise, and it cannot be applied in a realistic CVQKD scenario, because it does not properly describe the noise characteristic of the quantum channel. The probability density function of ${\mathrm{\Delta }}_j\mathrm{\in }{\mathbb{R}}^d$ is:
\begin{equation} \label{73)} 
f\left({\mathrm{\Delta }}_j\right)\mathrm{=}\frac{\mathrm{1}}{{\left(\sqrt{\mathrm{2}\pi }\right)}^d\sqrt{\mathrm{det}{\boldsymbol{\mathrm{A}}}_j{\boldsymbol{\mathrm{A}}}^T_j}}e^{\mathrm{-}\frac{\mathrm{1}}{\mathrm{2}}{\mathrm{\Delta }}^T_j{\left({\boldsymbol{\mathrm{A}}}_j{\boldsymbol{\mathrm{A}}}^T_j\right)}^{\mathrm{-}\mathrm{1}}{\mathrm{\Delta }}_j},                           
\end{equation} 
where ${\boldsymbol{\mathrm{A}}}_j{\boldsymbol{\mathrm{A}}}^T_j$ stands for the $\mathfrak{C}\left({\mathrm{\Delta }}_j\right)$ covariance matrix of ${\mathrm{\Delta }}_j$, and it analogous of ${\sigma }^{\mathrm{2}}_{{\mathcal{N}}_{\mathrm{2}}}$, i.e., in a more precise form $\mathfrak{C}\left({\mathrm{\Delta }}_j\right)\mathrm{=}\mathbb{E}\left({\mathrm{\Delta }}_j{\mathrm{\Delta }}^T_j\right)\mathrm{=}{\boldsymbol{\mathrm{A}}}_j{\boldsymbol{\mathrm{A}}}^T_j$. The noise on the units ${{{X}'_{j,i}}}$ of ${{{\boldsymbol{\mathrm{X}}}'_{j}}}$ at Bob's side arises from the quantum-level transmission of the combined phase space states $\left|\left.{\phi }_{j,i}\right\rangle \right.\mathrm{\in }{\mathcal{S}}_{A\mathrm{\times }B}$, and vectors ${\mathrm{\Lambda }}_j\mathrm{\in }\mathcal{N}{\left(\mathrm{0,1}\right)}_d$ and ${\mathrm{\Delta }}_j\mathrm{\in }\mathcal{N}{\left(0,{\sigma }^{\mathrm{2}}_{{\mathcal{N}}_{\mathrm{2}}}\right)}_d$ is built up by \textit{d} components, ${\mathrm{\Lambda }}_{j,i}$$\mathrm{\in }\mathcal{N}\left(\mathrm{0,1}\right)\mathrm{\in }\mathbb{R}$ and ${\mathrm{\Delta }}_{j,i}\mathrm{\in }\mathcal{N}\left(0,{\sigma }^{\mathrm{2}}_{{\mathcal{N}}_{\mathrm{2}}}\right)\mathrm{\in }\mathbb{R}$. The error ${\mathrm{\Delta }}_{j,i}$ on the \textit{i}-th unit ${{{X}'_{j,i}}}$ is as follows:
\begin{equation} \label{74)} 
{\mathrm{\Delta }}_{j,i}\mathrm{=}{\boldsymbol{\mathrm{A}}}_{j,i}{\mathrm{\Lambda }}_{j,i}, for\ i\mathrm{=0,}\text{ }i\mathrm{\le }d\mathrm{-}\mathrm{1},                                
\end{equation} 
where ${\boldsymbol{\mathrm{A}}}_{j,i}$ is a linear transformation that scales ${\mathrm{\Lambda }}_{j,i}$. The probability density function of ${\mathrm{\Lambda }}_{j,i}$ is: 
\begin{equation} \label{75)} 
f\left({\mathrm{\Lambda }}_{j,i}\right)\mathrm{=}\frac{\mathrm{1}}{\sqrt{\mathrm{2}\pi }}e^{\mathrm{-}\frac{{\left\|{\mathrm{\Lambda }}_{j,i}\right\|}^{\mathrm{2}}}{\mathrm{2}}},                                           
\end{equation} 
where $\left\|{\mathrm{\Lambda }}_{j,i}\right\|\mathrm{=}\sqrt{{\mathrm{\Lambda }}^{\mathrm{2}}_{j,i}}$ is the magnitude of ${\mathrm{\Lambda }}_{j,i}$. The probability density function of ${\mathrm{\Delta }}_{j,i}$ is:
\begin{equation} \label{76)} 
f\left({\mathrm{\Delta }}_{j,i}\right)\mathrm{=}\frac{\mathrm{1}}{\sqrt{\mathrm{2}\pi }\sqrt{\mathrm{det}{\boldsymbol{\mathrm{A}}}_{j,i}{\boldsymbol{\mathrm{A}}}^T_{j,i}}}e^{\mathrm{-}\frac{\mathrm{1}}{\mathrm{2}}{\mathrm{\Delta }}^T_{j,i}{\left({\boldsymbol{\mathrm{A}}}_{j,i}{\boldsymbol{\mathrm{A}}}^T_{j,i}\right)}^{\mathrm{-}\mathrm{1}}{\mathrm{\Delta }}_{j,i}}, 
\end{equation} 
where ${\boldsymbol{\mathrm{A}}}_{j,i}{\boldsymbol{\mathrm{A}}}^T_{j,i}\mathrm{=}\mathbb{E}\left({\mathrm{\Delta }}_{j,i}{\mathrm{\Delta }}^T_{j,i}\right)=\mathfrak{C}\left({\mathrm{\Delta }}_{j,i}\right)$. 

 From ${\mathrm{\Lambda }}_{j,i}$ and ${\mathrm{\Delta }}_{j,i}$, the correction of Bob's noisy secret ${\boldsymbol{\mathrm{U}}}_j$ can be approached by the units $\left\{{{{U}'_{j,0}}}\mathrm{,\dots ,}{{{U}'_{j,d\mathrm{-}\mathrm{1}}}}\right\}$, because the noise of ${\mathcal{N}}_{\mathrm{2}}$ is survived in the raw data level and lives also on ${{{U}'_{j,i}}}$, but in a modified form, see \eqref{ZEqnNum549840}. 

 Let us denote by $\left|\left.{\phi }_{j,i}\right\rangle \right.$ the phase-space representation of Alice's noise-free raw data unit $X_{j,i}$ given by \eqref{ZEqnNum542216}, and by $\left|\left.{\xi }_{j,i}\right\rangle \right.$ the noisy raw data unit ${{{X}'_{j,i}}}$ of Bob, from \eqref{ZEqnNum576926}. (State $\left|\left.{\phi }_{j,i}\right\rangle \right.$ is the second mode of the combined beam, while $\left|\left.{\xi }_{j,i}\right\rangle \right.$ is its noisy version). 

\begin{center}
\begin{figure}[h!]
\begin{center}
\includegraphics[angle = 0,width=1\linewidth]{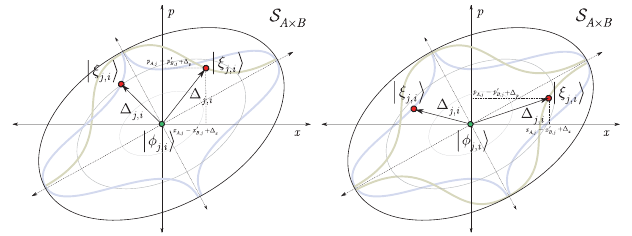}
\caption{The real Gaussian noise of the quantum channel ${\mathcal{N}}_{\mathrm{2}}$ causes a rotation and rescaled vector in the combined phase space ${\mathcal{S}}_{A\mathrm{\times }B}$ (\textit{x}: position quadrature, \textit{p}: momentum quadrature). The magnitude $\left\|{\mathrm{\Delta }}_{j,i}\right\|$ of the noise vector ${\mathrm{\Delta }}_{j,i}\mathrm{\in }\mathcal{N}\left(0,{\sigma }^{\mathrm{2}}_{{\mathcal{N}}_{\mathrm{2}}}\right)$ is not preserved, since the noise characteristic describes an ellipse in the combined phase space.} 
 \label{fig4}
 \end{center}
\end{figure}
\end{center}

 The effect of the real Gaussian noise of the quantum channel is shown in \fref{fig4}. The noise vector ${\mathrm{\Delta }}_j\mathrm{\in }\mathcal{N}{\left(0,{\sigma }^{\mathrm{2}}_{{\mathcal{N}}_{\mathrm{2}}}\right)}_d\mathrm{\in }{\mathcal{S}}_{A\mathrm{\times }B}$ of the quantum channel is a non-standard Gaussian random vector, which distorts the density. The circles of ${\mathrm{\Lambda }}_{j,i}\mathrm{\in }\mathcal{N}\left(\mathrm{0,1}\right)$ are scaled by ${\boldsymbol{\mathrm{A}}}_{j,i}$ resulting in ellipses. The magnitude $\left\|{\mathrm{\Delta }}_{j,i}\right\|$ of ${\mathrm{\Delta }}_{j,i}$ is not preserved in all directions, which leads to different density. The \textit{x} and \textit{p} quadratures of $\left|\left.{\phi }_{j,i}\right\rangle \right.\mathrm{\in }{\mathcal{S}}_{A\mathrm{\times }B}$ are modified by ${\mathrm{\Delta }}_x$ and ${\mathrm{\Delta }}_p$ in $\left|\left.{\xi }_{j,i}\right\rangle \right.\mathrm{\in }{\mathcal{S}}_{A\mathrm{\times }B}$.

\section{Theorems and Proofs}
\label{sec4}
 First we show that Alice's noisy secret can be corrected in the $\boldsymbol{\mathrm{v}}$ vector space of ${\mathbb{R}}^d$ by using an error-correction rule based on the apparatus provided by the maximum-likelihood decision [\cref{r15}-\cref{r19}], [\cref{r24},\cref{r25}], which renders unnecessary the use of the spherical space of ${\mathrm{\Gamma }}^{d\mathrm{-}\mathrm{1}}$. 

\begin{proposition}
(Vector reconciliation of correlated Gaussian variables).\textbf{ }The Gaussian noise ${\delta }_j$ on the received vector ${{{\boldsymbol{\mathrm{U}}}'_{j}}}\mathrm{\in }{\mathbb{R}}^d\mathrm{:}\left\{{{{U}'_{j,0}}}\mathrm{,\dots ,}{{{U}'_{j,d\mathrm{-}\mathrm{1}}}}\right\}$ can be corrected in the vector space $\boldsymbol{\mathrm{v}}$ of ${\mathbb{R}}^d$.
\end{proposition}
\begin{proof}
 First, Bob selects the \textit{d}-dimensional vector ${\boldsymbol{\mathrm{U}}}_j\mathrm{\in }\left\{U_{j,0}\mathrm{,\dots ,}U_{j,d\mathrm{-}\mathrm{1}}\right\}\mathrm{\in }{\mathbb{R}}^d$ where $\sum^{d\mathrm{-}\mathrm{1}}_{i\mathrm{=0}}{U_{j,i}}\mathrm{=}a$ or $\sum^{d\mathrm{-}\mathrm{1}}_{i\mathrm{=0}}{U_{j,i}}\mathrm{=}b$, $U_{j,i}\mathrm{\in }\mathcal{U}$ and sends $C\left({{{\boldsymbol{\mathrm{X}}}'_{j}}}\right){\boldsymbol{\mathrm{U}}}_j$ over the classical channel. Alice uses her CDF-transformed raw data $C\left({\boldsymbol{\mathrm{X}}}_j\right)\mathrm{=}\left\{C\left(X_{j,0}\right)\mathrm{,\dots ,}C\left(X_{j,d\mathrm{-}\mathrm{1}}\right)\right\}$ to obtain ${{{\boldsymbol{\mathrm{U}}}'_{j}}}\mathrm{\in }{\mathbb{R}}^d$\textit{. }Since Alice knows $a$, $b$ and $d$, she can draw two vectors $\boldsymbol{\mathrm{A}}\mathrm{=}{\left(A_0\mathrm{,\dots ,}A_{d\mathrm{-}\mathrm{1}}\right)}^T\mathrm{\in }{\mathbb{R}}^d,$ with norm $\left\|\boldsymbol{\mathrm{A}}\right\|\mathrm{=}\sqrt{\sum^{d\mathrm{-}\mathrm{1}}_{i\mathrm{=0}}{{\left(A_i\right)}^{\mathrm{2}}}}$, where $\left\{\sum^{d\mathrm{-}\mathrm{1}}_{i\mathrm{=0}}{A_i}\mathrm{=}a\right\}$, $A_i\mathrm{\in }\mathcal{U}$ and $\boldsymbol{\mathrm{B}}\mathrm{=}{\left(B_0\mathrm{,\dots ,}B_{d\mathrm{-}\mathrm{1}}\right)}^T\mathrm{\in }{\mathbb{R}}^d,$ with $\left\|\boldsymbol{\mathrm{B}}\right\|\mathrm{=}\sqrt{\sum^{d\mathrm{-}\mathrm{1}}_{i\mathrm{=0}}{{\left(B_i\right)}^{\mathrm{2}}}}$, where $\left\{\sum^{d\mathrm{-}\mathrm{1}}_{i\mathrm{=0}}{B_i}\mathrm{=}b\right\}$, ${{B}}_{\boldsymbol{i}}\boldsymbol{\mathrm{\in }}\mathcal{U}$. She then corrects the noise on ${{{\boldsymbol{\mathrm{U}}}'_{j}}}$ by the following error-correction rule [\cref{r15}-\cref{r19}]:
\begin{equation} \label{ZEqnNum121990} 
{\boldsymbol{\mathrm{U}}}_j\mathrm{=}\boldsymbol{\mathrm{A}}\mathrm{:}\left\|{{{\boldsymbol{\mathrm{U}}}'_{j}}}\mathrm{-}\boldsymbol{\mathrm{A}}\right\|\mathrm{<}\left\|{{{\boldsymbol{\mathrm{U}}}'_{j}}}\mathrm{-}\boldsymbol{\mathrm{B}}\right\|,                                  
\end{equation} 
\begin{equation} \label{ZEqnNum872486} 
{\boldsymbol{\mathrm{U}}}_j\mathrm{=}\boldsymbol{\mathrm{B}}\mathrm{:}\left\|{{{\boldsymbol{\mathrm{U}}}'_{j}}}\mathrm{-}\boldsymbol{\mathrm{A}}\right\|\mathrm{>}\left\|{{{\boldsymbol{\mathrm{U}}}'_{j}}}\mathrm{-}\boldsymbol{\mathrm{B}}\right\|,                                  
\end{equation} 
where the quantity $\left\|{{{\boldsymbol{\mathrm{U}}}'_{j}}}\mathrm{-}{\boldsymbol{\mathrm{U}}}_j\right\|$, ${\boldsymbol{\mathrm{U}}}_j\mathrm{\in }\left\{\boldsymbol{\mathrm{A}},\boldsymbol{\mathrm{B}}\right\}$ is evaluated as
\begin{equation} \label{ZEqnNum730241} 
 \begin{array}{l}
\begin{split}
\left\|{{{\boldsymbol{\mathrm{U}}}'_{j}}}\mathrm{-}{\boldsymbol{\mathrm{U}}}_j\right\|\mathrm{=}\sqrt{\sum^{d\mathrm{-}\mathrm{1}}_{i\mathrm{=0}}{{\left({{{U}'_{j,i}}}\mathrm{-}U_{j,i}\right)}^{\mathrm{2}}}}&\mathrm{=}\sqrt{\sum^{d\mathrm{-}\mathrm{1}}_{i\mathrm{=0}}{{\left(\frac{U_{j,i}}{C\left(X_{j,i}\right)}C\left({\Delta}_{i,j}\right)\right)}^{\mathrm{2}}}} \\ 
&\mathrm{=}\sqrt{\sum^{d\mathrm{-}\mathrm{1}}_{i\mathrm{=0}}{{\left({\delta }_{j,i}\right)}^{\mathrm{2}}}}\mathrm{=}\left\|{\overrightarrow{{{{\delta }_{j}}}}}\right\|, 
\end{split}
\end{array}
\end{equation} 
which precisely coincidences with the norm of the Gaussian noise in \eqref{ZEqnNum820535}. However, since Alice does not know Bob's $U_{j,i}$, in \eqref{ZEqnNum730241} an additional noise, ${\mathrm{\Upsilon }}_j$, also brings up, i.e., $\left\|{{{\boldsymbol{\mathrm{U}}}'_{j}}}\mathrm{-}{\boldsymbol{\mathrm{U}}}_j\right\|\mathrm{=}\left\|{\delta }_j\mathrm{+}{\mathrm{\Upsilon }}_j\right\|$. The noise vector ${\overrightarrow{{{{\Upsilon }_{j}}}}}$ with expected variance ${\sigma }^{\mathrm{2}}_{{\overrightarrow{{{{\Upsilon }_{j}}}}}}$ is independent from the real noise on ${{{U}'_{j,i}}}$. This problem will be resolved in Theorem 1 and will be shown that this quantity completely vanishes from the picture. 

 Alice receives the \textit{d}-dimensional vectors ${{{\boldsymbol{\mathrm{U}}}'_{j}}}\mathrm{\in }\left\{{{{U}'_{j,0}}}\mathrm{,\dots ,}{{{U}'_{j,d\mathrm{-}\mathrm{1}}}}\right\}\mathrm{\in }{\mathbb{R}}^d$, and corrects ${{{\boldsymbol{\mathrm{U}}}'_{j}}}$ into ${\boldsymbol{\mathrm{U}}}_j$ and then from the components she rebuilds the full key $\boldsymbol{\mathrm{K}}\mathrm{=}{\left(U_0\mathrm{,\dots ,}U_{\left({N}/{d}\right)\mathrm{-}\mathrm{1}}\right)}^T\mathrm{\in }{\mathbb{R}}^{{N}/{d}}$. The error-vector ${\overrightarrow{{{{\delta }_{j}}}}}\mathrm{\in }{\mathbb{R}}^d$ on a given noisy ${{{\boldsymbol{\mathrm{U}}}'_{j}}}$ is 
\begin{equation} \label{ZEqnNum300012} 
 \begin{array}{l}
\begin{split}
{\overrightarrow{{{{\delta }_{j}}}}}\mathrm{=}{\delta }_{j,i}\mathrm{=}{\left(\frac{{\boldsymbol{\mathrm{U}}}_j}{C\left({\boldsymbol{\mathrm{X}}}_j\right)}\right)}^TC\left({\mathrm{\Delta }}_j\right)&\mathrm{\in }\mathcal{N}{\left(0,{\sigma }^{\mathrm{2}}_{{\overrightarrow{{{{\delta }_{j}}}}}}=\mathfrak{C}\left({\left(\frac{{\boldsymbol{\mathrm{U}}}_j}{C\left({\boldsymbol{\mathrm{X}}}_j\right)}\right)}^TC\left({\mathrm{\Delta }}_j\right)\right)\right)}_d \\ 
&\mathrm{=}\mathcal{N}\left(0,{\sigma }^{\mathrm{2}}_{{\delta }_{j,i}}=\mathfrak{C}\left({\left(\frac{U_{j,i}}{C\left(X_{j,i}\right)}\right)}^TC\left({\mathrm{\Delta }}_{j,i}\right)\right)\right)\mathrm{\in }{\mathbb{R}}^d\mathrm{,0}\mathrm{\le }i\mathrm{\le }d\mathrm{-}\mathrm{1,} 
\end{split}
\end{array}
\end{equation} 
The covariance matrix of \eqref{ZEqnNum300012} is expressed as:
\begin{equation} \label{81)} 
\mathfrak{C}\left({\left(\frac{{\boldsymbol{\mathrm{U}}}_j}{C\left({\boldsymbol{\mathrm{X}}}_j\right)}\right)}^TC\left({\mathrm{\Delta }}_j\right)\right)\mathrm{=}\mathbb{E}\left({\left(\frac{{\boldsymbol{\mathrm{U}}}_j}{C\left({\boldsymbol{\mathrm{X}}}_j\right)}\right)}^TC\left({\mathrm{\Delta }}_j\right){\left({\left(\frac{{\boldsymbol{\mathrm{U}}}_j}{C\left({\boldsymbol{\mathrm{X}}}_{\delta}\right)}\right)}^TC\left({\mathrm{\Delta }}_j\right)\right)}^T\right)\mathrm{=}{\left({\sigma }^{\mathrm{2}}_{{\overrightarrow{{{{\delta }_{j}}}}}}\right)}_d 
\end{equation} 
along with
\begin{equation} \label{ZEqnNum614922} 
{\delta }_{j,i}\mathrm{=}{\left(\frac{U_{j,i}}{C\left(X_{j,i}\right)}\right)}^TC\left({\mathrm{\Delta }}_{j,i}\right)\mathrm{\in }\mathcal{N}\left(0,{\sigma }^{\mathrm{2}}_{{\delta}_{j,i}}=\mathfrak{C}\left({\left(\frac{U_{j,i}}{C\left(X_{j,i}\right)}\right)}^TC\left({\mathrm{\Delta }}_{j,i}\right)\right)\right)\mathrm{\in }\mathbb{R}\mathrm{,} 
\end{equation} 
and \eqref{ZEqnNum614922} is characterized by covariance matrix
\begin{equation} \label{83)} 
\mathfrak{C}\left(\frac{U_{j,i}}{C\left(X_{j,i}\right)}C\left({\mathrm{\Delta }}_{j,i}\right)\right)\mathrm{=}\mathbb{E}\left(\frac{U_{j,i}}{C\left(X_{j,i}\right)}C\left({\mathrm{\Delta }}_{j,i}\right){\left(\frac{U_{j,i}}{C\left(X_{j,i}\right)}C\left({\mathrm{\Delta }}_{j,i}\right)\right)}^T\right)\mathrm{=}{\sigma }^{\mathrm{2}}_{{\delta }_{j,i}}. 
\end{equation} 
The error-corrected ${\boldsymbol{\mathrm{U}}}_j$ can be expressed as:
\begin{equation} \label{84)} 
{\boldsymbol{\mathrm{U}}}_j\mathrm{=}{{{\boldsymbol{\mathrm{U}}}'_{j}}}\mathrm{-}{\overrightarrow{{{{\varsigma }_{j}}}}}\mathrm{\in }{\mathbb{R}}^d,                                            
\end{equation} 
where
\begin{equation} \label{ZEqnNum538290} 
 \begin{array}{l}
\begin{split}
{\overrightarrow{{{{\varsigma }_{j}}}}}\mathrm{=}{\left(\frac{{{{\boldsymbol{\mathrm{U}}}'_{j}}}}{C\left({\boldsymbol{\mathrm{X}}}_j\right)\mathrm{+}C\left({\mathrm{\Delta }}_j\right)}\right)}^TC\left({\mathrm{\Delta }}_j\right)&\mathrm{\in }\mathcal{N}{\left(0,{\sigma }^{\mathrm{2}}_{{\overrightarrow{{{{\varsigma }_{j}}}}}}=\mathfrak{C}\left({\left(\frac{{{{\boldsymbol{\mathrm{U}}}'_{j}}}}{C\left({\boldsymbol{\mathrm{X}}}_j\right)\mathrm{+}C\left({\mathrm{\Delta }}_j\right)}\right)}^TC\left({\mathrm{\Delta }}_j\right)\right)\right)}_d \\ 
&\mathrm{=}\mathcal{N}\left(0,{\sigma }^{\mathrm{2}}_{{\varsigma }_j}=\mathfrak{C}\left(\frac{{{{U}'_{j,i}}}}{C\left(X_{j,i}\right)\mathrm{+}C\left({\mathrm{\Delta }}_{j,i}\right)}C\left({\mathrm{\Delta }}_{j,i}\right)\right)\right)\mathrm{\in }{\mathbb{R}}^d\mathrm{,0}\mathrm{\le }i\mathrm{\le }d\mathrm{-}\mathrm{1.} 
\end{split}
\end{array}
\end{equation} 
The covariance matrix of \eqref{ZEqnNum538290} is as follows:
\begin{equation} \label{ZEqnNum615035} 
\begin{split}
\mathfrak{C}\left({\left(\frac{{{{\boldsymbol{\mathrm{U}}}'_{j}}}}{C\left({\boldsymbol{\mathrm{X}}}_j\right)\mathrm{+}C\left({\mathrm{\Delta }}_j\right)}\right)}^TC\left({\mathrm{\Delta }}_j\right)\right)&\mathrm{=}\mathbb{E}\left({\left(\frac{{{{\boldsymbol{\mathrm{U}}}'_{j}}}}{C\left({\boldsymbol{\mathrm{X}}}_j\right)\mathrm{+}C\left({\mathrm{\Delta }}_j\right)}\right)}^TC\left({\mathrm{\Delta }}_j\right){\left({\left(\frac{{{{\boldsymbol{\mathrm{U}}}'_{j}}}}{C\left({\boldsymbol{\mathrm{X}}}_j\right)\mathrm{+}C\left({\mathrm{\Delta }}_j\right)}\right)}^TC\left({\mathrm{\Delta }}_j\right)\right)}^T\right)\\
&\mathrm{=}{\left({\sigma }^{\mathrm{2}}_{{\overrightarrow{{{{\varsigma }_{j}}}}}}\right)}_d 
\end{split}
\end{equation} 
and
\begin{equation} \label{ZEqnNum506352} 
{\varsigma }_{j,i}\mathrm{=}\frac{{{{U}'_{j,i}}}}{C\left(X_{j,i}\right)\mathrm{+}C\left({\mathrm{\Delta }}_{j,i}\right)}C\left({\mathrm{\Delta }}_{j,i}\right)\mathrm{\in }\mathcal{N}\left(0,{\sigma }^{\mathrm{2}}_{{\varsigma }_{j,i}}=\mathfrak{C}\left(\frac{{{{U}'_{j,i}}}}{C\left(X_{j,i}\right)\mathrm{+}C\left({\mathrm{\Delta }}_{j,i}\right)}C\left({\mathrm{\Delta }}_{j,i}\right)\right)\right), 
\end{equation} 
along with
\begin{equation} \label{88)} 
\begin{split}
\mathfrak{C}\left(\frac{{{{U}'_{j,i}}}}{C\left(X_{j,i}\right)\mathrm{+}C\left({\mathrm{\Delta }}_{j,i}\right)}C\left({\mathrm{\Delta }}_{j,i}\right)\right)&\mathrm{=}\mathbb{E}\left(\frac{{{{U}'_{j,i}}}}{C\left(X_{j,i}\right)\mathrm{+}C\left({\mathrm{\Delta }}_{j,i}\right)}C\left({\mathrm{\Delta }}_{j,i}\right){\left(\frac{{{{U}'_{j,i}}}}{C\left(X_{j,i}\right)\mathrm{+}C\left({\mathrm{\Delta }}_{j,i}\right)}C\left({\mathrm{\Delta }}_{j,i}\right)\right)}^T\right)\\
&\mathrm{=}{\sigma }^{\mathrm{2}}_{{\varsigma }_{j,i}}. 
\end{split}
\end{equation} 
From \eqref{ZEqnNum614922} and \eqref{ZEqnNum506352} the quantities $U_{j,i}$ and ${{{U}'_{j,i}}}$ are evaluated as follows:
\begin{equation} \label{89)} 
U_{j,i}\mathrm{=}{{{U}'_{j,i}}}\mathrm{-}\frac{{{{U}'_{j,i}}}}{C\left({{{X}'_{j,i}}}\right)}C\left({\mathrm{\Delta }}_{j,i}\right)\mathrm{=}{{{U}'_{j,i}}}\mathrm{-}{\varsigma }_{j,i}\mathrm{\in }\mathbb{R},                        
\end{equation} 
and
\begin{equation} \label{90)} 
{{{U}'_{j,i}}}\mathrm{=}\frac{C\left({{{X}'_{j,i}}}\right)}{C\left(X_{j,i}\right)}U_{j,i}\mathrm{=}U_{j,i}\mathrm{+}{\delta }_{j,i}\mathrm{\in }\mathbb{R}\mathrm{.} 
\end{equation} 
Let us denote by $\nu $ the standard deviation of ${\overrightarrow{{{{\delta }_{j}}}}}\mathrm{+}{\overrightarrow{{{{\Upsilon }_{j}}}}}\mathrm{=}{\delta }_{j,i}\mathrm{+}{\mathrm{\Upsilon }}_{j,i},\mathrm{\ }\mathrm{0}\mathrm{\le }i\mathrm{\le }d\mathrm{-}\mathrm{1,}$ which is evaluated from \eqref{ZEqnNum615035} and ${\sigma }^{\mathrm{2}}_{{\overrightarrow{{{{\Upsilon }_{j}}}}}}$ as
\begin{equation} \label{91)} 
\nu \mathrm{=}\sqrt{{\left({\sigma }^{\mathrm{2}}_{{\overrightarrow{{{{\delta }_{j}}}}}}\mathrm{+}{\sigma }^{\mathrm{2}}_{{\overrightarrow{{{{\Upsilon }_{j}}}}}}\right)}_d}.                                              
\end{equation} 
The maximum-likelihood-based correction rules can be given in the form of: 
\begin{equation} \label{92)} 
{\boldsymbol{\mathrm{U}}}_j\mathrm{=}\boldsymbol{\mathrm{A}}\mathrm{:}\frac{\mathrm{1}}{{\left(\pi \mathrm{2}{\nu }^{\mathrm{2}}\right)}^{{d}/{\mathrm{2}}}}e^{\mathrm{-}\frac{{\left\|{{{\boldsymbol{\mathrm{U}}}'_{j}}}\mathrm{-}\boldsymbol{\mathrm{A}}\right\|}^{\mathrm{2}}}{\mathrm{2}{\nu }^{\mathrm{2}}}}\mathrm{\ge }\frac{\mathrm{1}}{{\left(\pi \mathrm{2}{\nu }^{\mathrm{2}}\right)}^{{d}/{\mathrm{2}}}}e^{\mathrm{-}\frac{{\left\|{{{\boldsymbol{\mathrm{U}}}'_{j}}}\mathrm{-}\boldsymbol{\mathrm{B}}\right\|}^{\mathrm{2}}}{\mathrm{2}{\nu }^{\mathrm{2}}}}, 
\end{equation} 
and:
\begin{equation} \label{93)} 
{\boldsymbol{\mathrm{U}}}_j\mathrm{=}\boldsymbol{\mathrm{B}}\mathrm{:}\frac{\mathrm{1}}{{\left(\pi \mathrm{2}{\nu }^{\mathrm{2}}\right)}^{{d}/{\mathrm{2}}}}e^{\mathrm{-}\frac{{\left\|{{{\boldsymbol{\mathrm{U}}}'_{j}}}\mathrm{-}\boldsymbol{\mathrm{A}}\right\|}^{\mathrm{2}}}{\mathrm{2}{\nu }^{\mathrm{2}}}}\mathrm{\le }\frac{\mathrm{1}}{{\left(\pi \mathrm{2}{\nu }^{\mathrm{2}}\right)}^{{d}/{\mathrm{2}}}}e^{\mathrm{-}\frac{{\left\|{{{\boldsymbol{\mathrm{U}}}'_{j}}}\mathrm{-}\boldsymbol{\mathrm{B}}\right\|}^{\mathrm{2}}}{\mathrm{2}{\nu }^{\mathrm{2}}}}.                          
\end{equation} 
The error probability for the case of decoding vector ${\boldsymbol{\mathrm{U}}}_j\mathrm{=}\boldsymbol{\mathrm{A}}$, is
\begin{equation} \label{94)} 
\mathrm{P}{\mathrm{r}}_e\left({\left\|{\overrightarrow{{{{\delta }_{j}}}}}\mathrm{+}{\overrightarrow{{{{\Upsilon }_{j}}}}}\right\|}^{\mathrm{2}}\mathrm{>}{\left\|\left(\boldsymbol{\mathrm{A}}\mathrm{+}{\overrightarrow{{{{\delta }_{j}}}}}\mathrm{+}{\overrightarrow{{{{\Upsilon }_{j}}}}}\right)\mathrm{-}\boldsymbol{\mathrm{B}}\right\|}^{\mathrm{2}}\right)\mathrm{=P}{\mathrm{r}}_e\left({\left(\boldsymbol{\mathrm{A}}\mathrm{-}\boldsymbol{\mathrm{B}}\right)}^T\left({\overrightarrow{{{{\delta }_{j}}}}}\mathrm{+}{\overrightarrow{{{{\Upsilon }_{j}}}}}\right)\mathrm{<-}\frac{{\left\|\boldsymbol{\mathrm{A}}\mathrm{-}\boldsymbol{\mathrm{B}}\right\|}^{\mathrm{2}}}{\mathrm{2}}\right). 
\end{equation} 
For the case of correction of ${\boldsymbol{\mathrm{U}}}_j\mathrm{=}\boldsymbol{\mathrm{B}}$, the error probabilities are evaluated as
\begin{equation} \label{95)} 
\mathrm{P}{\mathrm{r}}_e\left({\left\|{\overrightarrow{{{{\delta }_{j}}}}}\mathrm{+}{\overrightarrow{{{{\Upsilon }_{j}}}}}\right\|}^{\mathrm{2}}\mathrm{>}{\left\|\left(\boldsymbol{\mathrm{B}}\mathrm{+}{\overrightarrow{{{{\delta }_{j}}}}}\mathrm{+}{\overrightarrow{{{{\Upsilon }_{j}}}}}\right)\mathrm{-}\boldsymbol{\mathrm{A}}\right\|}^{\mathrm{2}}\right)\mathrm{=P}{\mathrm{r}}_e\left({\left(\boldsymbol{\mathrm{B}}\mathrm{-}\boldsymbol{\mathrm{A}}\right)}^T\left({\overrightarrow{{{{\delta }_{j}}}}}\mathrm{+}{\overrightarrow{{{{\Upsilon }_{j}}}}}\right)\mathrm{<-}\frac{{\left\|\boldsymbol{\mathrm{B}}\mathrm{-}\boldsymbol{\mathrm{A}}\right\|}^{\mathrm{2}}}{\mathrm{2}}\right). 
\end{equation} 
The decision regions can be separated into two hyperplanes ${\mathcal{H}}_{\mathrm{1}}$ and ${\mathcal{H}}_{\mathrm{2}}$ along $\boldsymbol{\mathrm{B}}\mathrm{-}\boldsymbol{\mathrm{A}}$, which separate ${\boldsymbol{\mathrm{U}}}_j\mathrm{=}\boldsymbol{\mathrm{A}}$ and ${\boldsymbol{\mathrm{U}}}_j\mathrm{=}\boldsymbol{\mathrm{B}}$. In other words, the correction-condition of a given noisy ${{{\boldsymbol{\mathrm{U}}}'_{j}}}$ is reduced to the following decision problem:
\begin{equation} \label{96)} 
{\boldsymbol{\mathrm{U}}}_j\mathrm{=}\left\{ \begin{array}{l}
\boldsymbol{\mathrm{A}},\text{if }{{{\boldsymbol{\mathrm{U}}}'_{j}}}\mathrm{\in }{\mathcal{H}}_{\mathrm{1}}, \\ 
\boldsymbol{\mathrm{B}},\text{if }{{{\boldsymbol{\mathrm{U}}}'_{j}}}\mathrm{\in }{\mathcal{H}}_{\mathrm{2}}. \end{array}
\right. 
\end{equation} 
As follows, by applying the procedure Alice can retrieve ${\boldsymbol{\mathrm{U}}}_j\mathrm{\in }\left\{\boldsymbol{\mathrm{A}},\boldsymbol{\mathrm{B}}\right\}$ from the noisy ${\boldsymbol{\mathrm{U}}}_j$ in the vector space$\boldsymbol{\mathrm{v}}$ of ${\mathbb{R}}^d$. From the error-corrected ${\boldsymbol{\mathrm{U}}}_j$ components, Alice finally rebuilds the full key vector $\boldsymbol{\mathrm{K}}\mathrm{=}{\left(U_0\mathrm{,\dots ,}U_{\left({N}/{d}\right)\mathrm{-}\mathrm{1}}\right)}^T\mathrm{\in }{\mathbb{R}}^{{N}/{d}}$, which concludes the proof. 
\end{proof}

 Proposition 1 demonstrated that there is no need for the use of ${\mathrm{\Gamma }}^{d\mathrm{-}\mathrm{1}}$ of ${\mathbb{R}}^d$ in the error correction, however the corrected noise is not precisely a Gaussian. Theorem 1 reveals that the reconciliation process, in fact, does not require vector operations in ${\mathbb{R}}^d$, and the noise is a real Gaussian noise in the scalar space $\mathbb{R}$. 

 \begin{theorem}\text{(Scalar reconciliation of correlated Gaussian variables).} The Gaussian noise ${\delta }_j$ on the received scalar ${{{U}'_{j}}}\mathrm{=}\sum^{d\mathrm{-}\mathrm{1}}_{i\mathrm{=}0}{{{{U}'_{j,i}}}}$ can be corrected in $\mathbb{R}$.
\end{theorem}
\begin{proof}
 We exploit that the noise on ${{{U}'_{j,i}}}$-s is ${\delta }_{j,i}\mathrm{=}\frac{U_{j,i}}{C\left(X_{j,i}\right)}C\left({\mathrm{\Delta }}_{j,i}\right)\mathrm{\in }\mathcal{N}\left(0,{\sigma }^{\mathrm{2}}_{{\delta }_{j,i}}\right)$, while on the sum of the noise of the \textit{d} units is a zero-mean Gaussian random variable $\sum^{d\mathrm{-}\mathrm{1}}_{i\mathrm{=0}}{{\delta }_{j,i}}\mathrm{\in }\mathcal{N}\left(0,{\sigma }^{\mathrm{2}}_{{\delta }_j}\right)$, that is justified by the CLT and the Lyapunov-condition. Alice will correct the units in the following form:\textit{}
\begin{equation} \label{97)} 
{{{U}'_{j}}}\mathrm{=}\sum^{d\mathrm{-}\mathrm{1}}_{i\mathrm{=0}}{{{{U}'_{j,i}}}}\mathrm{=}\frac{\sum^{d\mathrm{-}\mathrm{1}}_{i\mathrm{=0}}{C\left({{{X}'_{j,i}}}\right)U_{j,i}}}{\sum^{d\mathrm{-}\mathrm{1}}_{i\mathrm{=0}}{C\left(X_{j,i}\right)}}\mathrm{=}U_j\mathrm{+}{\delta }_j\mathrm{\in }\mathbb{R}\mathrm{.} 
\end{equation} 
First, expresses the secret vector ${\boldsymbol{\mathrm{U}}}_j\mathrm{\in }{\mathbb{R}}^d$ as follows:
\begin{equation} \label{ZEqnNum331821} 
{\boldsymbol{\mathrm{U}}}_j\mathrm{=}x\left(\boldsymbol{\mathrm{A}}\mathrm{-}\boldsymbol{\mathrm{B}}\right)\mathrm{+}\frac{\mathrm{1}}{\mathrm{2}}\left(\boldsymbol{\mathrm{A}}\mathrm{+}\boldsymbol{\mathrm{B}}\right),                                      
\end{equation} 
where $x\mathrm{\in }\left\{\mathrm{-}\mathrm{0.5,0.5}\right\}\mathrm{\in }\mathbb{R}$ is a scalar. From this, Alice can also rewrite the noisy ${{{\boldsymbol{\mathrm{U}}}'_{j}}}$ as:
\begin{equation} \label{ZEqnNum971030} 
{{{\boldsymbol{\mathrm{U}}}'_{j}}}\mathrm{=}x\left(\boldsymbol{\mathrm{A}}\mathrm{-}\boldsymbol{\mathrm{B}}\right)\mathrm{+}\frac{\mathrm{1}}{\mathrm{2}}\left(\boldsymbol{\mathrm{A}}\mathrm{+}\boldsymbol{\mathrm{B}}\right)\mathrm{+}{\overrightarrow{{{{\delta }_{j}}}}}.                                 
\end{equation} 
From \eqref{ZEqnNum971030} follows that:
\begin{equation} \label{ZEqnNum435931} 
 \begin{array}{l}
\begin{split}
{{{U}'_{j}}}&\mathrm{=}\sum^{d\mathrm{-}\mathrm{1}}_{i\mathrm{=0}}{\left(x\left(A_i\mathrm{-}B_i\right)\mathrm{+}\frac{\mathrm{1}}{\mathrm{2}}\left(A_i\mathrm{+}B_i\right)\mathrm{+}{\delta }_{j,i}\right)} \\ 
&\mathrm{=}\sum^{d\mathrm{-}\mathrm{1}}_{i\mathrm{=0}}{\left(x\left(A_i\mathrm{-}B_i\right)\mathrm{+}\frac{\mathrm{1}}{\mathrm{2}}\left(A_i\mathrm{+}B_i\right)\right)\mathrm{+}{\delta }_j} \\ 
&\mathrm{=}\sum^{d\mathrm{-}\mathrm{1}}_{i\mathrm{=0}}{{{{U}'_{j,i}}}} \\ 
&\mathrm{=}U_j\mathrm{+}\frac{U_j}{C\left(X_j\right)}C\left({\mathrm{\Delta }}_j\right), 
\end{split}
\end{array}
\end{equation} 
where $C\left(X_j\right)\mathrm{=}\sum^{d\mathrm{-}\mathrm{1}}_{i\mathrm{=0}}{C\left(X_{j,i}\right)},C\left({\mathrm{\Delta }}_j\right)\mathrm{=}\sum^{d\mathrm{-}\mathrm{1}}_{i\mathrm{=0}}{C\left({\mathrm{\Delta }}_{j,i}\right)}$, $U_j\mathrm{=}\sum^{d\mathrm{-}\mathrm{1}}_{i\mathrm{=0}}{U_{j,i}}$ and ${\delta}_j\mathrm{=}\sum^{d\mathrm{-}\mathrm{1}}_{i\mathrm{=0}}{{\delta }_{j,i}}$. 

 In fact, Alice does not have to use all elements from \eqref{ZEqnNum435931}, because she can apply a simpler process. For this purpose, she draws a new vector, $\boldsymbol{\mathrm{d}}$:
\begin{equation} \label{ZEqnNum441379} 
\boldsymbol{\mathrm{d}}\mathrm{=}\frac{\boldsymbol{\mathrm{A}}\mathrm{-}\boldsymbol{\mathrm{B}}}{\left\|\boldsymbol{\mathrm{A}}\mathrm{-}\boldsymbol{\mathrm{B}}\right\|},                                                    
\end{equation} 
where $\left\|\boldsymbol{\mathrm{A}}\mathrm{-}\boldsymbol{\mathrm{B}}\right\|\mathrm{=}\sqrt{\sum^{d\mathrm{-}\mathrm{1}}_{i\mathrm{=0}}{{\left(A_i\mathrm{-}B_i\right)}^{\mathrm{2}}}}$ is the effective distance of $\boldsymbol{\mathrm{A}}$ and $\boldsymbol{\mathrm{B}}$. A useful property of vector $\boldsymbol{\mathrm{d}}$ drawn in \eqref{ZEqnNum441379}, that any independent noise [\cref{r15}] (i.e., independent from the noise on ${{{\boldsymbol{\mathrm{U}}}'_{j}}}$) could live only in the orthogonal directions to $\boldsymbol{\mathrm{d}}$, i.e., $\left({\boldsymbol{\mathrm{n}}}_{\mathrm{1}}\mathrm{,\dots ,}{\boldsymbol{\mathrm{n}}}_l\right)\mathrm{\bot }\boldsymbol{\mathrm{d}}$. It immediately follows, that the ${\boldsymbol{\mathrm{n}}}_{\mathrm{1}}\mathrm{,\dots ,}{\boldsymbol{\mathrm{n}}}_l$ orthogonal directions will have no further importance for Alice in the decoding [\cref{r15}-\cref{r19}]. Since $x$ is a scalar and in \eqref{ZEqnNum971030} the term $\frac{\mathrm{1}}{\mathrm{2}}\left(\boldsymbol{\mathrm{A}}\mathrm{+}\boldsymbol{\mathrm{B}}\right)$ is a constant, Alice introduces vector $\chi \mathrm{\in }\boldsymbol{\mathrm{v}}$ as follows:
\begin{equation} \label{102)} 
\chi \mathrm{\equiv }{{{\boldsymbol{\mathrm{U}}}'_{j}}}\mathrm{-}\frac{\mathrm{1}}{\mathrm{2}}\left(\boldsymbol{\mathrm{A}}\mathrm{+}\boldsymbol{\mathrm{B}}\right)\mathrm{=}x\left(\boldsymbol{\mathrm{A}}\mathrm{-}\boldsymbol{\mathrm{B}}\right)\mathrm{+}{\overrightarrow{{{{\delta }_{j}}}}}.                              
\end{equation} 
She also draws an orthogonal matrix $\boldsymbol{\mathrm{M}}$, which contains $\boldsymbol{\mathrm{d}}$ and the orthogonal directions ${\boldsymbol{\mathrm{n}}}_{\mathrm{1}}\mathrm{,\dots ,}{\boldsymbol{\mathrm{n}}}_l$ with unit norm as:
\begin{equation} \label{103)} 
\mathbf{M}=\left( \begin{matrix}
   \mathbf{d}  \\
   {{\mathbf{n}}_{1}}  \\
   {{\mathbf{n}}_{2}}  \\
   \vdots   \\
   {{\mathbf{n}}_{l}}  \\
\end{matrix} \right)
\end{equation} 
By multiplying $\boldsymbol{\mathrm{M}}$ with $\chi $ leads to:
\begin{equation} \label{ZEqnNum611432} 
\mathbf{M}\chi =\left( \begin{matrix}
   x\left\| \mathbf{A}-\mathbf{B} \right\|  \\
   0  \\
   0  \\
   \vdots   \\
   0  \\
\end{matrix} \right)+\mathbf{M}{{\vec{\delta }}_{j}}.
\end{equation} 
From \eqref{ZEqnNum611432}, it clearly follows that only $x\left\|\boldsymbol{\mathrm{A}}\mathrm{-}\boldsymbol{\mathrm{B}}\right\|$ and the first component of $\boldsymbol{\mathrm{M}}{\overrightarrow{{{{\delta }_{j}}}}}$ have relevance in the error-correction process, because all of the other components are orthogonal to $\boldsymbol{\mathrm{d}}$ [\cref{r15}]. Since the evolution of $\boldsymbol{\mathrm{d}}$ is a trivial process on Alice's side, the received ${{{\boldsymbol{\mathrm{U}}}'_{j}}}$ can be projected by $\mathcal{P}$ onto the direction of $\boldsymbol{\mathrm{d}}$, since all valuable information including the real noise is carried only by this direction. The projection $\mathcal{P}$ on ${{{\boldsymbol{\mathrm{U}}}'_{j}}}$ is made by ${\boldsymbol{\mathrm{d}}}^T\chi $, which then results in:
\begin{equation} \label{105)} 
 \begin{array}{l}
\begin{split}
\mathcal{P}\left({{{\boldsymbol{\mathrm{U}}}'_{j}}}\right)&\mathrm{=}{\boldsymbol{\mathrm{d}}}^T\chi  \\ 
&\mathrm{=}{\left(\frac{\boldsymbol{\mathrm{A}}\mathrm{-}\boldsymbol{\mathrm{B}}}{\left\|\boldsymbol{\mathrm{A}}\mathrm{-}\boldsymbol{\mathrm{B}}\right\|}\right)}^T\left(x\left(\boldsymbol{\mathrm{A}}\mathrm{-}\boldsymbol{\mathrm{B}}\right)\mathrm{+}{\overrightarrow{{{{\delta }_{j}}}}}\right) \\ 
&\mathrm{=}{\boldsymbol{\mathrm{d}}}^T\left({{{\boldsymbol{\mathrm{U}}}'_{j}}}\mathrm{-}\frac{\mathrm{1}}{\mathrm{2}}\left(\boldsymbol{\mathrm{A}}\mathrm{+}\boldsymbol{\mathrm{B}}\right)\right). 
\end{split}
\end{array}
\end{equation} 
The projected vector $\mathcal{P}\left({{{\boldsymbol{\mathrm{U}}}'_{j}}}\right)$ is analogous to the scalar representation $U_j\mathrm{=}\sum^{d\mathrm{-}\mathrm{1}}_{i\mathrm{=0}}{U_{j,i}}$ in $\mathbb{R}$, and makes it possible to correct the noise in the scalar space $\mathbb{R}$. The received ${{{U}'_{j}}}\mathrm{=}U_j\mathrm{+}{\delta }_j$ has mean ${\mu }_a\mathrm{=}a$ or ${\mu }_b\mathrm{=}b$, and the decision boundary is $\frac{{\mu }_a\mathrm{+}{\mu }_b}{\mathrm{2}}$, which defines a separator in $\mathbb{R}$. 

 According to the previously obtained calculations, \eqref{ZEqnNum611432} can be rewritten as follows:
\begin{equation} \label{106)} 
\mathbf{M}\chi =\left( \begin{matrix}
   x\sqrt{\sum\nolimits_{i=0}^{d-1}{{{\left( {{A}_{i}}-{{B}_{i}} \right)}^{2}}}}  \\
   0  \\
   0  \\
   \vdots   \\
   0  \\
\end{matrix} \right)+{{\delta }_{j}}.
\end{equation} 
As follows, only the first component of $\boldsymbol{\mathrm{M}}{\overrightarrow{{{{\delta }_{j}}}}}$ has relevance in the error-correction, which in particular coincidences with the scalar quantity ${\delta }_j\mathrm{=}\sum^{d\mathrm{-}\mathrm{1}}_{i\mathrm{=0}}{{\delta }_{j,i}}\mathrm{=}\frac{\sum^{d\mathrm{-}\mathrm{1}}_{i\mathrm{=0}}{C\left({\mathrm{\Delta }}_{j,i}\right)U_{j,i}}}{\sum^{d\mathrm{-}\mathrm{1}}_{i\mathrm{=0}}{C\left(X_{j,i}\right)}}$ shown in \eqref{ZEqnNum730241}. Putting the pieces together, $\mathcal{P}\left({{{\boldsymbol{\mathrm{U}}}'_{j}}}\right)$ is evaluated as: 
\begin{equation} \label{107)} 
\mathcal{P}\left({{{\boldsymbol{\mathrm{U}}}'_{j}}}\right)\mathrm{=}x\sqrt{\sum^{d\mathrm{-}\mathrm{1}}_{i\mathrm{=0}}{{\left(A_i\mathrm{-}B_i\right)}^{\mathrm{2}}}}\mathrm{+}\sum^{d\mathrm{-}\mathrm{1}}_{i\mathrm{=0}}{{\delta }_{j,i}},                          
\end{equation} 
which contains all sufficient information for the error correction in $\mathbb{R}$; the proof is concluded here.
\end{proof}

 In Theorem 2 the error probability of scalar reconciliation is proposed in an exact form.
\begin{theorem} The error probability $\mathrm{Pr}\left(error\right)\mathrm{=}Q\left(\frac{\left|a\mathrm{-}b\right|}{\mathrm{2}}\frac{\mathrm{1}}{\eta }\right)$ of scalar reconciliation depends only on $\left|a\mathrm{-}b\right|$, where $Q\left(\frac{\left|a\mathrm{-}b\right|}{\mathrm{2}}\frac{\mathrm{1}}{\eta }\right)\mathrm{=Pr}\left(\frac{\left|a\mathrm{-}b\right|}{\mathrm{2}}\frac{\mathrm{1}}{\eta }\mathrm{<}g\right)$ is the Q-function (tail function), g  is a standard Gaussian random variable $g\mathrm{\in }\mathcal{N}\left(\mathrm{0,1}\right)$, and $\eta \mathrm{=}\sqrt{{\sigma }^{\mathrm{2}}_{{\delta }_j}}\mathrm{=}\sqrt{\sum^{d\mathrm{-}\mathrm{1}}_{i\mathrm{=0}}{{\sigma }^{\mathrm{2}}_{{\delta }_{j,i}}}}$ is the standard deviation of the Gaussian noise ${\delta }_j$. The $\mathrm{Pr}\left(error\right)$ exponentially converges to zero for any $\left|a\mathrm{-}b\right|\mathrm{>2}\eta $.
\end{theorem}
\begin{proof}
Let $U_j\mathrm{=}\sum^{d\mathrm{-}\mathrm{1}}_{i\mathrm{=0}}{U_{j,i}}$ from \eqref{ZEqnNum435931}, $C\left(X_j\right)\mathrm{=}\sum^{d\mathrm{-}\mathrm{1}}_{i\mathrm{=0}}{C\left(X_{j,i}\right)}$ and $C\left({\mathrm{\Delta }}_j\right)\mathrm{=}\sum^{d\mathrm{-}\mathrm{1}}_{i\mathrm{=0}}{C\left({\mathrm{\Delta }}_{j,i}\right)}$. Exploiting the result of Theorem 1, in the scalar reconciliation process Alice decides on the scalar quantity ${{{U}'_{j}}}\mathrm{=}a$, if:
\begin{equation} \label{108)} 
\mathrm{Pr}\left(\left.U_j\mathrm{=}a\right|{{{U}'_{j}}}\right)\mathrm{\ge }\mathrm{Pr}\left(\left.U_j\mathrm{=}b\right|{{{U}'_{j}}}\right).                                 
\end{equation} 
Similarly, she decides on ${{{U}'_{j}}}\mathrm{=}b$, if:
\begin{equation} \label{109)} 
\mathrm{Pr}\left(\left.U_j\mathrm{=}b\right|{{{U}'_{j}}}\right)\mathrm{\ge }\mathrm{Pr}\left(\left.U_j\mathrm{=}a\right|{{{U}'_{j}}}\right).                                 
\end{equation} 
Conditioned on $a$ or $b$, the received ${{{U}'_{j}}}$ has mean ${\mu }_a\mathrm{=}a$ or ${\mu }_b\mathrm{=}b$, with $\mathcal{N}\left({\mu }_a,{\eta }^{\mathrm{2}}\right)$ and $\mathcal{N}\left({\mu }_b,{\eta }^{\mathrm{2}}\right)$. Applying the maximum-likelihood-based correction rule [\cref{r15}-\cref{r19}], Alice calculates with the following inequalities:
\begin{equation} \label{110)} 
\frac{\mathrm{1}}{\sqrt{\mathrm{2}\pi {\eta }^{\mathrm{2}}}}e^{\left(\mathrm{-}\frac{{\left({{{U}'_{j}}}\mathrm{-}a\right)}^{\mathrm{2}}}{\mathrm{2}{\eta }^{\mathrm{2}}}\right)}\mathrm{\ge }\frac{\mathrm{1}}{\sqrt{\mathrm{2}\pi {\eta }^{\mathrm{2}}}}e^{\left(\mathrm{-}\frac{{\left({{{U}'_{j}}}\mathrm{-}b\right)}^{\mathrm{2}}}{\mathrm{2}{\eta }^{\mathrm{2}}}\right)} 
\end{equation} 
and:
\begin{equation} \label{111)} 
\frac{\mathrm{1}}{\sqrt{\mathrm{2}\pi {\eta }^{\mathrm{2}}}}e^{\left(\mathrm{-}\frac{{\left({{{U}'_{j}}}\mathrm{-}b\right)}^{\mathrm{2}}}{\mathrm{2}{\eta }^{\mathrm{2}}}\right)}\mathrm{\ge }\frac{\mathrm{1}}{\sqrt{\mathrm{2}\pi {\eta }^{\mathrm{2}}}}e^{\left(\mathrm{-}\frac{{\left({{{U}'_{j}}}\mathrm{-}a\right)}^{\mathrm{2}}}{\mathrm{2}{\eta }^{\mathrm{2}}}\right)},                                   
\end{equation} 
which then leads to (for a comparison see \eqref{ZEqnNum121990} and \eqref{ZEqnNum872486}):
\begin{equation} \label{112)} 
\left|{{{U}'_{j}}}\mathrm{-}a\right|\mathrm{<}\left|{{{U}'_{j}}}\mathrm{-}b\right| 
\end{equation} 
and:
\begin{equation} \label{113)} 
\left|{{{U}'_{j}}}\mathrm{-}a\right|\mathrm{>}\left|{{{U}'_{j}}}\mathrm{-}b\right|.                                             
\end{equation} 
The received ${{{U}'_{j}}}$ has mean ${\mu }_a\mathrm{=}a$ or ${\mu }_b\mathrm{=}b$, hence one obtains the following conditional probability for an error event, conditioned on Bob has sent $U_j\mathrm{=}a$:   
\begin{equation} \label{ZEqnNum699835} 
\mathrm{Pr}\left({{{U}'_{j}}}\mathrm{=}\frac{U_j}{C\left(X_j\right)}C\left({\mathrm{\Delta }}_j\right)\mathrm{<}\left.\frac{{\mu }_a\mathrm{+}{\mu }_b}{\mathrm{2}}\right|U_j\mathrm{=}a\right)\mathrm{=Pr}\left(\left({{{U}'_{j}}}\mathrm{-}U_j\right)\mathrm{>}\frac{\left|{\mu }_a\mathrm{-}{\mu }_b\right|}{\mathrm{2}}\right), 
\end{equation} 
where $\frac{\left|{\mu }_a\mathrm{-}{\mu }_b\right|}{\mathrm{2}}$ assigns a decision boundary. The tail function $Q\left(\frac{\left|a-b\right|}{2}\frac{1}{\eta }\right)=Pr\left(\frac{\left|a-b\right|}{2}\frac{1}{\eta }<g\right)$, where $g\in \mathcal{N}\left(0,1\right)$\textit{, }has exponential decay for any $\left|a-b\right|>2\eta $, hence: 
\begin{equation} \label{ZEqnNum849863} 
\frac{\mathrm{1}}{\sqrt{\mathrm{2}\pi }\left(\frac{\left|a\mathrm{-}b\right|}{\mathrm{2}}\frac{\mathrm{1}}{\eta }\right)}\left(\mathrm{1-}\frac{\mathrm{1}}{{\left(\frac{\left|a\mathrm{-}b\right|}{\mathrm{2}}\frac{\mathrm{1}}{\eta }\right)}^{\mathrm{2}}}\right)e^{\mathrm{-}\frac{{\left(\frac{\left|a\mathrm{-}b\right|}{\mathrm{2}}\frac{\mathrm{1}}{\eta }\right)}^{\mathrm{2}}}{\mathrm{2}}}\mathrm{<}Q\left(\frac{\left|a\mathrm{-}b\right|}{\mathrm{2}}\frac{\mathrm{1}}{\eta }\right)\mathrm{<}e^{\mathrm{-}\frac{{\left(\frac{\left|a\mathrm{-}b\right|}{\mathrm{2}}\frac{\mathrm{1}}{\eta }\right)}^{\mathrm{2}}}{\mathrm{2}}}, 
\end{equation} 
which clearly demonstrates that the error probability of scalar reconciliation exponentially converges to zero. As one can readily obtain from \eqref{ZEqnNum849863}, for arbitrary large differences between \textit{a} and \textit{b}, $Q\left(\frac{\left|a\mathrm{-}b\right|}{\mathrm{2}}\frac{\mathrm{1}}{\eta }\right)\mathrm{\to }\mathrm{0}$ [\cref{r15}-\cref{r17}]. Then, by applying the maximum-likelihood decision theory and the Bayes' rule [\cref{r15}-\cref{r19}], for a given ${U}_j$ one obtains error probability via the tail function: 
\begin{equation} \label{116)} 
 \begin{array}{l}
\begin{split}
\mathrm{Pr}\left({{{U}'_{j}}}\mathrm{<}\left.\frac{{\mu }_a\mathrm{+}{\mu }_b}{\mathrm{2}}\right|U_j\mathrm{=}a\right)&\mathrm{=}Q\left(\frac{\left|a\mathrm{-}b\right|}{\mathrm{2}}\frac{\mathrm{1}}{\eta }\right) \\ 
&\mathrm{=Pr}\left(\frac{\left|a\mathrm{-}b\right|}{\mathrm{2}}\frac{\mathrm{1}}{\eta }\mathrm{<}g\right) \\ 
&\mathrm{=Pr}\left(error\right), 
\end{split}
\end{array}
\end{equation} 
where $g\mathrm{\in }\mathcal{N}\left(\mathrm{0,1}\right)$ is a standard Gaussian random variable such that $Q\left(x\right)\mathrm{=Pr}\left(x\mathrm{<}g\right)$,which clearly demonstrates that $\mathrm{Pr}\left(error\right)$ depends only on the distance $\left|a\mathrm{-}b\right|$ of \textit{a} and \textit{b}. 

 The exponential decay of $\mathrm{Pr}\left(error\right)$ is depicted in \fref{fig5}.

   \begin{center}
\begin{figure}[h!]
\vspace{-0.4cm}
\begin{center}
\includegraphics[angle = 0,width=0.7\linewidth]{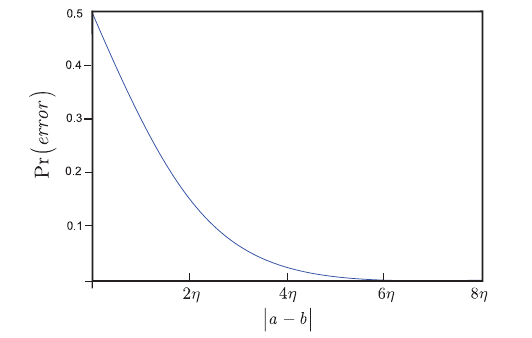}
\caption{The error probability of the scalar reconciliation process. It converges exponentially to zero as $\left|a\mathrm{-}b\right|\mathrm{>2}\eta $.} 
 \label{fig5}
 \end{center}
\end{figure}
\end{center}

 The condition $\left|a\mathrm{-}b\right|\mathrm{>2}\eta $ can be trivially satisfied by the parties in any practical CVQKD scenario; the proof is concluded here.  
\end{proof}
 
\section{Numerical Evidence and Noise Model}
\label{sec5}

\subsection{Reconciliation Characteristics}

 In this section, we analyze the performance of the proposed reconciliation for Gaussian modulation, in terms of secret key rates (bits/pulse) and distances. The excess noise $\mathfrak{N}$ of the Gaussian quantum channel is expressed as 
\begin{equation} \label{117)} 
\mathfrak{N}\mathrm{=}\left({\sigma }^{\mathrm{2}}_{{\omega }_E}\mathrm{-}\mathrm{1}\right)\left(\mathrm{1-}T\right)T^{\mathrm{-}\mathrm{1}},                                     
\end{equation} 
where $T$ is the transmission, and ${\sigma }^{\mathrm{2}}_{{\omega }_E}$ is Eve's modulation variance [\cref{r1}]. 

 Assuming reconciliation efficiency $\mathrm{0}\mathrm{\le }\beta \mathrm{\le }\mathrm{1}$, the key rate can be rewritten as
\begin{equation} \label{118)} 
R\mathrm{=}\beta I\left(A\mathrm{:}B\right)\mathrm{-}\chi \left(B\mathrm{:}E\right),                                      
\end{equation} 
where $I\left(A\mathrm{:}B\right)$ is the mutual information between Alice and Bob, while $\chi \left(B\mathrm{:}E\right)$ is the Holevo information between Bob and Eve, respectively, with relation
\begin{equation} \label{119)} 
\chi \left(B\mathrm{:}E\right)\mathrm{<}\chi \left(A\mathrm{:}E\right),                                           
\end{equation} 
where $\chi \left(A\mathrm{:}E\right)$ is the Holevo information between Alice and Eve at a direct reconciliation [\cref{r1}-\cref{r13}].

 At a given SNR, the mutual information of Alice and Bob is [\cref{r1}-\cref{r8}] 
\begin{equation} \label{120)} 
\chi \left(A\mathrm{:}B\right)\mathrm{\ge }{\mathrm{1}}/{\mathrm{2}}\mathrm{lo}{\mathrm{g}}_{\mathrm{2}}\left(\mathrm{1+SNR}\right),                                   
\end{equation} 
where
\begin{equation} \label{121)} 
\mathrm{SNR=}{{\sigma }^{\mathrm{2}}_{\phi }}/{{\sigma }^{\mathrm{2}}_{{\mathcal{N}}_{\mathrm{2}}}},                                               
\end{equation} 
where ${\sigma }^{\mathrm{2}}_{\phi }$ is the transmit signal's variance, ${\sigma }^{\mathrm{2}}_{{\mathcal{N}}_{\mathrm{2}}}$ is the variance of ${\mathcal{N}}_{\mathrm{2}}$, which has parameters that can be calculated from\textit{ T} and $\mathfrak{N}$. 

\begin{center}
\begin{figure}[h!]
\begin{center}
\includegraphics[angle = 0,width=1\linewidth]{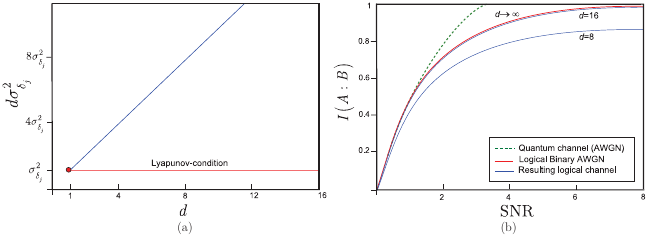}
\caption{(\textbf{a}): The SNR of the resulting logical binary channel is maximized by the Lyapunov-condition (red line). It makes possible to convert the physical Gaussian quantum channel to a logical channel with the same noise variance for arbitrary \textit{d}. For the blue line the Lyapunov-condition is not satisfied. (\textbf{b}): The capacity of the logical channel for various dimensions. At low SNRs the capacity of the physical Gaussian quantum channel (dashed line) coincidences with the capacity of the binary Gaussian channel (red). For $d\mathrm{=16}$, the capacity of the logical channel is very close to the capacity of a binary Gaussian channel, and at low SNRs it perfectly coincidences with the capacity of the Gaussian quantum channel. The reconciliation efficiency at $d\mathrm{=16}$ is $\beta \mathrm{=0.9}\mathrm{7}$. The curves for lower \textit{d}-s do not exist because the resulting logical channels are not Gaussian, since the Lyapunov-condition is not satisfied in the low-regimes.} 
 \label{fig6}
 \end{center}
\end{figure}
\end{center}

 In \fref{fig6}(a) the $d{\sigma }^{\mathrm{2}}_{{\delta }_j}$ quantities of the converted logical binary Gaussian channel for various dimensions are shown. As depicted by the red line, the Lyapunov-condition can be exploited to get variance 
\begin{equation} \label{122)} 
\mathop{\mathrm{lim}}_{{N}/{d}\mathrm{\to }\mathrm{\infty }}d\mathrm{var}\left[{\delta }_{\mathrm{0\dots }{N}/{d}}\right]\mathrm{=var}\left[{\delta }_{\mathrm{0\dots }{N}/{d}}\right]\mathrm{\approx }{\left({\sigma }^{\mathrm{2}}_{{\mathcal{N}}_{\mathrm{2}}}\right)}_d 
\end{equation} 
for arbitrary \textit{d} to maximize the 
\begin{equation} \label{123)} 
\mathrm{SNR=}{{\sigma }^{\mathrm{2}}_X}/{{\sigma }^{\mathrm{2}}_{{\delta }_j}} 
\end{equation} 
of the converted logical channel. 

 As depicted in \fref{fig6}(b), for $d\mathrm{\to }\mathrm{\infty }$, the efficiency converges to one, $\beta \mathrm{\to }\mathrm{1}$, because the noise perfectly converges to a zero-mean Gaussian random variable.

 The numerical analysis uses a PM-RR two-way CVQKD protocol, with homodyne measurements. The parameters are as follows. Excess noise $\mathfrak{N}\mathrm{=0.015}$, $T\mathrm{=0.8}$, variance ${\sigma }^{\mathrm{2}}_X\mathrm{=1.06}$, channel correlation $n_C\mathrm{=0.5}$, which parameter describes the correlation of the Gaussian attacks of Eve in the range of $\mathrm{0}\mathrm{\le }n_C\mathrm{\le }\mathrm{1}$ [\cref{r7}, \cref{r8}]. (\textit{Note}: If $n_C\mathrm{=0}$, there is no correlation between her attacks of ${\mathcal{N}}_{\mathrm{1}}$ and ${\mathcal{N}}_{\mathrm{2}}$). 

\begin{center}
\begin{figure}[h!]
\begin{center}
\includegraphics[angle = 0,width=0.65\linewidth]{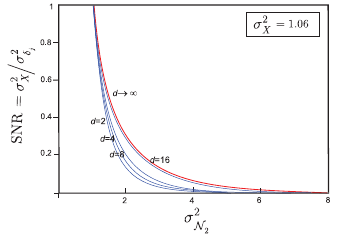}
\caption{The SNRs of the logical channel at variance ${\sigma }^{\mathrm{2}}_X\mathrm{=1.06}$. As the dimension increases the variance of the logical channel reaches the variance of the physical quantum channel. At $d\mathrm{=16}$ the variances perfectly coincidence.} 
 \label{fig7}
 \end{center}
\end{figure}
\end{center}

 In \fref{fig7} the SNR of the logical binary Gaussian are depicted for various dimensions.
 
\begin{center}
\begin{figure}[h!]
\begin{center}
\includegraphics[angle = 0,width=0.65\linewidth]{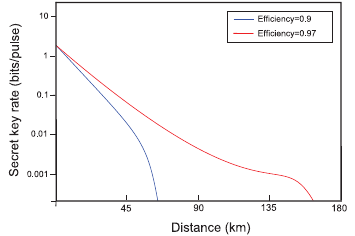}
\caption{The performance of scalar reconciliation in two-way PM-RR CVQKD at $d\mathrm{=16}$ (homodyne measurement at both sides). Excess noise: $\mathfrak{N}\mathrm{=0.015}$, transmittance: $T\mathrm{=0.8}$, Eve's variance ${\sigma }^{\mathrm{2}}_{{\omega }_E}\mathrm{=1.06}$, channel correlation: $n_C\mathrm{=0.5}$, signal variance ${\sigma }^{\mathrm{2}}_{\phi }\mathrm{=2}0$.} 
 \label{fig8}
 \end{center}
\end{figure}
\end{center}

 The performance of scalar reconciliation is summarized in \fref{fig8}. The performance of the simulated protocol without scalar reconciliation with reconciliation efficiency $\beta \mathrm{=0.9}$, is depicted by the blue curve [\cref{r7}, \cref{r8}]. At $d\mathrm{=16}$, improved the reconciliation efficiency to $\beta \mathrm{=0.97}$, which resulted in significantly higher transmission distances and secret key rates.

 The scalar reconciliation applied on the two-way CVQKD protocol resulted in approximately 160 km of achievable transmission distance (for the computations of the secret key rate, and the detection parameters see the derivations of [\cref{r1}], and [\cref{r7}, \cref{r8}]). The results indicate that the range of the current two-way CVQKD without our post-processing technique can be significantly extended, and the maximal 80.5 km range of the current one-way CVQKD systems [\cref{r12}] can be doubled, and almost tripled compared with existing two-way CVQKD systems [\cref{r7}, \cref{r8}]. The reason behind the phenomenon is the possibility of the conversion of the Gaussian quantum channel to a logical binary Gaussian channel, similar to the multidimensional reconciliation approaches developed for one-way CVQKD. 

 The favorable properties of the multidimensional solutions are preserved here, however the proposed scalar reconciliation does not require any multidimensional spherical calculations [\cref{r9}-\cref{r11}] and can be extended to arbitrary high dimensions thanks to the fact that it completely eliminates the spherical operations. From the use of higher dimensions a more precise approximation of the logical binary Gaussian channel has also become available which resulted in significantly higher reconciliation efficiency in comparison to current two-way CVQKD reconciliation methods. 

 The proposed scalar reconciliation is available at low SNRs, and the transmission ranges of experimental long-distance CVQKD can significantly be improved because at low SNRs the capacity of the logical binary Gaussian channel coincidences with the capacity of the Gaussian quantum channel, and the logical channel resulted from the conversion can approximate it with arbitrary-high precision.
 
\subsection{Noise Analysis}

\subsubsection{Noise on the Raw Data}

 The following example demonstrates the change of behavior of the probability distribution of raw data units and the CDF-transformed units, and serves only demonstration purposes. 

 For an illustrative example, let $N=1000$ units, the amount of sample raw data units $X_{j,i}$, ${{{X}'_{j,i}}}$ (the units are resulted from random quadrature measurements) taken from Alice's and Bob's raw data, respectively. The Gaussian random units $X_{j,i}$ are characterized with zero mean and variance ${\sigma }^{\mathrm{2}}_X\mathrm{=100}$. 

\begin{center}
\begin{figure}[h!]
\vspace{-0.4cm}
\begin{center}
\includegraphics[angle = 0,width=1\linewidth]{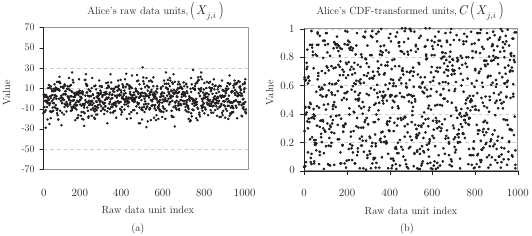}
\caption{(\textbf{a})\textbf{ }The distribution of Alice's raw data units. The units follow Gaussian random distribution. (\textbf{b}) The distribution of the CDF-transformed units. The probability distribution has changed into uniform in the range of $\left[\mathrm{0,1}\right]$.} 
 \label{fig9}
 \end{center}
\end{figure}
\end{center}

 In \fref{fig9}(a) the distribution of the $X_{j,i}$ Gaussian random raw data units is shown. \fref{fig9}(b) depicts the result of the $C\left(\mathrm{\cdot }\right)$ Gaussian CDF function applied on $X_{j,i}$. The Gaussian random behavior is eliminated and is changed into uniform.

\begin{center}
\begin{figure}[h!]
\begin{center}
\includegraphics[angle = 0,width=0.65\linewidth]{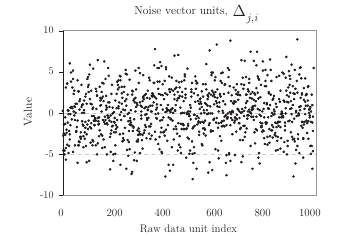}
\caption{The distribution of the units of the noise vector of the Gaussian quantum channel. The noise affects the combined state in the phase space and the resulting raw data units on Bob's side.} 
 \label{fig10}
 \end{center}
\end{figure}
\end{center}

 The distribution of the Gaussian noise vector ${\mathrm{\Delta }}_{j,i}\mathrm{\in }\mathcal{N}\left(0,{\sigma }^{\mathrm{2}}_{{\mathcal{N}}_{\mathrm{2}}}\right)$ of the quantum channel ${\mathcal{N}}_{\mathrm{2}}$, at  ${\sigma }^{\mathrm{2}}_{{\mathcal{N}}_{\mathrm{2}}}\mathrm{=4}$ is shown in \fref{fig10}. 

\begin{center}
\begin{figure}[h!]
\begin{center}
\includegraphics[angle = 0,width=1\linewidth]{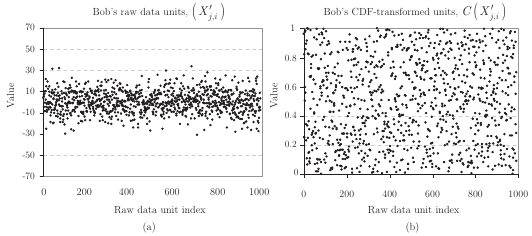}
\caption{(\textbf{a}) The distribution of the noisy raw data units on Bob's side. (\textbf{b}) The CDF-transformed raw data units have uniform distribution in $\left[\mathrm{0,1}\right]$.} 
 \label{fig11}
 \end{center}
\end{figure}
\end{center}

 At Bob's side, the received noisy units ${{{X}'_{j,i}}}$ and the CDF-transformed $C\left({{{X}'_{j,i}}}\right)$ units have a modified distribution with variance ${\sigma }^{\mathrm{2}}_{X\mathrm{'}}\mathrm{=}{\sigma }^{\mathrm{2}}_X\mathrm{+}{\sigma }^{\mathrm{2}}_{{\mathcal{N}}_{\mathrm{2}}}\mathrm{=104}$, as depicted in \fref{fig11}. The Gaussian noise on the units is added by ${\mathrm{\Delta }}_{j,i}\mathrm{\in }\mathcal{N}\left(0,{\sigma }^{\mathrm{2}}_{{\mathcal{N}}_{\mathrm{2}}}\right)$.

 This example showed that the uniform distribution of the Gaussian random raw data can be achieved by trivial operations, without any multidimensional calculations or coding.

\subsubsection{Noise on the Random Secret}

 This example demonstrates that the noise ${\delta }_j\mathrm{=}\frac{\sum^{d\mathrm{-}\mathrm{1}}_{i\mathrm{=0}}{C\left({\mathrm{\Delta }}_{j,i}\right)U_{j,i}}}{\sum^{d\mathrm{-}\mathrm{1}}_{i\mathrm{=0}}{C\left(X_{j,i}\right)}}$ on the secret ${{{U}'_{j}}}\mathrm{=}\sum^{d\mathrm{-}\mathrm{1}}_{i\mathrm{=0}}{{{{U}'_{j,i}}}}$ is inherited from the Gaussian quantum channel and by applying the Central Limit Theorem (CLT), the noise of the logical binary channel can be approximated by a Gaussian random variable ${\delta }_j\mathrm{=}\sum^{d\mathrm{-}\mathrm{1}}_{i\mathrm{=0}}{{\delta }_{j,i}}\mathrm{\in }\mathcal{N}\left(0,{\sigma }^{\mathrm{2}}_{{\delta }_j}\right)$. 

 Let $N=1000\ $ units, the amount of sample raw data units $X_{j,i}$, ${{{X}'_{j,i}}}$. The quantity $C\left({\mathrm{\Delta }}_j\right)\\\mathrm{=}C\left({{{\boldsymbol{\mathrm{X}}}'_{j}}}\right)\mathrm{-}C\left({\boldsymbol{\mathrm{X}}}_j\right)$ measures the difference of $C\left({{{\boldsymbol{\mathrm{X}}}'_{j}}}\right)$ and $C\left({\boldsymbol{\mathrm{X}}}_j\right)$, i.e., the noise of Bob's CDF-transformed data. Let ${\boldsymbol{\mathrm{X}}}_j\mathrm{\in }\mathcal{N}\left(0,{\sigma }^{\mathrm{2}}_X\mathrm{=100}\right)$ and ${{{\boldsymbol{\mathrm{X}}}'_{j}}}\mathrm{\in }\mathcal{N}\left(0,{\sigma }^{\mathrm{2}}_{X\mathrm{'}}\mathrm{=104}\right)$. The example uses an $d\mathrm{=16}$ dimensional approximation. 

\begin{center}
\begin{figure}[h!]
\vspace{-0.4cm}
\begin{center}
\includegraphics[angle = 0,width=0.65\linewidth]{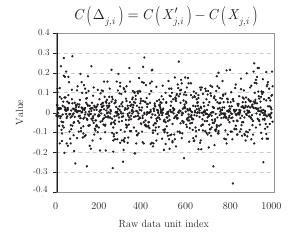}
\caption{The distribution of the error $C\left({\Delta}_{j,i}\right)\mathrm{=}C\left({{{X}'_{j,i}}}\right)\mathrm{-}C\left(X_{j,i}\right)$ on the CDF-transformed raw data units.} 
 \label{fig12}
 \end{center}
\end{figure}
\end{center}

 The distribution of the error $C\left({\mathrm{\Delta }}_{j,i}\right)$ of the CDF-transformed raw data units $C\left({{{X}'_{j,i}}}\right)$, $C\left(X_{j,i}\right)$ are depicted in \fref{fig12}. 

\begin{center}
\begin{figure}[h!]
\begin{center}
\includegraphics[angle = 0,width=1\linewidth]{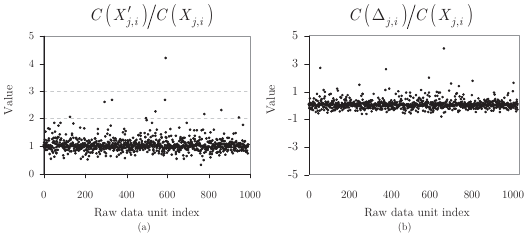}
\caption{(\textbf{a}) The distribution of the ratio of the raw data level noise and Alice's CDF-transformed raw data units. It equals to 1 for a noise-free case. (\textbf{b}) The distribution of quantity ${C\left({\Delta}_{j,i}\right)}/{C\left(X_{j,i}\right)}$.} 
 \label{fig13}
 \end{center}
\end{figure}
\end{center}

The ratio ${C\left({{{X}'_{j,i}}}\right)}/{C\left(X_{j,i}\right)}$ of the CDF-transformed units is shown in \fref{fig13}(a). In the ideal (noise-free) case the ratio equals to 1. In \fref{fig13}(b) the distribution of the quantity ${C\left({\Delta}_{j,i}\right)}/{C\left(X_{j,i}\right)}$ is shown. 

\begin{center}
\begin{figure}[h!]
\begin{center}
\includegraphics[angle = 0,width=1\linewidth]{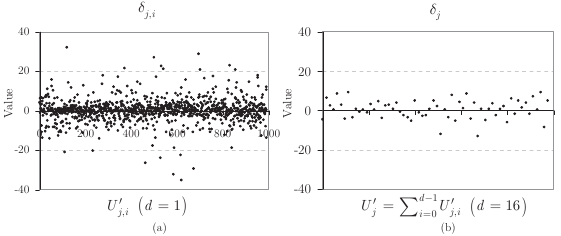}
\caption{(\textbf{a}) The distribution of the unit-level noise ${\delta }_{j,i}$ on ${{{U}'_{j,i}}}$, $U_{j,i}\mathrm{\in }\left\{\mathrm{-}\mathrm{25,25}\right\}$, ${\sigma }^{\mathrm{2}}_X\mathrm{=100}$, ${\sigma }^{\mathrm{2}}_{X\mathrm{'}}\mathrm{=10}\mathrm{4}$. (\textbf{b}) The noise ${\delta }_j\mathrm{=}\sum^{d\mathrm{-}\mathrm{1}}_{i\mathrm{=0}}{{\delta }_{j,i}}\mathrm{\in }\mathcal{N}\left(0,{\sigma }^{\mathrm{2}}_{{\delta }_j}\right)$ on ${{{U}'_{j}}}\mathrm{=}\sum^{d\mathrm{-}\mathrm{1}}_{i\mathrm{=0}}{{{{U}'_{j,i}}}}$ at $d\mathrm{=16}$. The precision of the physical-binary channel conversion gets closer to perfect as $d\mathrm{\to }\mathrm{\infty }$.} 
 \label{fig14}
 \end{center}
\end{figure}
\end{center}

 In \fref{fig14}(a) the distribution of noise ${\delta }_{j,i}$ on units ${{{U}'_{j,i}}}$ is shown, assuming that Bob selects $U_{j,i}\mathrm{\in }\left\{{\mathrm{-}\mathrm{400}}/{\mathrm{16}},{\mathrm{400}}/{\mathrm{16}}\right\}$. 

 In \fref{fig14}(b) the distribution of ${\delta }_j$ on ${{{U}'_{j}}}$, using $U_j\mathrm{=}\sum^{d\mathrm{-}\mathrm{1}}_{i\mathrm{=0}}{U_{j,i}}\mathrm{\in }\left\{\mathrm{-}\mathrm{400,400}\right\}$ is depicted. The distribution of ${\delta }_j$ is given by the formula of $\mathcal{N}\left(0,{\sigma }^{\mathrm{2}}_{{\delta }_j}\right)$, and the approximation of the binary Gaussian logical channel is justified by the CLT and the Lyapunov-condition.

 The results make it possible to achieve a high-precision conversion of the physical Gaussian quantum channel into a logical binary Gaussian channel. Precisely, only an approximation is possible by the logical layer manipulations, which gets closer to perfect as $d\mathrm{\to }\mathrm{\infty }$. At $d\mathrm{=16}$ the approximation is almost perfect, and the noise on ${{{U}'_{j}}}\mathrm{=}\sum^{d\mathrm{-}\mathrm{1}}_{i\mathrm{=0}}{{{{U}'_{j,i}}}}$ is a real Gaussian noise $\mathcal{N}\left(0,{\sigma }^{\mathrm{2}}_{{\delta }_j}\right)$.

 \section{Conclusions}
\label{sec6}
 The CVQKD protocols are based on Gaussian modulation, and powerful post-processing is needed to maximize the extractable valuable information from the correlated raw data. The physical layer solutions for the reconciliation of Gaussian variables require tomography that is intractable in a practical CVQKD scenario. The reconciliation is also possible in the level of the logical layer by a classical authenticated communication channel and by traditional algorithmical tools. The multidimensional approaches were developed for this purpose, however the use of complex multidimensional calculations is also not desirable in a practical scenario. The proposed scalar reconciliation eliminates the use of multidimensional spherical space along with the dimensional boundaries. The scalar reconciliation process neither requires any physical-layer tomography, and only standard operations and calculations needed in the level of raw data. The method provides an easy implementation to maximize the extractable valuable binary information from the correlated raw data to significantly boost up the key rates and to improve the distance ranges of CVQKD.

\section*{Acknowledgements}
This work was partially supported by the GOP-1.1.1-11-2012-0092 project sponsored by the EU and European Structural Fund, by the Hungarian Scientific Research Fund - OTKA K-112125, by the COST Action MP1006, and by the National Research Development and Innovation Office of Hungary (Project No. 2017-1.2.1-NKP-2017-00001). 

\section*{References}
\begin{enumerate}[ {[}1{]} ]
\item \label{r1} S. Pirandola, S. Mancini, S. Lloyd and S. L. Braunstein, Continuous-variable quantum cryptography using two-way quantum communication, \textit{Nature Phys.} 4~726, (2008).

\item \label{r2} S. Pirandola, R. Garcia-Patron, S. L. Braunstein and S. Lloyd, Direct and Reverse Secret-Key Capacities of a Quantum Channel, \textit{Phys. Rev. Lett.} 102 050503. (2009)

 \item \label{r3} S. Pirandola, A. Serafini and S. Lloyd, Correlation matrices of two-mode bosonic systems, \textit{Phys. Rev. A} 79 052327. (2009).

\item \label{r4} S. Pirandola, S. L. Braunstein and S. Lloyd, Characterization of Collective Gaussian Attacks and Security of Coherent-State Quantum Cryptography, \textit{Phys. Rev. Lett. }101 200504 (2008).

\item \label{r5} C. Weedbrook, S. Pirandola, S. Lloyd and T. Ralph, Quantum Cryptography Approaching the Classical Limit, \textit{Phys. Rev. Lett.} 105 110501 (2010).

\item \label{r6} C. Weedbrook, S. Pirandola, R. Garcia-Patron, N. J. Cerf, T. Ralph, J. Shapiro, and S. Lloyd, Gaussian quantum information, \textit{Rev. Mod. Phys.} 84, 621 (2012).

\item \label{r7} M. Sun, X. Peng, Y. Shen, H. Guo, Security of a new two-way continuous-variable quantum key distribution protocol,\textit{ Int. J. Quant. Inf.} 10 1250059 (2012).

\item \label{r8} M. Sun, Xiang Peng and Hong Guo, An improved two-way continuous-variable quantum key distribution protocol with added noise in homodyne detection, \textit{J. Phys. B: At. Mol. Opt. Phys.} 46 085501 (2013)

\item \label{r9} P. Jouguet, S. Kunz-Jacques, and A. Leverrier, Long-distance continuous-variable quantum key distribution with a Gaussian modulation, \textit{Phys. Rev. A} 84, 062317 (2011).

\item \label{r10} P. Jouguet, S. Kunz-Jacques, E. Diamanti, and A. Leverrier, Analysis of imperfections in practical continuous-variable quantum key distribution, \textit{Phys. Rev. A} 86, 032309 (2012).

\item \label{r11} A. Leverrier, R. Alleaume, J. Boutros, G. Zemor, and P. Grangier, Multidimensional reconciliation for a continuous-variable quantum key distribution, \textit{Phys. Rev. A} 77, 042325 (2008).

\item \label{r12} P. Jouguet, S. Kunz-Jacques, A. Leverrier,  P. Grangier, E. Diamanti, Experimental demonstration of long-distance continuous-variable quantum key distribution, arXiv:1210.6216v1 (2012).

\item \label{r13} A. Leverrier, R. Garcia-Patron, R. Renner, and N. J. Cerf, Security of continuous-variable quantum key distribution against general attacks, \textit{Phys. Rev. Lett.} 110, 030502 (2013).

\item \label{r14} S. Imre and L. Gyongyosi. \textit{Advanced Quantum Communications - An Engineering Approach}. Wiley-IEEE Press (StateNew Jersey, USA), (2012).

\item \label{r15} D. Tse and P. Viswanath. \textit{Fundamentals of Wireless Communication}, Cambridge University Press, (2005).

\item \label{r16} J. Hamkins and K. Zeger. Asymptotically efficient spherical codes---Part I: Wrapped spherical codes, \textit{IEEE Trans. Inform. Theory}, vol. 43, pp. 1774--1785, (1997).

\item \label{r17} P. F. Swaszek and J. B. Thomas. Multidimensional spherical coordinates quantization, \textit{IEEE Trans. Inform. Theory}, vol. IT-29, pp. 570--576, (1983).

\item \label{r18} K. Miller. \textit{Multidimensional Gaussian Distributions}. New York: Wiley, (1964).

\item \label{r19} J. Hamkins. \textit{Design and analysis of spherical codes}, Ph.D. dissertation, Univ. Illinois at Urbana-Champaign, (1996).

\item \label{r20} J. H. Conway and D. A. Smith, \textit{On Quaternions and Octonions: Their Geometry, Arithmetic, and Symmetry}, A K Peters/CRC Press, (2003).

\item \label{r21} T. Richardson and R. Urbanke, \textit{Modern Coding Theory}, (Cambridge University Press, New York, NY, USA), (2008).

\item \label{r22} A. Gersho. Asymptotically optimal block quantization, \textit{IEEE Trans. Inform. Theory}, vol. IT-25, pp. 373--380, (1979).

\item \label{r23} D. J. Sakrison. A geometric treatment of the source encoding of a Gaussian random variable, \textit{IEEE Trans. Inform. Theory}, vol. IT-14, pp. 481--486, (1968).

\item \label{r24} J. Hamkins and K. Zeger. Gaussian Source Coding With Spherical Codes, \textit{IEEE Trans. Inform. Theory}, vol. 48, no 11, (2002).

\item \label{r25} L. Hanzo, H. Haas, S. Imre, D. O'Brien, M. Rupp, L. Gyongyosi. Wireless Myths, Realities, and Futures: From 3G/4G to Optical and Quantum Wireless, \textit{Proceedings of the IEEE}, Volume: 100, \textit{Issue: Special Centennial Issue}, pp. 1853-1888. (2012).

\item \label{r26} J. Rice. \textit{Mathematical Statistics and Data Analysis} (Second ed.), Duxbury Press, ISBN 0-534-20934-3) (1995).

\item \label{r27} Botsinis, Panagiotis, Alanis, Dimitrios, Ng, Soon Xin and Hanzo, Lajos  Low-Complexity Soft-Output Quantum-Assisted Multi-User Detection for Direct-Sequence Spreading and Slow Subcarrier-Hopping Aided SDMA-OFDM Systems. \textit{IEEE Access}, PP, (99) (2014).

\item \label{r28} Botsinis, Panagiotis, Ng, Soon Xin and Hanzo, Lajos Fixed-complexity quantum-assisted multi-user detection for CDMA and SDMA. \textit{IEEE Transactions on Communications}, vol. 62, (no. 3), pp. 990-1000  (2014).

\item \label{r29} L. Gyongyosi, S. Imre, Geometrical Analysis of Physically Allowed Quantum Cloning Transformations for Quantum Cryptography, \textit{Information Sciences, }Elsevier, DOI: 10.1016/j.ins.2014.07.010 (2014).

\item \label{r30} L. Gyongyosi, S. Imre: Algorithmic Superactivation of Asymptotic Quantum Capacity of Zero-Capacity Quantum Channels, \textit{Information Sciences}, Elsevier, ISSN: 0020-0255; (2011).

\item \label{r31} L. Gyongyosi, S. Imre: Superactivation of Quantum Channels is Limited by the Quantum Relative Entropy Function, \textit{Quantum Information Processing}, Springer, ISSN: 1570-0755, ISSN: 1573-1332, (2012).

\item \label{r32} L. Gyongyosi, S. Imre, Adaptive multicarrier quadrature division modulation for long-distance continuous-variable quantum key distribution, \textit{Proc. SPIE 9123, Quantum Information and Computation XII}, 912307; doi:10.1117/12.2050095, From Conference Volume 9123, Quantum Information and Computation XII, Baltimore, Maryland, USA (2014).

\item \label{r33} S. Imre, F. Balazs: \textit{Quantum Computing and Communications -- An Engineering Approach}, John Wiley and Sons Ltd, ISBN 0-470-86902-X, 283 pages (2005).

\item \label{r34} D. Petz, \textit{Quantum Information Theory and Quantum Statistics}, Springer-Verlag, Heidelberg, Hiv: 6. (2008).

\item \label{r35} R. V. Meter,\textit{Quantum Networking}, John Wiley and Sons Ltd, ISBN 1118648927, 9781118648926 (2014).

\item \label{r36} L. Ruppert, V. C. Usenko, R. Filip, Long-distance continuous-variable quantum key distribution with efficient channel estimation, \textit{Physical Review A} 90, 062310 (2014).

\item \label{r37} Lloyd, S. Capacity of the noisy quantum channel. \textit{Physical Rev. A 55}, 1613–1622 (1997).

\item \label{r38} R. Renner and J. I. Cirac , de Finetti Representation Theorem for Infinite-Dimensional Quantum Systems and Applications to Quantum Cryptography, \textit{Physcal Review Letters }102, 110504 (2009). 

\item \label{r39} F. Furrer, T. Franz, M. Berta, A. Leverrier, V. B. Scholz, M. Tomamichel, and R. F. Werner, Continuous Variable Quantum Key Distribution: Finite-Key Analysis of Composable Security against Coherent Attacks, \textit{Physcal Review Letters} 109, 100502 (2012). 

\item \label{r40} A. Leverrier, Composable Security Proof for Continuous-Variable Quantum Key Distribution with Coherent States, \textit{Physcal Review Letters} 114, 070501 (2015).

\item \label{r41} G. Van Assche, J. Cardinal, N. J. Cerf, Reconciliation of a quantum-distributed Gaussian key, \textit{IEEE Transactions on Information Theory} 50, 394 (2004).

\item \label{r42} A. Leverrier, P. Grangier, Continuous-variable quantum-key-distribution protocols with a non-Gaussian modulation, \textit{Physical Review A} 83, 042312 (2011).

\item \label{r43} D. Zwillinger, S. Kokoska,  \textit{Standard Probability and Statistics Tables and Formulae}, CRC Press. ISBN 978-1-58488-059-2 (2010).

\item \label{r44} J. E. Gentle, \textit{Computational Statistics}, Springer. ISBN 978-0-387-98145-1. Retrieved 2010-08-06 (2009).

\item \label{r45} P. Billingsley, \textit{Probability and Measure} (Third ed.), John Wiley \& sons, ISBN 0-471-00710-2 (1995).

\item \label{r46} A. Leverrier, P. Grangier, Unconditional Security Proof of Long-Distance Continuous-Variable Quantum Key Distribution with Discrete Modulation, \textit{Physical Review Letters} 102, 180504 (2009).

\item \label{r47} L. Gyongyosi, Improved Long-Distance Two-way Continuous Variable Quantum Key Distribution over Optical Fiber, \textit{2013 Frontiers in Optics/Laser Science XXIX (FiO/LS)}, 6-10 October 2013, Orlando, Florida, USA.

\item \label{r48} L. Gyongyosi, S. Imre, Proceedings Volume 8997, Advances in Photonics of Quantum Computing, Memory, and Communication VII; 89970C; doi: 10.1117/12.2038532 (2014).

\item \label{r49} Bennett, C. H., Brassard, G., Quantum Cryptography: public key distribution and coin tossing, \textit{Proceeding of IEEE International Conference on Computer, Systems Signal Processing} 175--179  (1984). 

\item \label{r50} Scarani, V. et al., The security of practical quantum key distribution. \textit{Rev. Mod. Phys.} 81, 1301-1350 (2009).

\item \label{r51} Inoue, K., Waks, E. \& Yamamoto, Y. Differential-phase-shift quantum key distribution using coherent light. \textit{Phys. Rev. A} 68, 022317 (2003).

\item \label{r52} Stucki, D., Brunner, N., Gisin, N., Scarani, V. \& Zbinden, H. Fast and simple one-way quantum key distribution. \textit{Appl. Phys. Lett.} 87, 194108 (2005).

\item \label{r53} Bacco D. et al., Two-dimensional distributed-phase-reference protocol for quantum key distribution, \textit{Sci. Reports} 6:36756 (2016).
\end{enumerate}

\newpage
\appendix
\setcounter{table}{0}
\setcounter{figure}{0}
\setcounter{equation}{0}
\setcounter{algocf}{0}
\renewcommand{\thetable}{\Alph{section}.\arabic{table}}
\renewcommand{\thefigure}{\Alph{section}.\arabic{figure}}
\renewcommand{\theequation}{\Alph{section}.\arabic{equation}}
\renewcommand{\thealgocf}{\Alph{section}.\arabic{algocf}}

\setlength{\arrayrulewidth}{0.1mm}
\setlength{\tabcolsep}{5pt}
\renewcommand{\arraystretch}{1.5}
\section{Appendix}
\subsection{Spherical Code}
 A \textit{d}-dimensional spherical code $\mathcal{X}$ is defined over the \textit{d}-dimensional unit sphere ${\mathrm{\Gamma }}^{d\mathrm{-}\mathrm{1}}$, given by ${\mathrm{\Gamma }}^{d\mathrm{-}\mathrm{1}}\mathrm{=}\left(x\mathrm{=}\left(x_0,x_{\mathrm{1}}\mathrm{,\dots ,}x_{d\mathrm{-}\mathrm{1}}\right)\mathrm{\in }{\mathbb{R}}^d\mathrm{:}\left\|x\right\|\mathrm{=1}\right)$, and $\left\|x\right\|\mathrm{=1}$ is the unit norm. The $\left(d\mathrm{-}\mathrm{1}\right)$-dimensional surface $\mathrm{S}\left({\mathrm{\Gamma }}^{d\mathrm{-}\mathrm{1}}\right)$ of ${\mathrm{\Gamma }}^{d\mathrm{-}\mathrm{1}}$ is defined as $\mathrm{S}\left({\mathrm{\Gamma }}^{d\mathrm{-}\mathrm{1}}\right)\mathrm{=}{\mathrm{2}{\pi }^{d\mathrm{/2}}}/{\mathcal{G}\left({d}/{\mathrm{2}}\right)}$, where $\mathcal{G}\left({d}/{\mathrm{2}}\right)\mathrm{=}\int^{\mathrm{\infty }}_0{t^{\left({d}/{\mathrm{2}}\right)\mathrm{-}\mathrm{1}}}e^{\mathrm{-}t}dt$ is the gamma function [\cref{r24}]. The number of codewords of the code is $\left|\mathcal{X}\right|$, the smallest dimension $d_{\mathrm{min}}$ of any Euclidean space for the spherical code $\mathcal{X}$ is $d_{\mathrm{min}}\mathrm{=dim}\left|\mathcal{X}\right|$, while the minimum distance between any two elements $x$ and $y$ of $\mathcal{X}\mathrm{\subseteq }{\mathrm{\Gamma }}^{d\mathrm{-}\mathrm{1}}$, $x\mathrm{\neq }y$,  is $D\mathrm{=min}\left\{{\left\|x\mathrm{-}y\right\|}^{\mathrm{2}}\right\}.$
 
\subsection{Gaussian Random Spherical Vectors}
Let $\mathfrak{X}={\left(X_0\mathrm{,\dots ,}X_{d\mathrm{-}\mathrm{1}}\right)}^T\mathrm{\in }{\mathbb{R}}^d$ be a Gaussian random vector with independent components, and with norm $\left\|\mathfrak{X}\right\|$ drawn from an $\mathcal{N}\left(0,{\sigma }^{\mathrm{2}}\right)$ memoryless Gaussian source. Over the \textit{d}-dimensional unit sphere ${\mathrm{\Gamma }}^{d\mathrm{-}\mathrm{1}}$, spherical Gaussian random vector $\mathbb{E}\left[\left\|\mathfrak{X}\right\|\right]\left({\mathfrak{X}}/{\left\|\mathfrak{X}\right\|}\right)\mathrm{\in }{\mathrm{\Gamma }}^{d\mathrm{-}\mathrm{1}}\mathrm{\in }{\mathbb{R}}^d$ has radius $r\mathrm{=}\mathbb{E}\left\|\mathfrak{X}\right\|$, where $\mathbb{E}$ is the mean of the norm $\left\|\mathfrak{X}\right\|$, defined [\cref{r24}] as
\begin{equation} \label{1)} 
\mathbb{E}\left[\left\|\mathfrak{X}\right\|\right]\mathrm{=}\frac{\sqrt{\mathrm{2}{\sigma }^{\mathrm{2}}}\mathcal{G}\left(\frac{d\mathrm{+1}}{\mathrm{2}}\right)}{\mathcal{G}\left(\frac{d}{\mathrm{2}}\right)}\mathrm{=}\frac{\sqrt{\mathrm{2}\pi {\sigma }^{\mathrm{2}}}}{\beta \left(\frac{d}{\mathrm{2}},\frac{\mathrm{1}}{\mathrm{2}}\right)},                                        
\end{equation} 
where $\beta \left(x,y\right)\mathrm{=}\frac{\mathcal{G}\left(x\right)\mathcal{G}\left(y\right)}{\mathcal{G}\left(x\mathrm{+}y\right)}$, is the beta function, while $\mathbb{E}\left[{\left\|\mathfrak{X}\right\|}^{\mathrm{2}}\right]\mathrm{=}d{\sigma }^{\mathrm{2}}$. The Gaussian random vector $\mathfrak{X}\mathrm{\in }{\mathbb{R}}^d$ over ${\mathrm{\Gamma }}^{d\mathrm{-}\mathrm{1}}$ has a probability density function
\begin{equation} \label{2)} 
f\left(\mathfrak{X}\right)\mathrm{=}\frac{\mathrm{2}r^{d\mathrm{-}\mathrm{1}}e^{\frac{\mathrm{-}r^{\mathrm{2}}}{\mathrm{2}{\sigma }^{\mathrm{2}}}}}{\mathcal{G}\left(\frac{k}{\mathrm{2}}\right){\left(\mathrm{2}{\sigma }^{\mathrm{2}}\right)}^{{k}/{\mathrm{2}}}}, 
\end{equation} 
and variance
\begin{equation} \label{3)} 
\mathrm{var}\left[\mathfrak{X}\right]\mathrm{=}d{\sigma }^{\mathrm{2}}\mathrm{-}\frac{\mathrm{2}\pi {\sigma }^{\mathrm{2}}}{{\beta }^{\mathrm{2}}\left(\frac{d}{\mathrm{2}},\frac{\mathrm{1}}{\mathrm{2}}\right)}.                                           
\end{equation} 
For $d\mathrm{\to }\mathrm{\infty }$, $\mathbb{E}\left\|{\mathfrak{X}}/{\sqrt{d{\sigma }^{\mathrm{2}}}}\right\|\mathrm{\to }\mathrm{1}$, and $r\mathrm{=}\mathop{\mathrm{lim}}_{d\mathrm{\to }\mathrm{\infty }}\left\|{\mathfrak{X}}/{\sqrt{d{\sigma }^{\mathrm{2}}}}\right\|\mathrm{\to }\mathrm{1}$. The distribution of $r$ approximates the Dirac distribution ${\mathcal{D}}_d\left(x\right)$, and gets to arbitrary close for $d\mathrm{\to }\mathrm{\infty }$. 

\subsection{Abbreviations}
\begin{description}
\item[AWGN] Additive White Gaussian Noise
\item[BAWGN] Binary Additive White Gaussian Noise
\item[BS] Beam Splitter
\item[BSC] Binary Symmetric Channel
\item[CDF] Cumulative Distribution Function
\item[CLT] Central Limit Theorem
\item[CV] Continuous-Variable
\item[DPR] Differential Phase Reference
\item[DV] Discrete-Variable
\item[LDPC] Low Density Parity Check
\item[PM] Prepare-and-Measure: entanglement-free protocol
\item[RR] Reverse Reconciliation
\item[SNR] Signal-to-Noise Ratio
\end{description}

\subsection{Notations}
The notations of the manuscript are summarized in  \tref{tab1}.
\small
\begin{center}
\begin{longtable}{|p{1in}|p{4.5in}|}
\caption{Summary of notations.}
\label{tab1}
\endfirsthead
\endhead
\hline
\textit{Notation} & \textit{Description} \\ \hline
$\left|\left.{\varphi }_i\right\rangle \right.$ & The first mode of the combined beam, phase space vector, expressed as\newline $\left|\left.{\varphi }_i\right\rangle \right.\mathrm{=}\left|\left.x_{A,i}\mathrm{+}{{{x}'_{B,i}}}\mathrm{+}i\left(p_{A,i}\mathrm{+}{{{p}'_{B,i}}}\right)\right\rangle \right.$,\newline where $x_{A,i},{{{x}'_{B,i}}}$ and $p_{A,i}, {{{p}'_{B,i}}}$ are the position and momentum quadratures. \\ \hline 
$\left|\left.{\phi }_i\right\rangle \right.$ & The second mode of the combined beam, phase space vector, expressed as \newline $\left|\left.{\phi }_i\right\rangle \right.\mathrm{=}\left|\left.x_{A,i}\mathrm{-}{{{x}'_{B,i}}}\mathrm{+}i\left(p_{A,i}\mathrm{-}{{{p}'_{B,i}}}\right)\right\rangle \right.$,\newline where $x_{A,i},{{{x}'_{B,i}}}$ and $p_{A,i},{{{p}'_{B,i}}}$ are the position and momentum quadratures.  \\ \hline 
$\left|\left.{\xi }_i\right\rangle \right.$ & The noisy version of phase space state $\left|\left.{\phi }_i\right\rangle \right.$, with the noisy quadratures, \newline $\left|\left.{\xi }_i\right\rangle \right.\mathrm{=}\left|\left.{{{x}'_{A,i}}}\mathrm{-}{{{x}''_{B,i}}}\mathrm{+}i\left({{{p}'_{A,i}}}\mathrm{-}{{{p}''_{B,i}}}\right)\right\rangle \right.$. \\ \hline 
$X$ & Alice's \textit{N}-unit length raw data generated by \textit{N} random quadrature measurements. Binary string, consists of ${N}/{d}$ number of \textit{d}-dimensional Gaussian random vectors, ${\boldsymbol{\mathrm{X}}}_j\mathrm{\in }{\mathbb{R}}^d$.  \\ \hline 
$X\mathrm{'}$ & Bob's \textit{N}-unit length raw data generated by \textit{N} random quadrature measurements. Binary string, consists of ${N}/{d}$ number of noisy \textit{d}-dimensional Gaussian random vectors ${{{\boldsymbol{\mathrm{X}}}'_{j}}}\mathrm{\in }{\mathbb{R}}^d$. \\ \hline 
$X_i$ & Alice's raw data \textit{unit}, obtained from a random quadrature measurement,\newline $ \begin{array}{l}
X_i\mathrm{=}x_{A,i}\mathrm{+}{{{x}'_{B,i}}}, \\ 
X_i\mathrm{=}p_{A,i}\mathrm{+}{{{p}'_{B,i}}}, \end{array}
$\newline where $x_{A,i},{{{x}'_{B,i}}}$ and $p_{A,i},{{{p}'_{B,i}}}$ are the position and momentum quadratures.  \\ \hline 
${{{X}'_{i}}}$ & Bob's noisy raw data \textit{unit}, obtained from a random quadrature measurement and by a correction $\mathrm{+2}x_{B,i}$ or $\mathrm{+2}p_{B,i}$, \newline $ \begin{array}{l}
{{{X}'_{i}}}\mathrm{=}{{{x}'_{A,i}}}\mathrm{+}{{{x}''_{B,i}}}, \\ 
{{{X}'_{i}}}\mathrm{=}{{{p}'_{A,i}}}\mathrm{+}{{{p}''_{B,i}}}, \end{array}
$\newline while ${{{x}'_{A,i}}},{{{x}''_{B,i}}}$ and ${{{p}'_{A,i}}},{{{p}''_{B,i}}}$ are the noisy position and momentum quadratures.  \\ \hline 
${\boldsymbol{\mathrm{X}}}_j$ & Alice's \textit{d}-dimensional Gaussian random \textit{vector} (\textit{d} unit length Gaussian random vector), \newline ${\boldsymbol{\mathrm{X}}}_j\mathrm{\in }{\mathbb{R}}^d\mathrm{:}\left\{X_{j,0},X_{j\mathrm{,1}}\mathrm{,\dots }X_{j,d\mathrm{-}\mathrm{1}}\right\}$,\newline where $X_{j,i}$ is a Gaussian random variable\textit{.} \\ \hline 
$X_{j,i}\mathrm{\in }\mathbb{R}$, ${{{X}'_{j,i}}}\mathrm{\in }\mathbb{R}$ & The \textit{i}-th unit of \textit{j}-th vector ${\boldsymbol{\mathrm{X}}}_j$ and ${{{\boldsymbol{\mathrm{X}}}'_{j}}}$. \\ \hline 
${{{\boldsymbol{\mathrm{X}}}'_{j}}}$ & Bob's noisy \textit{d}-dimensional Gaussian random vector (\textit{d} unit length vector), \newline ${{{\boldsymbol{\mathrm{X}}}'_{j}}}\mathrm{\in }{\mathbb{R}}^d\mathrm{:}\left\{{{{X}'_{j,0}}},{{{X}'_{j\mathrm{,1}}}}\mathrm{,\dots }{{{X}'_{j,d\mathrm{-}\mathrm{1}}}}\right\}$,\newline where ${{{X}'_{j,i}}}\mathrm{=}{{{x}'_{A,i}}}\mathrm{+}{{{x}''_{B,i}}}$ or ${{{X}'_{j,i}}}\mathrm{=}{{{p}'_{A,i}}}\mathrm{+}{{{p}''_{B,i}}}$ is a Gaussian random \textit{units} obtained from a quadrature measurement. \\ \hline 
$\boldsymbol{\mathrm{K}}$ & Bob's secret key vector,\newline $\boldsymbol{\mathrm{K}}\mathrm{=}\left\{{\boldsymbol{\mathrm{U}}}_0\mathrm{,\dots }{\boldsymbol{\mathrm{U}}}_{\left({N}/{d}\right)\mathrm{-}\mathrm{1}}\right\}\mathrm{\in }{\mathbb{R}}^{{N}/{d}}$.\newline The full key is granulated into ${N}/{d}$ number of ${\boldsymbol{\mathrm{U}}}_j\mathrm{\in }{\mathbb{R}}^d$ vectors, \newline ${\boldsymbol{\mathrm{U}}}_j\mathrm{=}\left\{U_{j,0},U_{j\mathrm{,1}}\mathrm{,\dots }U_{j,d\mathrm{-}\mathrm{1}}\right\}\mathrm{\in }{\mathbb{R}}^d,$\newline where $U_j\mathrm{\in }\left\{a,b\right\}\mathrm{\in }\mathbb{R}$. \\ \hline 
${{{\boldsymbol{\mathrm{X}}}'_{j}}}{\boldsymbol{\mathrm{U}}}_j\mathrm{\in }{\mathbb{R}}^d$ & Bob's \textit{d}-dimensional vector sent to the classical channel. \\ \hline 
${{{X}'_{j,i}}}U_{j,i}\mathrm{\in }\mathbb{R}$ & A unit of Bob's \textit{d}-dimensional message sent to the classical channel. \\ \hline 
$C\left(\mathrm{\cdot }\right)$ & The Gaussian CDF function. \\ \hline 
$\mathfrak{C}\left(\mathrm{\cdot }\right)$ & Covariance matrix. \\ \hline 
${\mathcal{D}}_d\left(\mathrm{\cdot }\right)$ & Dirac distribution of a \textit{d}-dimensional vector. \\ \hline 
$\mathfrak{L}$ & Lyapunov coefficient, $\mathfrak{L}\mathrm{\rhd }0$. \\ \hline 
${{{U}'_{j}}}$  & The noisy version of Bob's secret $U_j$,\newline ${{{U}'_{j}}}\mathrm{=}\sum^{d\mathrm{-}\mathrm{1}}_{i\mathrm{=0}}{{{{U}'_{j,i}}}}$,\newline where a unit $U_{j,i}$ is as \newline ${{{U}'_{j,i}}}\mathrm{=}\left(C\left({{{X}'_{j,i}}}\right)U_{j,i}\right)\frac{\mathrm{1}}{C\left(X_{j,i}\right)}$. \\ \hline 
${\delta }_j$\textit{, }${\delta }_{j,i}$ & Noise on ${{{U}'_{j}}}\mathrm{=}\sum^{d\mathrm{-}\mathrm{1}}_{i\mathrm{=0}}{{{{U}'_{j,i}}}}$, and on unit $U_{j,i}$. \\ \hline 
$\eta $\textit{} & Standard deviation of the noise vector ${\overrightarrow{{{{\delta }_{j}}}}}$\textit{, }$\eta =\sqrt{{\left({\sigma }^2_{{\delta }_j}\right)}_d}$\textit{.} \\ \hline 
${\mathrm{\Lambda }}_j$, ${\mathrm{\Lambda }}_{j,i}$ & Standard Gaussian random noise vector, and the noise of the \textit{i}-th unit of the \textit{j}-th block $X_{j,i}$, ${\mathrm{\Lambda }}_j\mathrm{=}\mathcal{N}{\left(\mathrm{0,1}\right)}_d\mathrm{\in }{\mathbb{R}}^d$, and ${\mathrm{\Lambda }}_{j,i}\mathrm{=}\mathcal{N}\left(\mathrm{0,1}\right)\mathrm{\in }\mathbb{R}$. \\ \hline 
${\mathrm{\Delta }}_j$ & Gaussian random noise vector of the quantum channel ${\mathcal{N}}_{\mathrm{2}}$ on ${\boldsymbol{\mathrm{X}}}_j$, ${\mathrm{\Delta }}_j\mathrm{=}\mathcal{N}{\left(0,{\sigma }^{\mathrm{2}}_{\mathrm{2}}\right)}_d\mathrm{\in }{\mathbb{R}}^d$. \\ \hline 
${\mathrm{\Delta }}_{j,i}$ & The \textit{i}-th unit of \textit{j}-th noise vector, that results raw data unit ${{{X}'_{j,i}}}\mathrm{=}X_{j,i}\mathrm{+}{\mathrm{\Delta }}_{j,i}$, where ${\mathrm{\Delta }}_{j,i}\mathrm{=}\mathcal{N}\left(0,{\sigma }^{\mathrm{2}}_{\mathrm{2}}\right)\mathrm{\in }\mathbb{R}$. \\ \hline  
\end{longtable}
\end{center}
\end{document}